\documentclass[epj]{svjour}
\pdfoutput=1
\usepackage{graphicx}
\usepackage{upgreek}
\usepackage{multirow}
\usepackage{bm}        
\usepackage{amssymb}   
\usepackage{amsmath}   
\usepackage[english=usenglishmax]{hyphsubst}
\begin{document}
\title{Dynamical measurements of deviations from Newton's $1/r^2$ law}

\author{J. Baeza-Ballesteros\inst{1,2} \and A. Donini\inst{1} \and S. Nadal-Gisbert\inst{1,2} 
}                     
\institute{Instituto de F\'{\i}sica Corpuscular, CSIC-Universitat de Val\`encia, 46980 Paterna, Spain \and Departamento de Física Teórica, Facultad de Física, Universitat de Val\`encia, 45100 Burjassot, Spain}
\date{Received: date / Revised version: date}
%
\abstract{
In Ref.~\cite{Donini:2016kgu}, an experimental setup aiming at the measurement of deviations from the Newtonian  $1/r^2$ distance dependence of gravitational interactions was proposed. The theoretical idea behind this setup was to study the trajectories of a ``Satellite'' with a mass
$m_{\rm S} \sim {\cal O}(10^{-9})$ g around a ``Planet'' with mass $m_{\rm  P} \in [10^{-7},10^{-5} ]$ g, looking for precession of the orbit. The observation of such feature induced by gravitational interactions would be an unambiguous indication of a gravitational potential with terms different from $1/r$ and, thus, a powerful tool 
to detect deviations from Newton's $1/r^2$ law. In this paper we optimize the proposed setup in order to achieve maximal sensitivity to look for such {\em Beyond-Newtonian} corrections. 
We then study in detail possible background sources that could induce precession and quantify their impact on the achievable sensitivity. 
We finally conclude that a dynamical measurement of deviations from newtonianity can test Yukawa-like corrections to 
the $1/r$ potential with strength as low as $\alpha \sim 10^{-2}$ for distances as small as $\lambda \sim 10 \, \upmu$m.
   } 
\maketitle

\section{Introduction}
\label{sec:intro}

Several experimental observations clearly point out that  the Standard Model (SM) is not the ``ultimate'' theory, but just a low-energy effective theory 
that must be extended to a more fundamental one at energies higher than those currently tested. 
This is the case of the huge amount of data that suggests the existence of some matter that gravitates but does not emit light 
(called, unambiguously, Dark Matter), of the so-called Dark Energy (responsible for the accelerated expansion of the Universe), of the observed asymmetry
between Matter and Anti-Matter, and of the origin of neutrino masses (the first compelling evidence so far for physics beyond the Standard Model).  
In an effective theory, massive particles have ``natural'' masses \cite{'tHooft:1979bh} of the order of the scale at which the theory must be replaced by a more fundamental one, 
with much lighter masses only allowed if some symmetry protects them ({\em i.e.}, if a symmetry is restored when some order parameter of the theory vanishes). 
In the case of the SM, most of the particles have masses much lighter than the scale of the electroweak symmetry breaking: this is the case of all of the fermion
masses (with the notable exception of the top quark). However, all of these particles are ``protected'' by chiral symmetry: a very large global symmetry is
restored if fermion masses vanish, thus making their smallness ``natural''. 
This is also the case of the gluon and the photon, that are massless, as they are ``protected'' by the exact SU$(3)_\text{c}$ color and U$(1)_Q$ electromagnetic gauge symmetries. 
Other particles, such as the $W$ and $Z$ bosons, on the other hand, have masses at the typical scale where the $\text{SU}(2)_\text{L}\times \text{U}(1)_Y$ symmetry is broken, 
$\mathit{\Lambda}_{\rm EW} \sim 245$ GeV.  

The existence of the Higgs boson, confirmed in the last decade at the LHC \cite{Aad:2012tfa}, 
poses a new theoretical problem, though. Its mass, $ m_{ H}=125.10\pm 0.14$ GeV \cite{Zyla:2020zbs}, is reasonably ``natural'' according to the
't Hooft naturalness criterium, {\it i.e.,} of the order of $\mathit{\Lambda}_{\rm EW}$. 
However, just by computing loop corrections to the Higgs mass we discover that scalar particles in a quantum field theory 
have the very peculiar feature that their masses get additively renormalized by quadratic divergent terms. These terms must be cut-off at an ultraviolet scale $\mathit{\Lambda}_{\text{UV}}$
and, thus, are sensitive to a scale much higher than $\mathit{\Lambda}_{\rm EW}$. Once the SM is assumed to be a low-energy effective theory, 
the Higgs boson mass should ``naturally'' be much larger than what has been measured:  as large as the scale at which the SM should be replaced by a more fundamental
theory. This is the so-called  ``hierarchy problem''. If we believe, for example, that the fundamental theory should include quantum gravity, then the Higgs mass
should be as large as the Planck mass, $M_{\rm Pl} \sim 10^{19}$ GeV. In order to recover the experimentally measured value of $m_{\rm H}$, thus, a huge cancellation between 
loop corrections from different massive particles must occur. For this reason, it is usually believed that some extension of the SM will be 
discovered at an energy not much larger than $\mathit{\Lambda}_{\rm EW}$, to minimize the amount of fine-tuning needed to keep $m_H$ at the experimental 
value after loop corrections are taken into account.

Many proposals to solve the hierarchy problem have been advanced in the last 40 years, most popular among them being Supersymmetry \cite{Dimopoulos:1981zb} 
and Technicolor \cite{Susskind:1978ms,Farhi:1980xs}. However, these involve new particles not yet found at the LHC. An alternative is offered by models of extra-dimensions.
One of these is the Large Extra-Dimensions model ~\cite{Antoniadis:1990ew,Antoniadis:1998ig}, the main idea of which is to solve the hierarchy problem by lowering the Planck scale down 
to an energy not much larger than $\mathit{\Lambda}_\text{EW}$. This is achieved assuming $n$ extra spatial dimensions which are compactified in a volume $V_n$. Considering the simplest case of toroidal compactification, $V_n = \mathrm{\Pi}_{i=1}^n (2 \pi R_i)$, with $R_i$ the radius of the $i$-th extra-dimension,  the following relation holds at distances much larger than the mean compactification radius:
\begin{equation}
M_{\rm Pl}^2 \equiv V_n \times M_{D}^{2 + n} \, ,
\end{equation}
being $D = 4 + n$ the total number of dimensions of the space-time, ${\cal M}_{ D} = {\cal M}_4 \times {\cal M}_n$ (where ${\cal M}_4$ is the Minkowski 4-dimensional space-time 
and ${\cal M}_n$ is a compact $n$-torus). If $V_n$ is large enough, the actual fundamental scale
of gravity $M_{D}$ can be much smaller than $M_{\rm Pl}$, that is then only an effective scale and not a fundamental parameter of the theory.  
Other extra-dimensional scenarios solve the hierarchy problem by means of the curvature of the extra-dimensions (see, {\em e.g.}, the Randall-Sundrum model \cite{Randall:1999ee,Randall:1999vf})
or by a combination of volume and curvature (as in the Clockwork/Linear Dilaton model \cite{Giudice:2016yja,Giudice:2017fmj}).

In the simplest extra-dimensional extensions of the SM, SM particles are confined to topological defects of the space-time called ``{\em branes}'' 
(the concept of D$3$-brane can be found in Ref.~\cite{Polchinski:1996na}) and only gravity can freely propagate along the new extra dimensions. 
A common feature of all these extensions is, thus, the fact that gravity is modified around some length scale $\lambda$ (even though it
may enormously differ between any two different models). For weak gravitational fields, bounds on the behaviour of gravity may be put
by looking for deviations from the Newtonian potential between two test masses (see, {\em e.g.}, Ref.~\cite{Buisseret:2007qd}). 
A summary of relatively recent experimental bounds can be found in Ref.~\cite{Adelberger:2009zz} (where the results of different techniques from 
Refs.~\cite{Hoyle:2004,Kapner:2006si,Spero:1980,Hoskins:1985,Tu:2007,Long:2003,Chiaverini:2003,Smullin:2005} are shown together), 
with the most recent results published in Ref.~\cite{Lee:2020zjt} giving $\lambda < 38.6$ $\upmu$m at 95$\%$ confidence level (CL). 
Other results can also be found in  Refs.~\cite{Perivolaropoulos:2016ucs,Antoniou:2017mhs,Perivolaropoulos:2019vkb}.
All of these experiments were performed by measuring the absolute strength of the 
gravitational force acting between two bodies at a given distance $r$. The shorter the distance, the larger the intensity of the gravitational attraction. However, 
it is also true that the shorter the distance, the larger the unavoidable electrically-induced forces between the two bodies. In particular, 
Coulombian, dipolar, and Van der Waals forces induce a $1/r$ potential that acts as a  hardly removable background to the measurement of the strength of the gravitational force. 
Thus, improving the present bounds using the same techniques from the aforementioned literature is extremely difficult. 

If we really want to test a whole class of extensions of the SM, as the extra-dimensional ones, it is therefore of great interest  to look for new methods that may allow to bypass 
the problems related to the presence of electrical backgrounds and, thus, to break the barrier of the tens of microns. 
A proposal to attain this goal was advanced in Ref.~\cite{Donini:2016kgu}, where it was suggested that looking to the geometrical features of the {\bf orbit} of a microscopic test body around a
heavier one (that acts as the source of gravitational field) may mostly surmount the problem of eletrically-induced backgrounds. The motivation for this was that
most of electrically-induced backgrounds behave with a potential that goes as $1/r$, just like Newton's potential. As central potentials with an $r$-dependence 
proportional to $1/r$ or $r$ induce closed orbits, as stated by the Bertrand's theorem (see, {\em e.g.}, Ref.~\cite{romero1997contemporary}),  looking for {\bf precession of the orbit} 
is a smoking gun for deviations from the $1/r$-dependence of the Newtonian potential independent of (dominant) electrically-induced backgrounds.

In this paper we further analyse that proposal. It is organized as follows: in Sect.~\ref{sec:theo} we remind the very simple classical mechanics that we are going to use throughout the paper; in Sect.~\ref{sec:setup} we introduce
the (gedanken) experiment that was sketched in Ref.~\cite{Donini:2016kgu}; in Sect.~\ref{sec:optimization} we optimize the setup in order to maximize its sensitivity to
deviations from Newton's $1/r^2$ law; in Sect.~\ref{sec:backgrounds} we study the impact of possible backgrounds on the sensitivity; in Sect.~\ref{sec:sens} we present
the sensitivity of our setup in the presence of backgrounds and study the attainable precision, in case of a positive signal of deviations from Newton's law, 
with particular interest to signals corresponding to extra-dimensional models; and, in Sect.~\ref{sec:concl}, we eventually come to a conclusion.

\section{Theoretical framework}
\label{sec:theo}

Consider a gravitating system composed of two bodies  called the Planet (P) with mass $m_{\rm P}$ and the Satellite (S) of mass $m_{\rm S}$, 
with $m_{\rm P} \gg m_{\rm S}$, so that $m_{\rm S}$ is small enough that the motion of P under the effect of S can be neglected. Then let P be at rest at $\bm{r}_\text{P}=\textbf{ }0$ and let $\bm{r}$ be the position of S.

\subsection{One-dimensional motion}
\label{sec:onedimmotion}

\begin{figure*}[h!]
\centering
\begin{minipage}{0.45\textwidth}
\centering
\includegraphics[angle=0,width=0.95\textwidth]{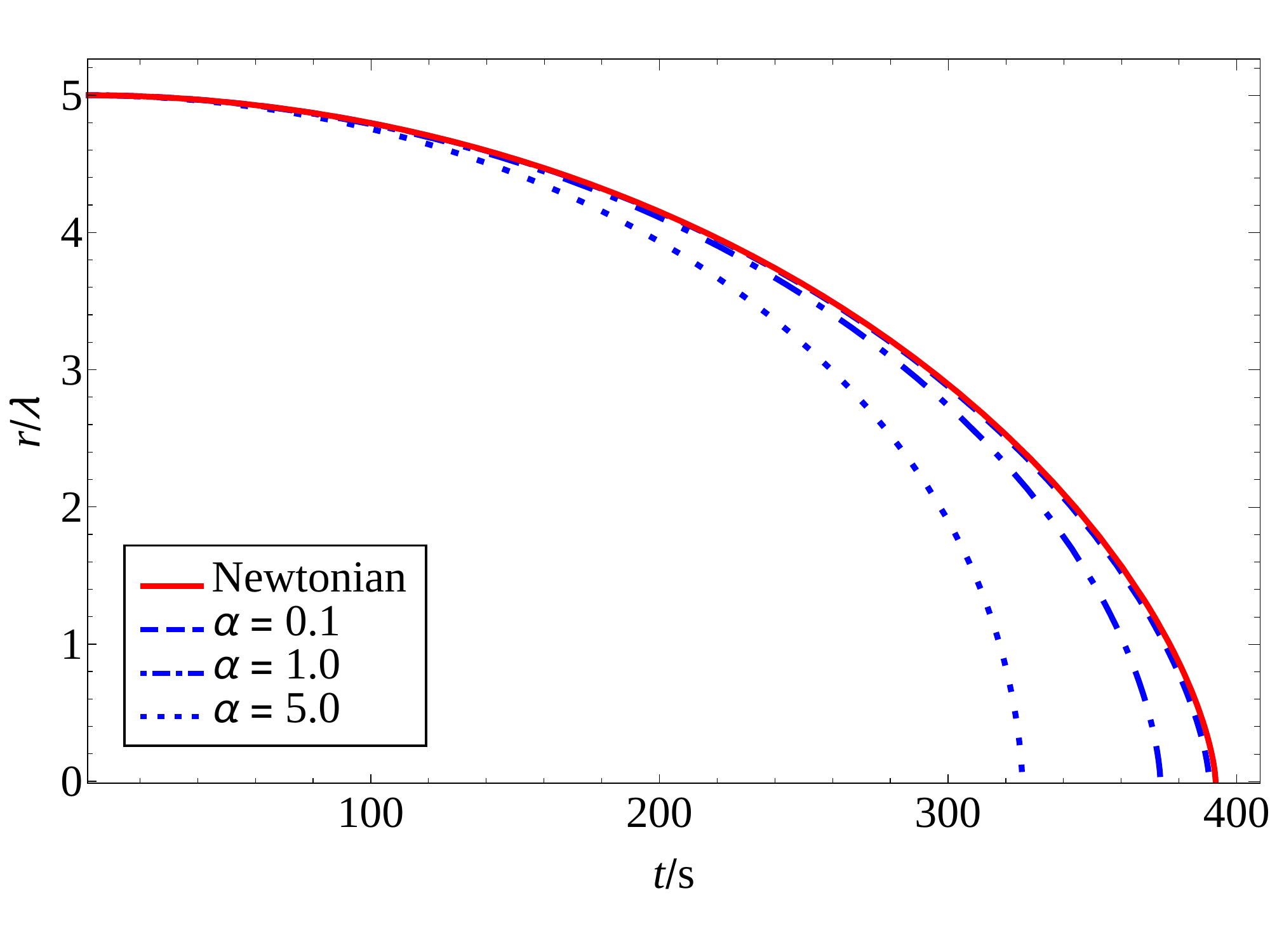} 
\end{minipage}\hspace{1cm}
\begin{minipage}{0.45\textwidth}
\centering
\includegraphics[angle=0,width=1\textwidth]{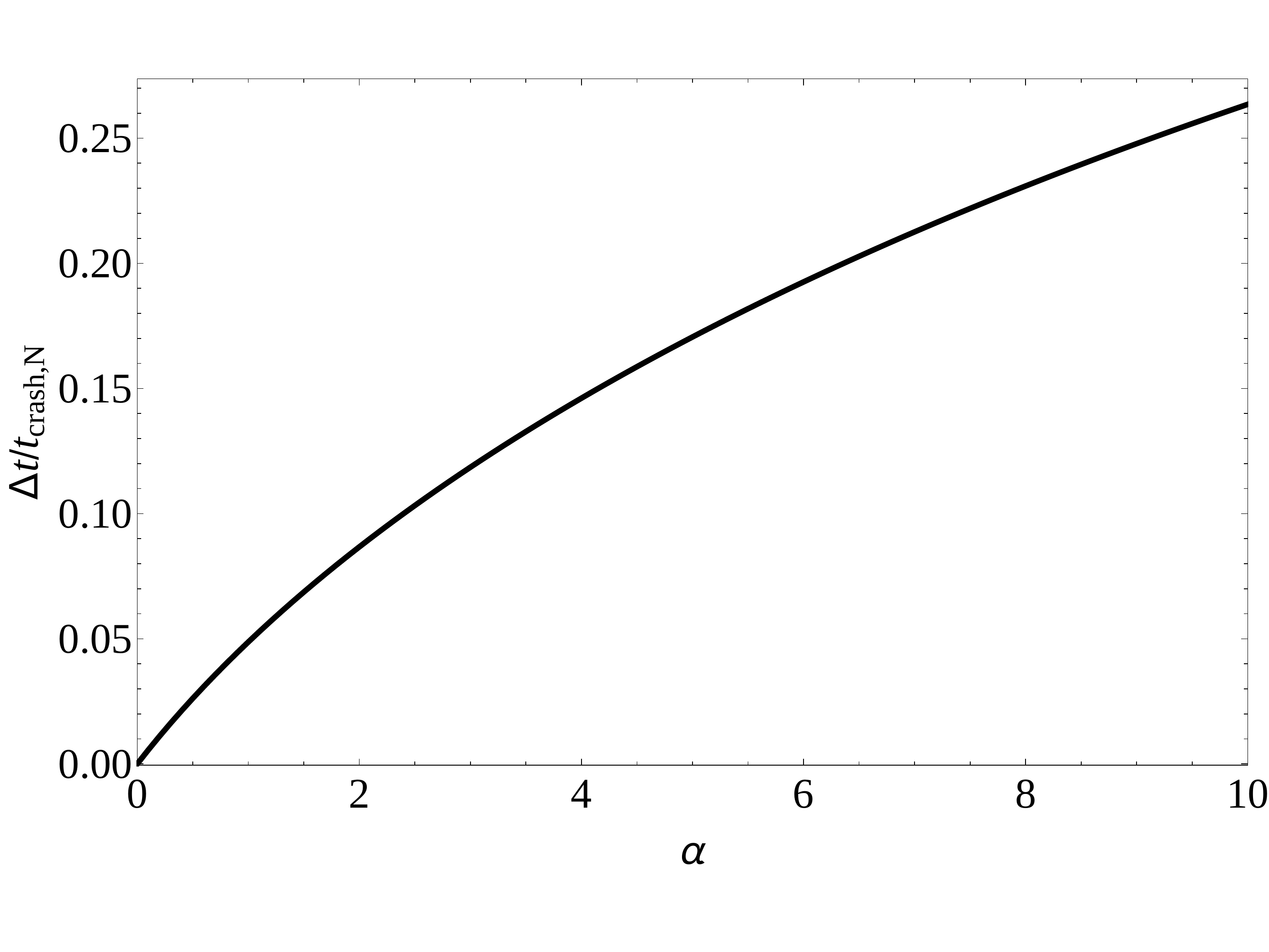}
\end{minipage}
\caption{\it 
Left: normalized distance $a=r/\lambda$ between the Satellite, \textup{S}, and the source of the gravitational field, \textup{P}, as a function of time, 
in the case of Newtonian (red solid line) and Beyond-Newtonian motion (blue non-solid lines). 
The initial conditions are $a_0 = 5$ and $\dot a_0 = 0$, 
with $k  = G_{\rm N} m_{\rm P}/\lambda^3 = 10^{-3}$ \textup{s}$^{-2}$ and $\lambda = 10$ $\upmu$\textup{m} for different $\alpha$: 
$\alpha = 0.1$ (blue dashed line), $\alpha = 1.0$ (blue dot-dashed line), and $\alpha = 5.0$ (blue dotted line). 
Right: Relative reduction of the time it takes for S to crash onto P with respect to the Newtonian case, as a function of $\alpha$. The result is found to be independent of $\lambda$ if we use the normalized initial conditions stated before ({\it i.e.,} $a_0=5$ and $\dot{a}_0=0$).
}
\label{fig:linearmotion}
\end{figure*}

Consider S initially located at a position $\bm{r}_0$, different from the origin, with initial radial  and angular velocities $\dot{r}_0$ and $\dot \theta_0 = 0$, respectively. The resulting motion of the system is then one-dimensional, with S straightly falling onto P. 
The equation of motion is:
\begin{equation}\label{eq:EOMgeneral}
\ddot{r} = \frac{F(r) }{m_{\rm S}} \, ,
\end{equation}
where $r=|\bm{r}|$, and $F (r)$ is either $F_{\rm N} (r)$ in the case of the 4-dimensional Newtonian force or $F_{\rm BN} (r)$ for a central {\em Beyond-Newtonian} (BN) force. 
This expression is quite general: for extra-dimensional models, $F_{\rm BN} (r)$ may be a $D$-dimensional version of Newton's law (see Refs.~\cite{Kehagias:1999my,Floratos:1999bv}) or  the brane-to-brane force, $F_{\rm BB} (r,d)$, of Refs.~\cite{Donini:2016kgu,Liu:2003rq} if S and P are onto different branes (being $d$  the brane-to-brane distance). 
It can also account for modification of gravity such as, for example, in Refs.~\cite{Perivolaropoulos:2016ucs,Edholm:2016hbt}.
In the case of Newton's force, eq.~(\ref{eq:EOMgeneral}) reduces to:
\begin{equation}
\ddot{r}(t) + G_{\rm N} \frac{m_{\rm P}}{r^2(t)}=0 \, ,
\end{equation}
where $G_\text{N}$ is the 4-dimensional Newton's constant. We can now normalize the distance to the characteristic length scale $\lambda$ at which New Physics corrections arise, introducing the normalized quantity $a = r/\lambda$, 
such that {\em natural} values of $a$ are ${\cal O}(1)$. The differential equation to be solved is now:
\begin{equation}
\ddot{a}(t) + \frac{k}{a^2 (t)} = 0 \, ,
\end{equation}
where $k = G_{\rm N} m_{\rm P} / \lambda^3$ is a coefficient with dimensions of time$^{-2}$. Since the present experimental bound on $\lambda$ 
is $\lambda < 38.6$ $\upmu$m at 95$\%$ CL (from Ref.~\cite{Lee:2020zjt}), the typical range of distances we aim at studying is $ r \in [1,200]$ $\upmu$m.
A reasonable choice for $m_{\rm P}$  could be $m_{\rm P} \sim 1.5 \times 10^{-5}$ g, 
for which we get $k \sim 10^{-3}$ s$^{-2}$ if $\lambda = 10$ $\upmu$m. For this, two-dimensional trajectories of S would lie in the desired range of $r$, as we will later see.

When looking for New Physics, it is typical to parameterize deviations from Newton's law in terms of a Yukawa potential depending on two parameters, a length scale, $\lambda$, and an adimensional coupling constant, $\alpha$,
\begin{equation}
\label{eq:YukawaPotential}
V_{\rm BN} (r) = - \frac{G_{\rm N} m_{\rm P}}{r} \left [ 1 + \alpha \, \text{e}^{-r/\lambda} \right] \,  .
\end{equation}
For distances $r \gg \lambda$, this potential exponentially reduces to the Newtonian one. On the other hand, 
for $r \lesssim \lambda$, New Physics becomes relevant. Notice that, conventionally, $\alpha$ is considered to be positive. However, in general terms, the sign of the leading correction 
to the Newtonian potential should not be fixed, and a negative $\alpha$ could also be considered. In the rest of the paper, though, we will take $\alpha \geq 0$ except when explicitly mentioned.

The corresponding one-dimensional central force can thus be written as 
\begin{equation}
\label{eq:modifiedNewton}
F_{\rm BN} (r) = - G_{\rm N} \frac{m_{\rm P} m_{\rm S}}{r^2} f_{\rm BN} (r/\lambda)\,,
\end{equation}
with
\begin{equation}
 f_{\rm BN} (r/\lambda)=1+\alpha\,\text{e}^{-r/\lambda}+\frac{\alpha r}{\lambda}\text{e}^{-r/\lambda}\,.
\end{equation}
Then, the equation of motion in terms of the normalized distance is 
\begin{equation}
\ddot{a (t)} + \frac{k}{a^2 (t)} \, f_{\rm BN} (a) =  0 \, .
\end{equation} 

In Fig.~\ref{fig:linearmotion} we can see the different time evolution of the distance between P and S in the case of Newtonian and Beyond-Newtonian
attractive forces, for $k  = G_{\rm N} m_{\rm P}/\lambda^3 = 10^{-3}$ s$^{-2}$, $\lambda = 10$ $\upmu$m and several values of $\alpha$,
with initial conditions $a_0 = r_0/\lambda = 5$ ({\em i.e} $r_0 = 50$ $\upmu$m) and  null initial velocity. 
We observe that the time needed for S to crash onto P gets smaller
in the case of Beyond-Newtonian motion with respect to the Newtonian one as $\alpha$ increases.

\subsection{Two-dimensional motion}
\label{sec:twodimsmotion}

The case of two-dimensional motion is far more interesting. Now, as it is well-known from classical mechanics, we can have bounded or unbounded trajectories, 
depending on the initial conditions. We will focus on the former case: consider P is located at the origin of coordinates, the initial position of S is $\bm{r}_0\neq\bm{0}$, and the initial radial and angular velocities of S are tuned such that a bounded orbit of S around P  is observed. 

In the Newtonian case the equation of motion can be written as
\begin{equation}
m_{\rm S} \ddot{{\bm{r}}} = {\bm{ F}}_{\rm N} (\bm{r}) = - \bm{ \nabla} V (r)  = - G_{\rm N} \frac{m_{\rm P} \, m_{\rm S}}{r^3} \, {\bm{r}} \, ,
\end{equation}
where $V (r)$ is the potential energy due to the gravitational field. The total energy is
\begin{equation}
\label{eq:energeticbalance}
{\cal E} = T + V = m_{\rm S} \left (\frac{|{\dot{\bm{r}}}|^2}{2} - G_{\rm N} \frac{ m_{\rm P} }{r} \right) \, ,
\end{equation}
where $T$ is the kinetic energy of S (recall we neglect the movement of P). As the movement of a body under a central potential is restricted to a plane, the velocity of S can be decomposed in polar coordinates in that plane to get
\begin{equation}
{\dot{\bm{r}}} = \dot{r} \, {\bm{e}}_r + r \dot{\theta} \, {\bm{e}}_\theta \, ,
\end{equation}
where $({\bm{e}}_r, {\bm{e}}_\theta)$ are two orthonormal vectors that define the coordinate frame at each point in the plane. 
Expressed in Cartesian coordinates, ${\bm{e}}_r = (\cos \theta, \sin \theta, 0)$ and ${\bm{e}}_\theta = (- \sin \theta, \cos \theta, 0)$, where the $z$-axis is taken to be normal to the plane of the orbit and $\theta=0$ on the positive $x$-axis. 
In this basis, the acceleration becomes:
\begin{equation}
{\ddot{\bm{r}}} = \left ( \ddot{r} - r \dot{\theta}^2 \right ) \, {\bm{e}}_r 
                              + \left ( r \ddot{\theta} + 2 \dot{r} \, \dot{\theta}  \right )  \, {\bm{e}}_\theta \, .
\end{equation}

Using this, it is now trivial to write a system of equations of motion for S in polar coordinates, 
\begin{equation}
\label{eq:newtonianeom}
\left \{ 
\begin{array}{lll}
\ddot{r} - r \, \dot{\theta}^2 &=& - G_{\rm N} \frac{ m_{\rm P}}{r^2} \, , \\
 \\
r \, \ddot{\theta} + 2 \dot{r} \, \dot{\theta} & = & 0 \, ,
\end{array}
\right .
 \longrightarrow 
\left \{ 
\begin{array}{lll}
\ddot{a} - a \, \dot{\theta}^2 &=& - \frac{k}{a^2} \, , \\
\\
a \, \ddot{\theta} + 2 \dot{a} \, \dot{\theta} & = & 0 \, ,
\end{array}
\right .
\end{equation}
where we again introduced the normalized distance, $a = r/\lambda$, and $k$ is defined as in the previous subsection. 

If we now replace the Newtonian 4-dimensional force with the BN force defined in eq.~(\ref{eq:modifiedNewton}) we get, instead:
\begin{equation}
\label{eq:BNeom}
\left\{ 
\begin{array}{lll}
\ddot{a} - a \, \dot{\theta}^2 &=& - \displaystyle{\frac{k}{a^2 }}  \, f_{\rm BN} (a)  \, , \\
\\
a \, \ddot{\theta} + 2 \dot{a} \, \dot{\theta} & = & 0 \, .
\end{array}
\right.
\end{equation}

In both the Newtonian and BN cases, the equation for $\dot{\theta}$ implies that angular momentum, defined as:
\begin{equation}
h(t) = r^2 (t) \, \dot {\theta (t)} = h_0 \, ,
\end{equation}
where $h_0=h(t=0)$, is a constant of motion. Using this result, the radial equation can be rewritten as:
\begin{equation}
\label{eq:eombeyondnewtonian}
\ddot{a} - \frac{h_0^2}{a^3} = \left \{ \begin{array}{l} - \frac{k}{a^2} \, , \\ \\  - \frac{k}{a^2} \, f_{\rm BN} (a) \, , \end{array} \right . 
\end{equation}
for the Newtonian (above) and Beyond-Newtonian (below) cases. We get different results in the two: for the Newtonian case, solutions of the first of eqs.~(\ref{eq:newtonianeom}) are conic sections. Possible trajectories are then circular, elliptic, parabolic or hyperbolic, and in all cases, they can be described by a simple function,
\begin{equation}
\label{eq:newtonianorbit}
r (\theta) = \frac{r_{\rm c}}{1 - e \, \cos \theta} \, ,
\end{equation}
where $r_{\rm c} = h_0^2/ G_{\rm N} m_{\rm P}$ and the eccentricity $e$ is given for closed orbits by
\begin{equation}
e = \frac{r_\text{a} - r_\text{p}}{r_\text{a} + r_\text{p}} \, ,
\end{equation}
being $r_{\rm a}$ and $r_{\rm p}$ the largest and smallest distances of S from P, respectively. The points of the orbit to which these correspond are known as apoapsis and periapsis, in this same order. 
For $e<1$ the orbit is closed, being circular for $e= 0$ and elliptic for $0<e<1$. For $ e \geq1$ the trajectory is open: it is parabolic for $e = 1$ 
and hyperbolic for $e > 1$. In case of a closed orbit, the Newtonian period of the Satellite around the Planet can easily be computed applying the third Kepler's law:
\begin{equation}
\label{eq:3rdKeplerLaw}
T_{\rm N} =  \pi \frac{(r_{\rm a} + r_{\rm p})^{3/2}}{\sqrt{2 G_{\rm N} \, m_{\rm P}}} \, .
\end{equation}

\begin{figure*}[t]
\centering
\begin{minipage}{0.333\textwidth}
\centering
\includegraphics[angle=0,width=0.9\textwidth]{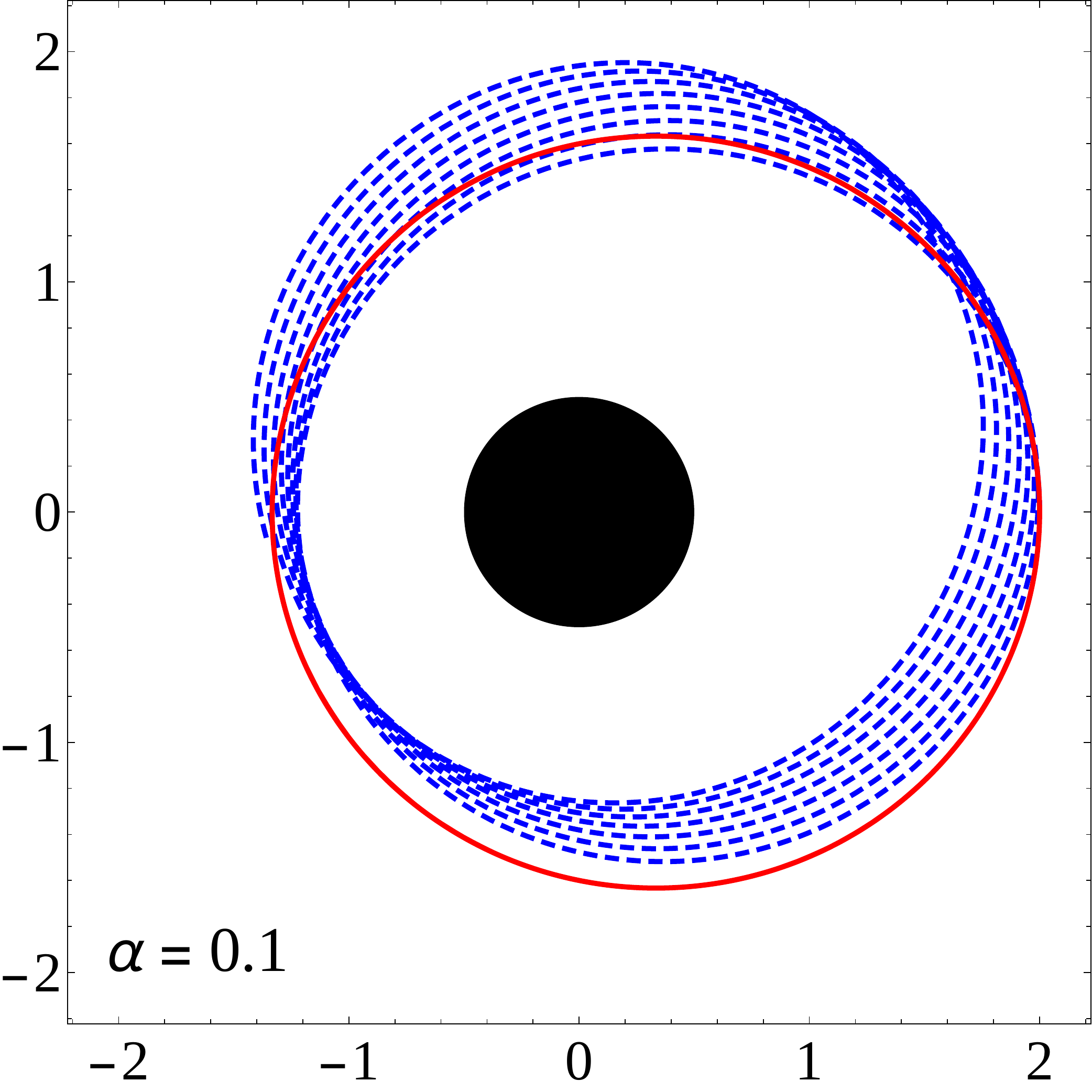} 
\end{minipage}\hfill
\begin{minipage}{0.333\textwidth}
\centering
\includegraphics[angle=0,width=0.9\textwidth]{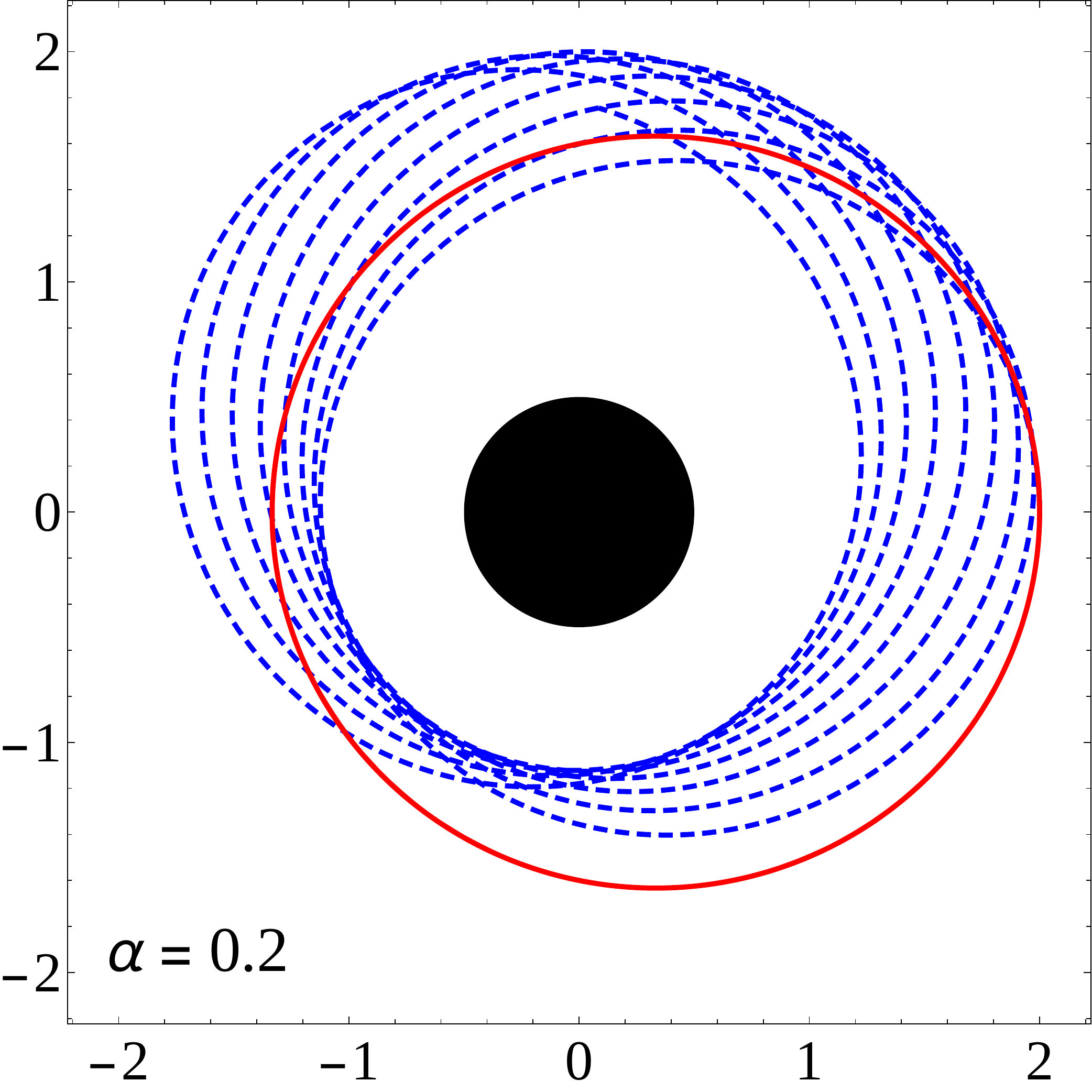} 
\end{minipage}\hfill
\begin{minipage}{0.333\textwidth}
\centering
\includegraphics[angle=0,width=0.9\textwidth]{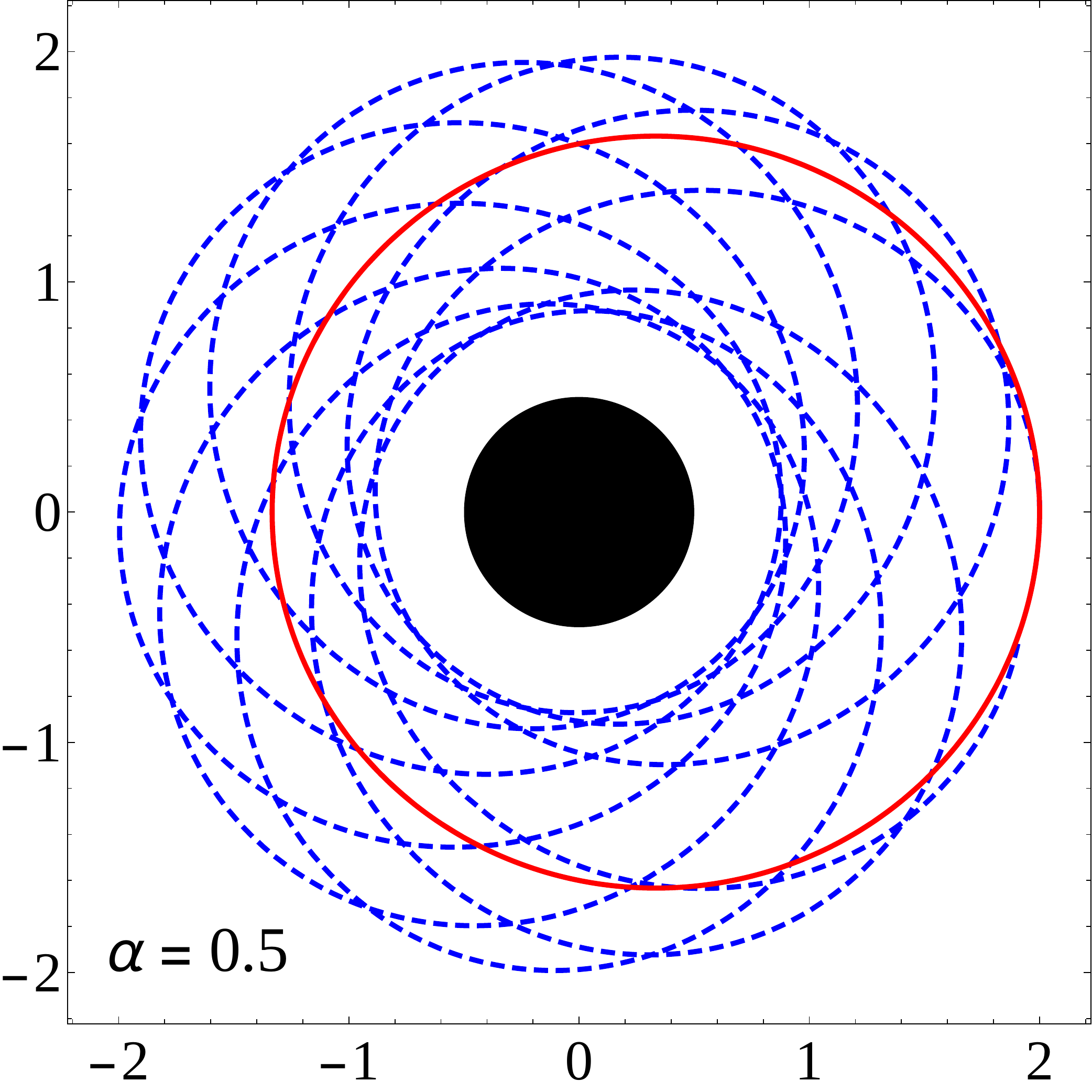}
\end{minipage}\hfill
\caption{
\it  First ten revolutions of the trajectory described by \textup{S} around \textup{P}, located at the origin of coordinates and represented by a black circle.
In solid red, we show the Newtonian orbit. In dashed blue, the motion under the effect of the Beyond-Newtonian force in eq.~(\ref{eq:modifiedNewton})
for $\lambda = 10$ $\upmu$\textup{m}, and $\alpha = 0.1$ (left panel), $\alpha = 0.2$ (middle panel) and $\alpha = 0.5$ (right panel). 
The initial conditions are: $k  = G_{\rm N} m_{\rm P}/\lambda^3 = 10^{-3}$ \textup{s}$^{-2}$, $a_0 = 2$ ($r_0 = 20$ $\upmu$\textup{m}), $\dot{\theta}_0 = 5 \times 10^{-2}$ \textup{rad s}$^{-1}$ and null radial velocity. 
All figures are represented using normalized adimensional distances.
}
\label{fig:orbits}
\end{figure*}

The results in the case of a Beyond-Newtonian force are very different. Recall that, according to Bertrand's theorem ~\cite{romero1997contemporary}, closed orbits are only possible for central forces with a radial dependence of the form $1/r^2$ or $r$. Any deviation from these two implies that the resulting trajectories are neither stable nor closed. 
A typical example of this is the General Relativity correction to the orbit of Mercury around the Sun: the leading corrections to the force are of the form $1/r^4$ and so produce an observable precession 
of the perihelion of the planet. This is precisely the case of the Beyond-Newtonian force in eq.~(\ref{eq:modifiedNewton}): the $r$-dependence of the central force is not proportional to $1/r^2$.  As a consequence, we generally do not expect closed orbits (they may be bounded, though). This is indeed depicted in Fig.~\ref{fig:orbits}, where we show the trajectory of S around P, which is located at the origin of coordinates and represented by a black circle. For the Beyond-Newtonian motion, we represent (in dashed blue) the first 10 revolutions, with $\alpha = 0.1$ (left panel), $0.2$ (middle panel) and $0.5$ (right panel), respectively. 
In all cases, we have chosen the initial conditions such that the Newtonian orbit (depicted in red) is elliptic (albeit with small eccentricity): $k = G_{\rm N} m_{\rm P}/\lambda^3 = 10^{-3}$ s$^{-2}$; 
$\lambda = 10$ $\upmu$m; $a_0 = 2$ ({\em{}i.e.} $r_0 = 20$ $\upmu$m); $\dot{a}_0 = 0$; $\dot{\theta}_0 = 5 \times 10^{-3}$ rad s$^{-1}$ ( {\em i.e.} $h_0 = 2$ $\upmu$m$^2$ rad s$^{-1}$). The initial angle, $\theta_0$, can be chosen arbitrarily, so we simply set $\theta_0 = 0$ rad. Since the initial radial velocity, $\dot{a}_0$, is also set to be zero, the starting point $(a = a_0, \theta = 0)$ is then necessarily either the periapsis or the apoapsis of the orbit. For the chosen initial condition it is indeed the latter.
 
In all panels, we can observe precession of the apoapsis of the trajectory of S around P: we have rotation of the major axis of the orbit.
However, depending on the value of $\alpha$, the orbits can be very different even for the same choice of the initial conditions. 
In the left panel (corresponding to $\alpha = 0.1$) we observe that S moves along nearly circular orbits with a slow 
counter-clockwise precession of the apoapsis. For $\alpha = 0.2$ (middle panel), precession is still rather slow, with the apoapsis moving approximately $90^\circ$ after 10 revolutions.
On the other hand, for $\alpha = 0.5$ (right panel), the apoapsis goes under almost one full circle around P after 10 revolutions.

Notice that the sense of precession is crucially related to the sign of $\alpha$. If, as we have considered here, $\alpha > 0$, the major axis of the orbit precedes in the same direction as the angular 
velocity ({\em i.e.,} counter-clockwise for the initial conditions considered in Fig.~\ref{fig:orbits}). Conversely, for $\alpha < 0$, precession of the major axis would proceed in the opposite direction with
respect to the angular velocity ({\em i.e.,} clockwise for the considered example). 
 
As it is clearly shown by comparing Fig.~\ref{fig:linearmotion} and \ref{fig:orbits}, studying the two-dimensional dynamics of a microscopic gravitational system offers a 
very interesting feature to detect small deviations from Newton's law: precession of the orbit of S around P. In the next section we present a method to exploit this feature.

\section{The Experimental Setup}
\label{sec:setup}

As it was explained in Sect. \ref{sec:intro}, deviations from Newton's law are usually tested by measuring the absolute strength of the attractive force between two bodies
at short distances, using the phenomenological Yukawa potential in eq.~(\ref{eq:YukawaPotential}). One of the most important limitations of these experiments is represented by  unavoidable electrically induced backgrounds. When two (microscopic) bodies are put one near the other, 
they are not only  sensitive to gravitational interactions, but also to electric forces due to the distribution of charge within them. 
In particular, Coulombian, Van der Waals and dipolar forces can affect the measurement, acting as additional attractive or repulsive corrections
on top of the would-be leading gravitational interaction. Many of these (dominant) backgrounds should be included in eq.~(\ref{eq:YukawaPotential}) in the form of 
additional $1/r$ terms, thus effectively modifying the strength of the gravitational interaction to be measured, $G_{\rm N}^\prime \equiv  G_{\rm N} (1 + B)$
(being $B$ the relative size of the electrically-induced $1/r$-dependent potential compared to the gravitational one).  Other terms with an $r$-dependence different from $1/r$ may also be produced: one such example is the Casimir force. Their $r$-dependence may depend on the particular shapes of the two bodies or the electric distribution inside them. In these ``static'' experiments, such additional terms also contribute to the strength of the interactions, although they are expected to be sub-dominant. Therefore, measurements of the absolute strength of the force
are limited by backgrounds that may depend on the size of the bodies and  the material of which they are made. In the absence of a precise knowledge of these
electrical effects, it is thus impossible to fully separate the ``true'' value of $G_{\rm N}$ from the effective one, although some methods allow to minimize their impact
(see,  {\em e.g.,} the ``iso-electronic'' or ``Casimir-less'' technique first introduced in Ref.~\cite{PhysRevLett.94.240401}). 

\begin{figure}[h]
\centering
\includegraphics[width=0.45 \textwidth]{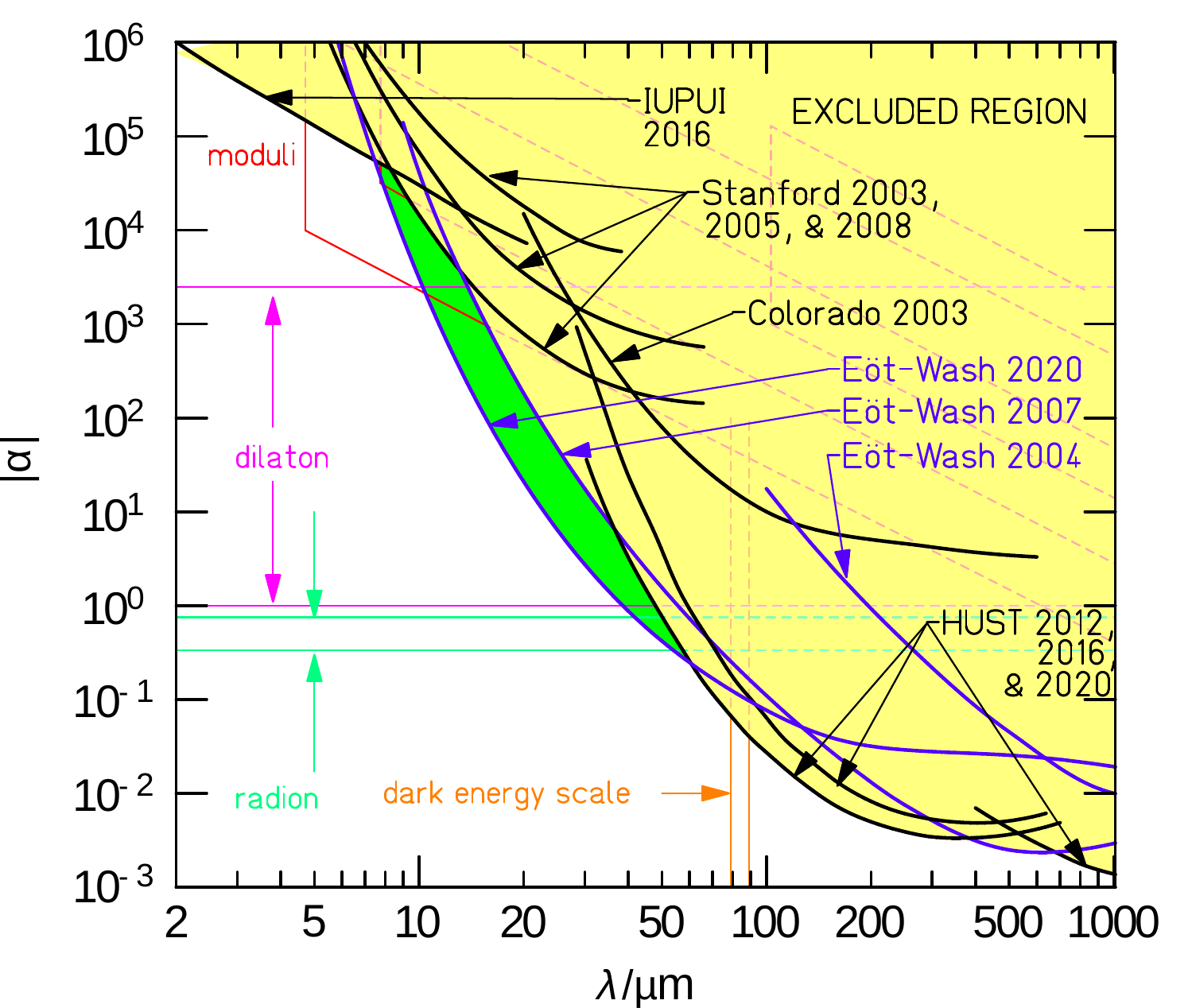}
\caption{Present 95\% CL exclusion contours (in yellow) from several experiments on Yukawa-like deviations from  
Newton's $1/r^2$ law in the $(\lambda, |\alpha|)$ plane (from Refs.~\cite{Adelberger:2009zz,Hoyle:2004,Kapner:2006si,Spero:1980,Hoskins:1985,Tu:2007,Long:2003,Chiaverini:2003,Smullin:2005}). 
In green, the region excluded by the recent analysis in Ref.~\cite{Lee:2020zjt}. The figure is adapted from this last reference.}
\label{fig:AlphaLambdaBounds}
\end{figure}

Experimental results from abovementioned experiments on Yukawa-like deviations from Newton's law are typically presented as exclusion bounds in the $(\lambda,|\alpha|)$ plane. 
A recently updated excluded region is shown in Fig.~\ref{fig:AlphaLambdaBounds}, where the yellow region represent the excluded bound
at 95\% CL from Refs.~\cite{Adelberger:2009zz,Hoyle:2004,Kapner:2006si,Spero:1980,Hoskins:1985,Tu:2007,Long:2003,Chiaverini:2003,Smullin:2005}
and the green region a recent update from Ref.~\cite{Lee:2020zjt}.

Notice that extra-dimensional extensions of the Standard Model can indeed be expressed in terms of the Yukawa potential in eq.~(\ref{eq:YukawaPotential}). 
In the case of Large Extra-Dimensions (LED) models, we get $\lambda = R$ (being $R$ the compactification radius) and $\alpha = 2 n$ (where $n$ is the number
of spatial extra-dimensions). 
For the Randall-Sundrum (RS) model, typical values of $\lambda$ to solve the hierarchy problem are much smaller than the Fermi scale (see Ref.~\cite{Randall:1999ee}) 
and, thus, untestable when looking for deviations from Newton's law with present (and foreseeable) experimental methods. On the other hand, the 
Clockwork/Linear Dilation (CW/LD) model has a wider range of applicability. Similarly to RS, the hierarchy problem is only solvable in this model ({\em i.e.} with a fundamental scale of gravity $M_{D}$ not much above the electroweak scale) 
 with values of the characteristic length scale much smaller than the present bound. However, if one gives up the requirement that $M_{ D}$ is at the TeV scale, we can have
a viable CW/LD model that can be cast in the form of a Yukawa-like potential with a length scale not much below the present bounds on $\lambda$. The main 
difference between the LED and the CW/LD models in this regime is the value of $\alpha$: whereas in LED $\alpha$ is fixed to be twice the number of extra-dimensions, 
in the CW/LD $\alpha$ is $\lambda$-dependent. Its value can be orders of magnitude larger than the LED limiting value, $\alpha \geq 2 n$, which is reached asymptotically
for vanishing curvature. 
The range of applicability of these models is delimited by the two pink horizontal lines in Fig.~\ref{fig:AlphaLambdaBounds}. A common feature of all extra-dimensional models, though, is the fact that $\alpha > 0$. 
At short distances, $r < \lambda$, the gravitational interaction feels more than 4-dimensions and gets stronger. For this reason, we mostly consider in this work the case of a positive $\alpha$.

Motivated by the theoretical arguments exposed in Sect.~\ref{sec:theo}, it was shown in Ref.~\cite{Donini:2016kgu} that a different way to study deviations from Newton's law could be based on the measurement of dynamical features of a micrometre-size system consisting of a small body orbiting around a relatively bigger one. We will keep the notation from last Section hereon and denote the lighter body as the Satellite (S) of mass $m_\text{S}$ and the heavier one the Planet (P) of mass $m_\text{P}\gg m_\text{S}$, so the movement of the latter can be neglected.  As explained before, central potentials that differ from $1/r$ and $r$ have no closed orbits, and so induce precession of the orbit of S around P. This is the case of the Beyond-Newtonian potential from eq.~(\ref{eq:YukawaPotential}). An experiment searching for this feature in a microscopic gravitational system would have the huge advantage when compared to ``static'' experiments that most of the unavoidable electric backgrounds would be irrelevant,  as they do depend on the distance as $1/r$ and thus do not induce precession. 


We now sketch a possible experimental setup to look for orbit precession in a microscopic gravitational system, following the outline of Ref.~\cite{Donini:2016kgu}:
\begin{enumerate}
\item Consider a 1 mm$^3$-wide laboratory, with a platinum sphere (the Planet) with radius $R_{\rm P} = 10.3$ $\upmu$m and mass $m_\text{P} = 0.75 \times 10^{-5}$ g, located 
         at the center of the setup in a fixed position\footnote{In Ref. \cite{Donini:2016kgu} the mass of the Planet was erroneously reported to be $m_\text{P} = 10^{-7}$ g, even if all
         computations were performed with $m_\text{P} = 10^{-5}$ g.}; we choose platinum in order to have a material with the highest possible density at room temperature, so as to have the smallest    
         possible planet radius to prevent collisions between S and P.
\item Insert the laboratory between magnets in a configuration such that we can levitate a diamagnetic Satellite to cancel the Earth's gravitational field\footnote{Possible alternatives 
may be to use an optically-cooled levitating dielectric satellite \cite{Geraci:2010,Yin:2013lqa,Geraci:2014gya}, or to move the mm$^3$-size lab into a zero-gravity environment.};
\item Introduce a diamagnetic sphere (the Satellite) with mass $m_\text{S} = 1.2\times 10^{-9}$ g in the lab.  The diamagnetic sphere could be made of pyrolitic graphite, 
         with a density $\rho_{\rm PG} = 2.2$ g/cm$^3$
         (for which the radius of the sphere would be $R_{\rm S} = 5$ $\upmu$m). Magnets producing a magnetic field $B \sim 1.25$ T may be used to levitate
         it, given the diamagnetic susceptibility of pyrolitic graphite, $\chi = - 1.6 \times 10^{-4}$  \cite{Simon:2000}. 
         Some details on a feasible magnet configuration are be given in App.~\ref{app:MC}.
\item Eventually, put the diamagnetic sphere into motion with appropriate initial conditions to have a bounded orbit of S around P, avoiding 
           crashes, and measure a conveniently defined observable (to be later introduced).  Introducing the satellite with given initial conditions is the most difficult task to achieve in this experimental proposal.  
         However, relatively recent results \cite{Kobayashi:2012} show that a levitating pyrolitic graphite sphere may be put into motion by means of photo-irradiation.
\end{enumerate}

Let's make some comparisons. As we told above, we want to study the classical dynamics of a microscopical gravitational system designed in close resemblance
to usual gravitational systems of astronomical sizes. However, we may ask ourselves which kind of system we have at hand. First of all, consider
the ratio between the mass of the Planet and that of the Satellite, which is $m_{\rm P}/m_{\rm S} = 6250$, more than six times larger than the ratio of the Sun and Jupiter masses. However, 
the Newtonian gravitational force acting on the center of mass of Jupiter due to the Sun at the aphelion is $|F_{\rm N}| = 3.8 \times 10^{22}$ N, whereas the force felt by S due to P at the apoapsis 
is only $|F_{\rm N}| = 2.6 \times 10^{-23}$ N (for a centre-to-centre distance at the apoapsis of $150$ $\upmu$m). Of course, a very intense gravitational force is needed to attract a body the size of Jupiter
at a distance at the aphelion of $817 \times 10^6$ km. For a better comparison, it is useful to compute the gravitational acceleration of the two systems: Jupiter is ``falling'' onto the Sun 
with $g = 2.0 \times 10^{-5}$ m/s$^2$, whereas our Satellite ``falls'' onto the Planet with $g = 2.2 \times 10^{-11}$ m/s$^2$. The (Newtonian) period of the orbit of Jupiter around the Sun, 
computed according to the third Kepler's law, is $T = 11.86$ years, whereas for the orbit of S around P (with an angular velocity at the apoapsis of $\dot{\theta}_0= 273.0$ 
$\upmu$rad s$^{-1}$) it is $T \approx 2$ h $30$ min. The average orbital angular 
velocities of the two systems are $\langle \dot{\theta} \rangle_{\rm J} = 16.8$ nrad s$^{-1}$ (for Jupiter around the Sun), and $\langle \dot{\theta} \rangle_{\rm S} = 705$ $\upmu$rad s$^{-1}$ (for the Satellite around the Planet). Our experimental 
setup is thus a much faster gravitational system, as a consequence of its much smaller size. In order to find a macroscopic gravitational system somehow more resemblant
to our experimental setup, we must look for a somewhat different one. Notice that the eccentricity of the orbit of S around P in our setup is $e=0.497$, much more similar to that of a comet around the Sun 
than to that of one of the eight major planets. For example, the orbit of the periodic comet P/2009 Q1 \cite{Hill2009} has $e=0.4965$. It is remarkable, however, that the distance from this comet to the Sun at the periapsis is $r_{\text{a}} = 4.2 \times 10^8$ km, much larger than the size of the Sun, $R_\odot=6.96\times10^5$ km, whereas in the proposed setup $r_{\text{a}}=50.4$ $\upmu$m (measured between the geometrical centre of the two bodies in the Newtonian case), avoiding collision by just 1.4 $\upmu$m.

Once the diamagnetic satellite S is put into motion around the platinum planet P, we may connect a trigger to a clock in such a way that every 
time S crosses the positive $x$-axis at any point, the measure of the time needed to S to perform a $2 \pi$-revolution around P is taken\footnote{The reason why such a simple measure was suggested was to have data taking the less intrusive as possible. Clearly, a more detailed information on the orbit, that could be obtained by a continuous image of the orbit, would lead to a larger sensitivity.}.
The error in the measurement of each revolution time $T^i_{\rm BN}$ is the clock precision $\sigma_T$, neglecting the delay between the trigger and the clock (revolution times for the considered system typically
ranges from minutes to several hours). In Ref.~\cite{Donini:2016kgu} we  performed a simple statistical analysis using a very conservative $\sigma_T = 1$ s precision. 
In that analysis, the data sample was the collection of $N_{\rm rev}$ consecutive revolution times, $T^i_{\rm BN}$, and we compared them with the hypothesis that they reproduced
a constant revolution time $T_{\rm BN}^i = T_{\rm N}$, being $T_{\rm N}$ the period for a Newtonian potential. We computed the following $\chi^2$ 
\begin{equation}
\label{eq:chi2revolution}
\chi^2 = \sum_{i =1}^{N_{\rm rev}} \frac{\left[T_{\rm BN}^i(\lambda, \alpha) - T_{\rm N}\right]^2}{\sigma_T^2} \, .
\end{equation}

In Ref.~\cite{Donini:2016kgu}  we considered $N_{\rm rev} = 30$, that would correspond approximately to a couple of days of data taking in the case of Newtonian orbits. 
Using as initial conditions $r_0 = 190.0$ $\upmu$m, $ \dot \theta_0 = 180.0$ $\upmu$rad s$^{-1}$ and null $\dot{r}_0$ (corresponding to starting our orbit at the apoapsis with $ \dot \theta_0$ initial angular velocity), 
we found that the bound on $\lambda$ could be pushed down to a few microns for any value of $\alpha \geq 10^{-1}$, whereas we got $\lambda \lesssim 10$ $\upmu$m for $\alpha$ as low as $5 \times 10^{-3}$. 
Below $\lambda = 1$ $\upmu$m sensitivity was lost, as the exponential factor in the Yukawa potential ${\rm exp} (- r/\lambda)$ was rapidly killing the effect.
For this particular choice of the mass of the source, $m_{\rm P}$, of the Satellite, $m_{\rm S}$, and of the initial conditions, we found maximal sensitivity for $\lambda$ in the range of interest,  $\lambda \in [10, 100]$ $\upmu$m. 
These results were, however, obtained without a detailed optimization of the setup, that is carried out in Sect.~\ref{sec:optimization}. More importantly, we did not consider the existence of possible
 background sources  that could be cast in the form of additional terms in the potential and that may affect our sensitivity. A thorough study of the impact of possible backgrounds
on the sensitivity of the proposed setup is performed in Sects.~\ref{sec:backgrounds} and \ref{sec:sens}.

\section{Optimization of the setup without backgrounds}
\label{sec:optimization}

In this Section, we first study the dependence of the setup outlined in Sect.~\ref{sec:setup} on the mass of the Planet and the initial conditions. Then, we introduce the difference between collisional trajectories and bounded orbits and analyze the sensitivity of the proposed experimental setup (in the absence of
backgrounds).

\subsection{Optimization in the absence of backgrounds}
\label{sec:optinoback}

\begin{figure*}[h]
\centering
\begin{minipage}{0.45\textwidth}
\centering
\includegraphics[width=0.95\textwidth]{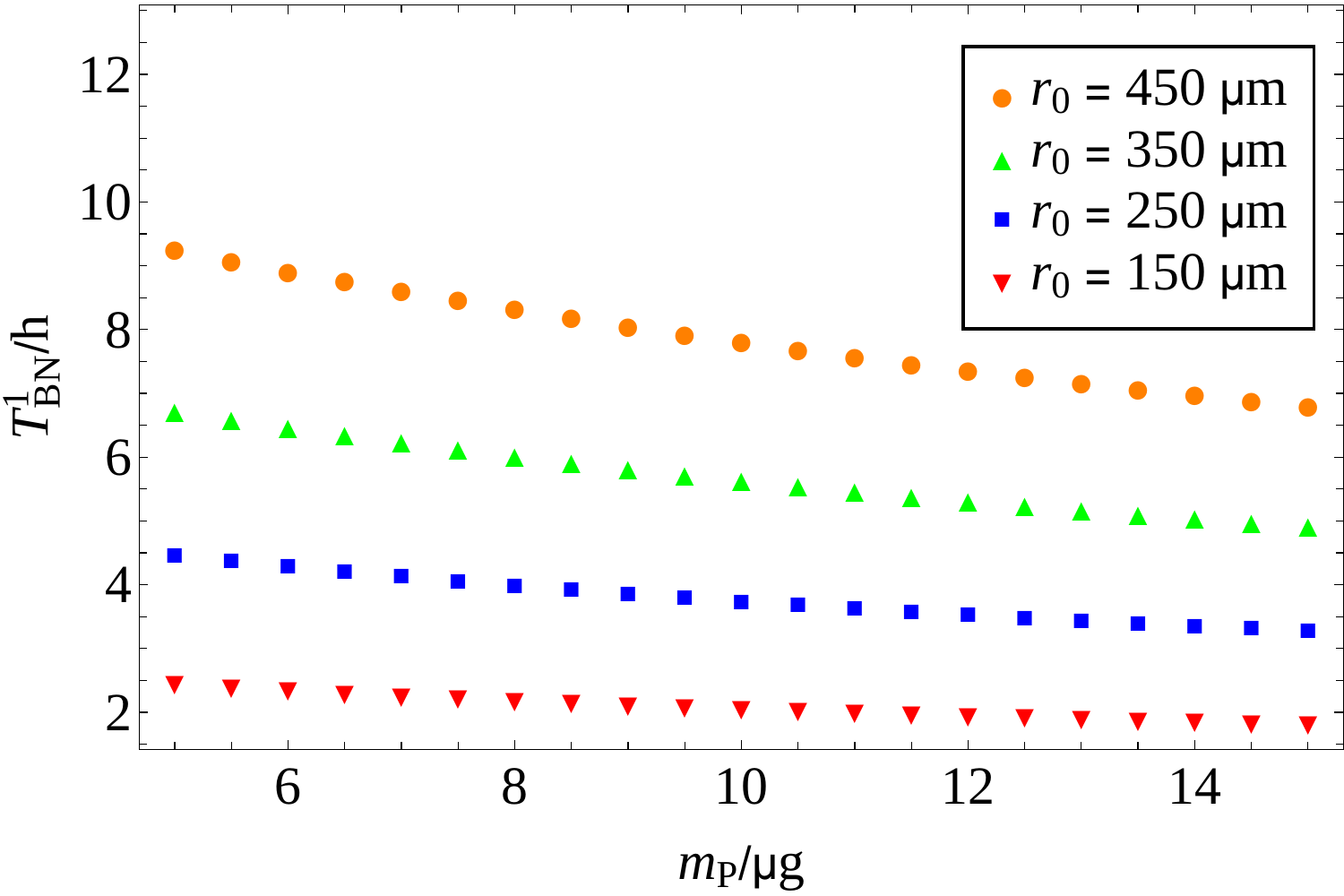}
\end{minipage}\hfill
\begin{minipage}{0.45\textwidth}
\centering
\includegraphics[width=0.985\textwidth]{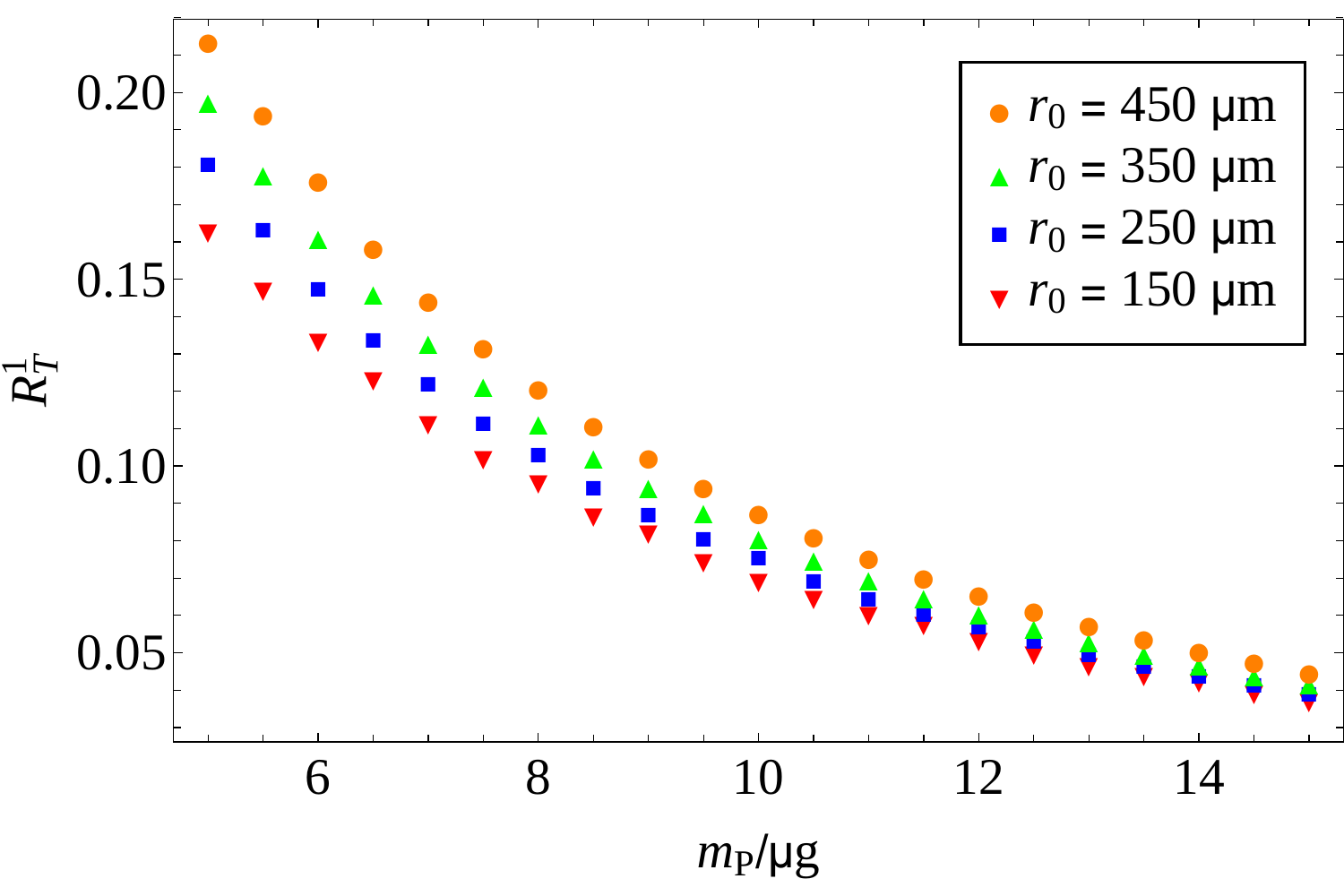}
\end{minipage}
\caption{\it 
Left panel: Time needed for \textup{S} to perform the first $2 \pi$-revolution around \textup{P}, $T_{\rm BN}^1$ (in hours), as a function of the mass of the source, $m_{\rm P}$, for different values 
of the initial distance at the apoapsis, $r_0$. 
Right panel: The relative variation of the orbital period for the first revolution, $R_T^1 = \left ( T_{\rm BN}^1 - T_{\rm N} \right )/T_{\rm N}$, 
as a function of the mass of the source $m_{\rm P}$ for different choices of the initial distance at the apoapsis, $r_0$.
In both panels, the color code is: $r_0 = 450$ $\upmu$\textup{m} (orange circles), $r_0 = 350$ $\upmu$\textup{m} (yellow triangles), $r_0 = 250$ $\upmu$\textup{m} (blue squares), and $r_0 = 150$ $\upmu$\textup{m} (red inverted triangles).
Notice that for each choice of $r_0$, $\dot \theta_0$ has been optimized to maximize the signal. 
}
\label{fig:PeriodMass}
\end{figure*}

The proposed setup was designed to have maximal sensitivity to deviations from Newton's $1/r^2$ law at distances $r\sim 10$ $\upmu$m. 
In order to optimize the system, we must study the dependence of the observable, chosen to detect orbit precession, on the mass of the Planet, $m_{\rm P}$, 
and on the initial conditions of the system (the initial distance, $r_0$, and the initial radial and angular velocities, $\dot r_0$ and $\dot \theta_0$, respectively). 
The initial position of the Satellite cannot be much larger than the value of $\lambda$ for which we want maximal sensitivity, $\lambda_{\rm max}$. However, it must be  several times larger than the sum of the
radius of the Planet, $R_{\rm P}$, and of the Satellite, $R_{\rm S}$, so as to avoid a fast crashing of the Satellite onto the Planet. We therefore need, approximately, 
$r_0 \in [2(R_{\rm P} + R_{\rm S}), 10^2 \times \lambda_{\rm max}]$. 
For example, for a platinum Planet with mass $m_{\rm P} = 10^{-5}$ g and a pyrolitic graphite Satellite with mass $m_{\rm S} = 1.2\times 10^{-9}$ g, we have $R_{\rm P} \simeq 48.1$ $\upmu$m and $R_{\rm S} \simeq 5.0$ $\upmu$m. 
Therefore, choosing $\lambda_\text{max}\approx 10$ $\upmu$m, a reasonable range for the initial radial distance of S to P would be  $r_0 \in [100, 1000]$ $\upmu$m.

Once we fix the initial position, we can compute the Newtonian and Beyond-Newtonian escape velocities (for a range of values of $(\lambda, \alpha)$ of interest), 
in order to determine for which range of values of $\dot r_0$ and $\dot \theta_0$ we have bounded orbits. Recall that this is the desired situation to perform the data gathering, as we can measure
the period over a large number of $2 \pi$-revolutions of S around P. Despite the fact that open trajectories will differ between Newtonian and BN cases, it is much more difficult
to detect the differences under conditions that do not allow for consecutive measurements.

Consider now the case in which we fix $\dot r_0 = 0$. Once the starting distance between S and P is fixed, the only remaining initial condition to be chosen is the initial angular velocity, 
that must be smaller than the escape velocity and larger than the value for which the trajectory of S brings it to crash onto P. For such a  value of $\dot \theta_0$ within these two bounds, 
we will have a closed Newtonian orbit with period $T_{\rm N}$ given by the third Kepler's law, eq.~(\ref{eq:3rdKeplerLaw}), and expectedly, a bounded non-collisional Beyond-Newtonian one.

In order to optimize the setup, we must first choose an observable that is able to reflect changes in the system and in the initial conditions. 
Following Ref.~\cite{Donini:2016kgu}, our first choice is the average (over $N_{\rm rev}$ revolutions) of the relative difference between the time needed for the Satellite to perform a $2 \pi$-revolution around 
the Planet  and the expected Newtonian period, $T_{\rm N}$:
\begin{equation}
\label{eq:meanRt}
\bar{R}_T = \frac{1}{N_{\rm rev}}\sum_{i=1}^{N_{\rm rev}} \left | \frac{T_{\rm BN}^{i}(\lambda, \alpha)- T_{\rm N}}{T_{\rm N}} \right | = \frac{1}{N_{\rm rev}}\sum_{i=1}^{N_{\rm rev}} R_T^i \, .\\
\end{equation}
Recall that $T_\text{BN}^i$ is the period of the $i$-th revolution in the BN scenario (thus depends on $\lambda$ and $\alpha$), which is the one we would measure in a real experiment. We remind that whether $T_{\rm BN}$ is lower or greater than $T_{\rm N}$ depends crucially on the sign of $\alpha$. For positive values, the BN force is stronger than the Newtonian one, and  the period is shorter. The opposite would happen for negative $\alpha$. To maximize the signal, $\bar R_T$ should be as large as possible, whereas to keep stability over a large number of revolutions a small period is desirable. These two constraints  condition the most the setup optimization. Due to the initial motivation in Ref.~\cite{Donini:2016kgu}, based on a particular $\mathcal{M}_4\times S^1$ LED model, we set $\lambda=10$ $\upmu$m and $\alpha=2$ for the following optimization. 

We first analyse the effect of varying $m_{\rm P}$, $r_0$ and $\dot \theta_0$, (keeping $\dot r_0 = 0$), and focus only on the first revolution, $N_{\rm rev}=1$. The dependence of $R^1_T$ on the mass of the source is studied in Fig.~\ref{fig:PeriodMass}. In the left panel, we show how the time needed for S to  perform the first $2 \pi$-revolution around P, $T_{\rm BN}^1$, depends on the mass of the source $m_{\rm P}$ for different values of  $r_0$. Notice that, for each point, we scan over $\dot \theta_0$ to look for the orbit with the largest eccentricity that still avoids crash of S onto P, to guarantee maximum precession. This ensures that we are studying the scenario of largest signal for the given value of $m_{\rm P}$. We can see the conjunction of two different effects. First $T_{\rm BN}^1$ slowly decreases for increasing $m_{\rm P}$ which is due to two different reasons: on the one hand, the gravitational force increases linearly with $m_\text{P}$; on the other, the radius of P grows as $m_\text{P}^{1/3}$, restricting the initial conditions that produce bounded non-collisional orbits. Second, the first measured period increases very significantly for growing $r_0$, which is related to the fact that the larger the initial distance, the longer the path to be covered to perform a revolution, and so the bigger $T_{\rm BN}^1$. After comparing both, we conclude that changes in the orbital period due to a small increase in $m_{\rm P}$ are much less relevant than those related to the choice of $r_0$. In the right panel we can see, however, that the relative difference after the first revolution, $\mathrm{\Delta} T^1/T_{\rm N}$ with $\mathrm{\Delta} T^i=T_{\rm BN}^i-T_{\rm N}$,  depends less on the choice of the initial distance than on $m_{\rm P}$. Therefore, to maximize the signal, we should choose the smallest $r_0$ possible  and the largest $m_{\rm P}$ (avoiding that S crashes onto P). A tentative value of  $m_{\rm P} = 1.4 \times 10^{-5}$ g is considered for now to study other possible optimizations.

\begin{figure*}[t]
\centering
\begin{minipage}{0.45\textwidth}
\centering
\includegraphics[width=0.95\textwidth]{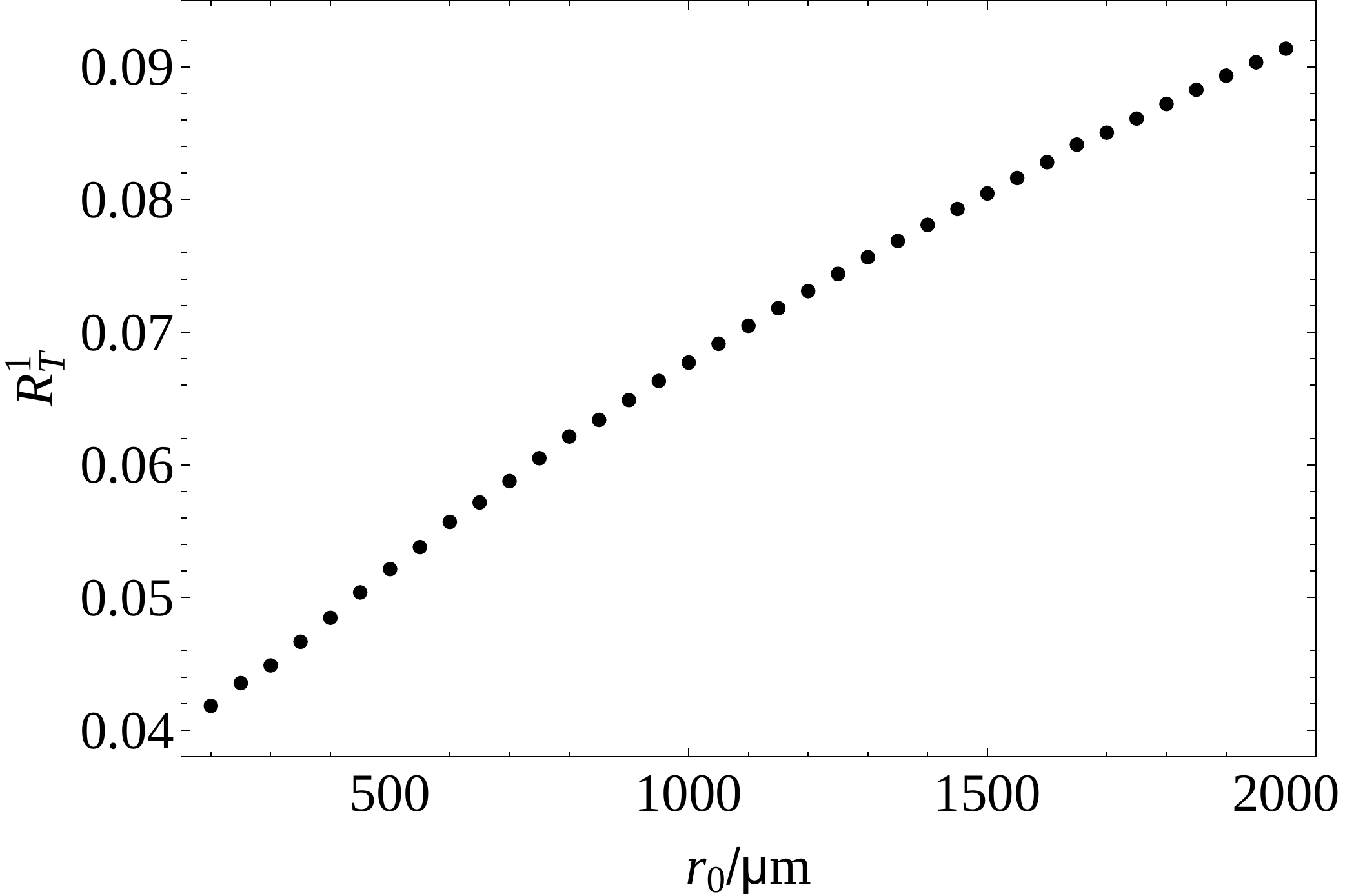} 
\end{minipage}\hfill
\begin{minipage}{0.45\textwidth}
\centering
\includegraphics[width=0.985\textwidth]{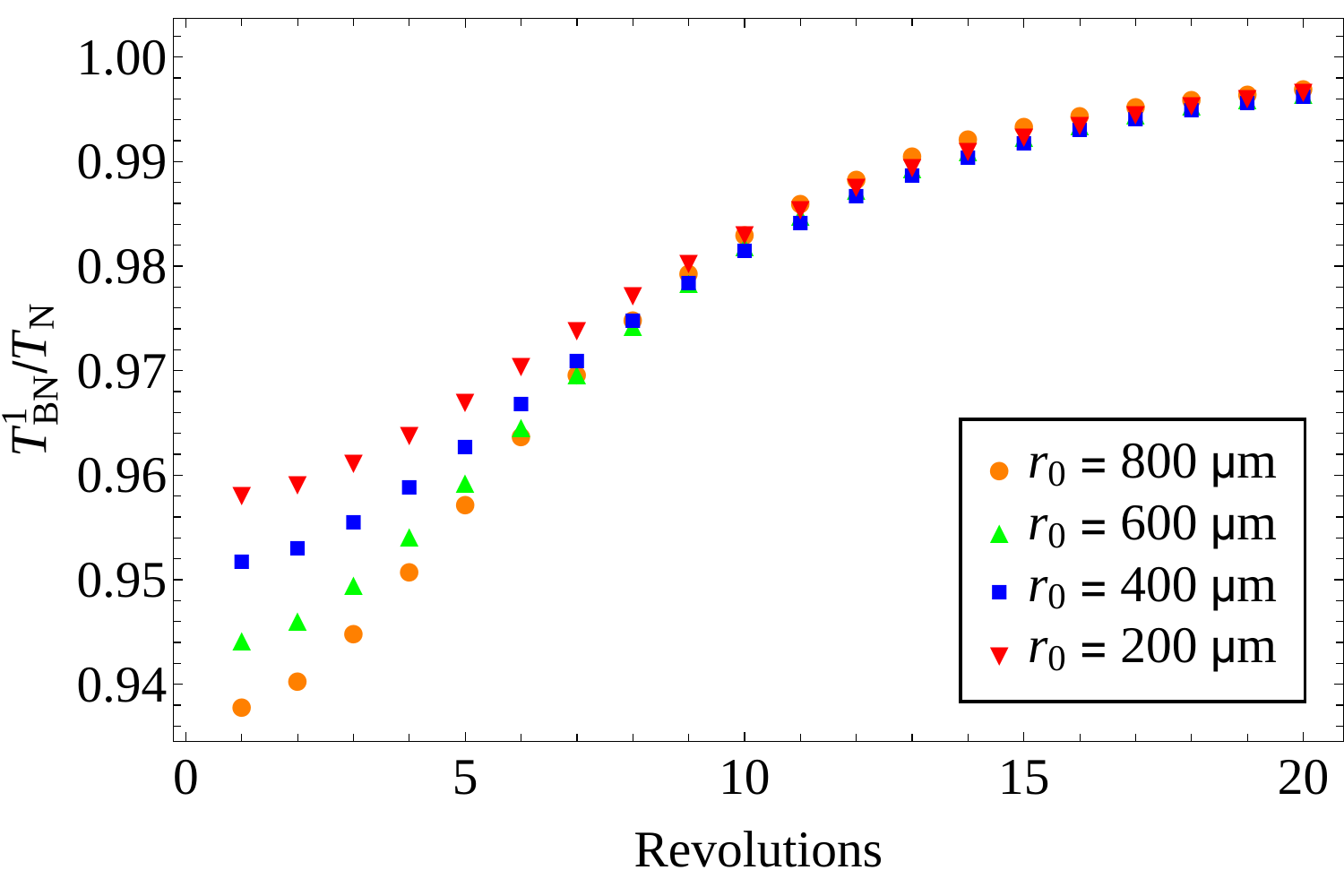}
\end{minipage}
\caption{\it
Left panel: Relative difference between $T_{\rm BN}^1$ and $T_{\rm N}$ as a function of the initial distance $r_0$ for $m_{\rm P} = 1.4 \times 10^{-4}$ \textup{g}. The initial angular velocity $\dot \theta_0$
has been optimized for each $r_0$ so as to maximize $R_T^1$.
Right panel: The ratio $T_{\rm BN}^i/T_{\rm N}$ for the first 20 revolutions for different initial conditions, where
$r_0 = 800$ $\upmu$\textup{m} (orange circles), $r_0 = 600$ $\upmu$\textup{m} (green triangles), $r_0 = 400$ $\upmu$\textup{m} (blue squares), and $r_0 = 200$ $\upmu$\textup{m} (red inverted triangles).}
\label{fig:PeriodVSr}
\end{figure*}

In Fig.~\ref{fig:PeriodVSr} we show a finer scan of the dependence of the sensitivity of the setup on $r_0$ for fixed $m_{\rm P}$, still optimizing $\dot{\theta}_0$ to obtain the maximum eccentricity. In the left panel we see how changing $r_0$ 
 modifies the time needed for S to perform the first revolution around P, $T_{\rm BN}^1$. This increases
monotonically with $r_0$, with a relative difference that can be as large as a factor two when going from $r_0 = 200$ $\upmu$m to $r_0 = 2000$ $\upmu$m. This hints that increasing $r_0$
at fixed $m_{\rm P}$ has an impact on the observable. In the right panel we show how the ratio $T_{\rm BN}^i/T_{\rm N}$ changes for consecutive revolutions 
for several choices of  $r_0$.


Before continuing, it is worth commenting that both the choice of observable and the optimization have been done considering a perfect noiseless system in which the initial conditions are moreover set without error, which is not realistic. 
In App.~\ref{app:IC} the effects of uncertainties in the initial position, $r_0$, and initial velocity, $v_0=(\dot{r}_0^2 + r_0^2 \, \dot{\theta}_0^2)^{1/2}$, are studied in detail. 
In particular, in Tab.~\ref{tab:icvariations} it is shown how an error in fixing the initial conditions would affect $T_{\rm N}$, finding that a $\sim$1\% error in the former may result in a $\sim$5\% variation of the observed period with respect to the expected result. 
This uncertainty would be much larger than the clock precision, which is conservatively taken to be $\sigma_T=1$ s, and so it would make impossible to obtain conclusive results with the proposed observable. For example, for $m_{\rm P} = 1.4 \times 10^{-5}$ g and $r_0 = 200$ $\upmu$m,  we have $T_{\rm BN}^1 \sim 2$ h and $R_T^1 \sim 0.05$ (corresponding to $T_\text{N}-T_\text{BN}^1\sim 360$ s). 
Since such a 5\% difference between the observed period and the expected value of $T_{\rm N}$ can be the consequence of a 1\% error in the choice of the initial conditions, it is clear that neither measuring $T_{\rm BN}^1$, nor repeating this measure over several revolutions would be enough to have a clear signal of non-Newtonianity of the orbit. The main reason is that we compare $T^i_\text{BN}$ to the expected Newtonian period, which is known with large uncertainties due to a possible wrong assignment of the initial conditions. Nevertheless, a wrong choice of initial conditions only changes the value of $T_{\rm N}$ with respect to what is expected a priori, but does not induce any precession. This suggest that a new observable related to the variation of $T_{\rm BN}^i$ over several revolutions is more appealing. Remember, however, that revolutions for gravitational system this size have a typical duration of hours. Even though a long period is an advantage for the sensitivity of the setup, as was discussed before, it is also a drawback since measuring  many revolutions of S around P requires to keep the experimental conditions stable for a long time. 

From this analysis, we draw the following conclusions:
\begin{itemize}
\item The dependence of the observable on the mass of the Planet is mild, whereas its dependence on the initial distance is rather significant.
\item  Measurements of several consecutive revolutions are required.  As, ultimately, we want to be able to repeat the  
         measurement for the largest number of revolutions possible, we may choose a relatively small value of $r_0$ and the smallest possible value of $m_{\rm P}$, in order to reduce 
         $T_{\rm BN}^i$ whilst maximizing the precession. 
\item We need a new observable tightly related to precession.  
\end{itemize}

With this in mind, we  fix the platinum Planet mass to $m_{\rm P} = 0.75 \times 10^{-5} $ g, and take the initial conditions for now to 
         be $r_0=150.0$ $\upmu$m, $\dot{\theta}_0=273.0$ $\upmu$rad s$^{-1}$ and zero initial radial velocity. These should maximize the signal  in a background-less scenario for $\lambda=10$ $\upmu$m and $\alpha=2$.
         In what follows we will show that a slight modification of these initial conditions allows to further increase the sensitivity without affecting any of the features of the orbit. In addition, we will introduce in the next Section another ``safer'' set of initial conditions, more robust against background effects.  

\begin{figure*}[h]
\centering
\begin{minipage}{0.45\textwidth}
\centering
\includegraphics[width=1\textwidth]{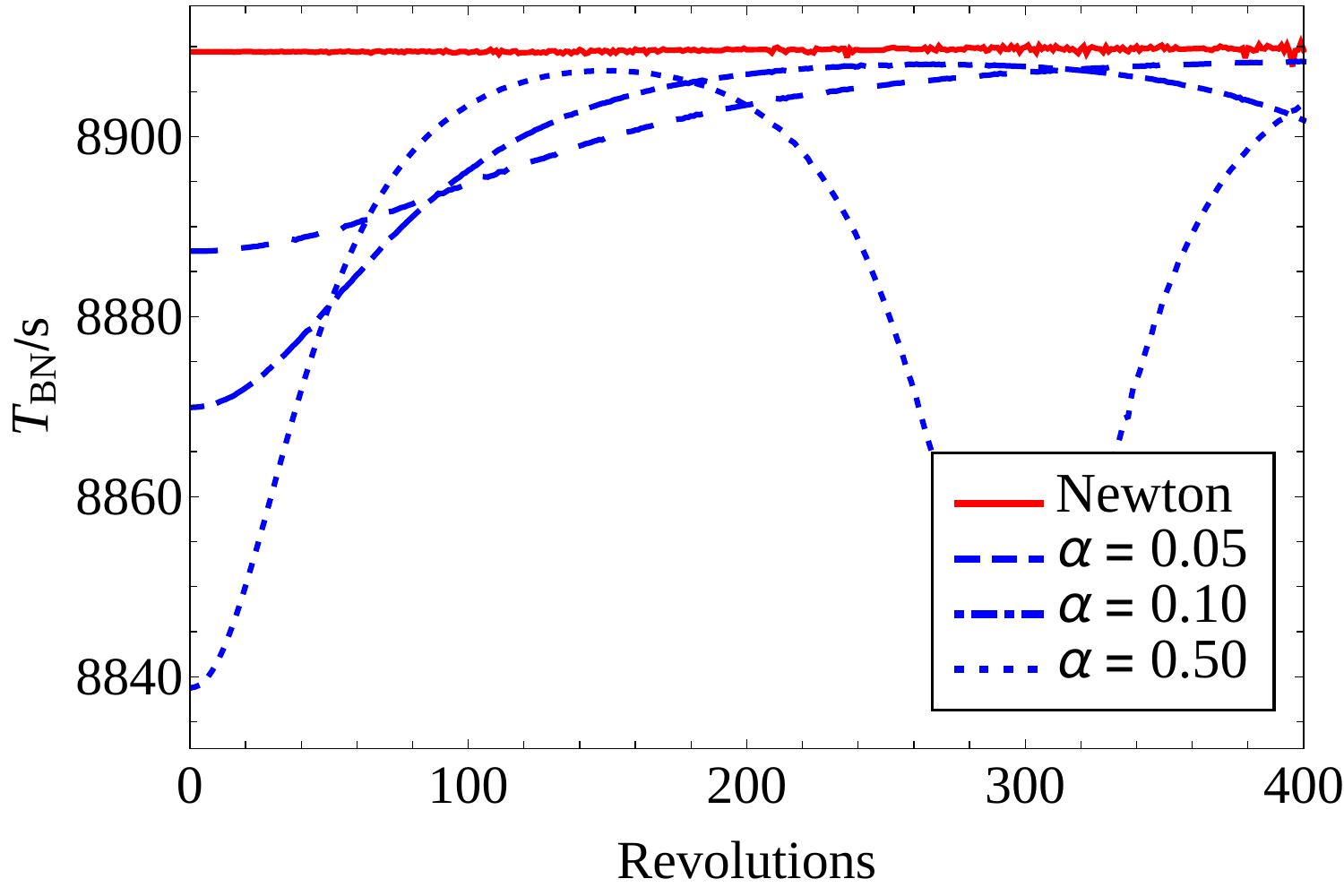} 
\end{minipage}\hspace{1cm}
\begin{minipage}{0.45\textwidth}
\centering
\includegraphics[width=1\textwidth]{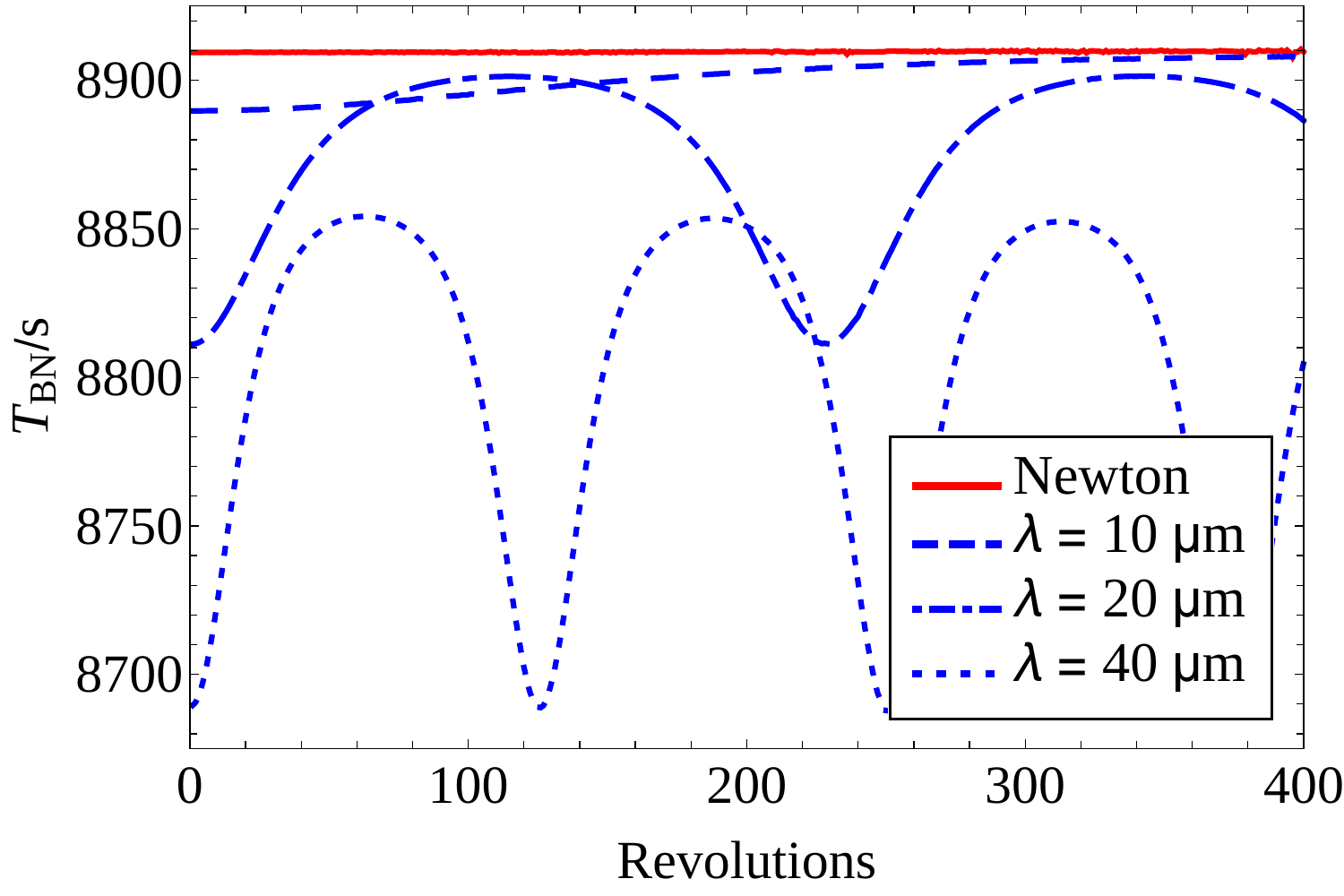}
\end{minipage}

\caption{\it
Revolution period of \textup{S} around \textup{P} (in seconds) for the first 400 revolutions, for different values of ($\lambda, \alpha$). Left panel: $\lambda = 10$ $\upmu$\textup{m}, $\alpha = 0.05$ $\upmu$m, $0.10$ $\upmu$m and $0.50$ $\upmu$m
    (blue dashed, dot-dashed and dotted lines, respectively). Right panel: $\alpha = 0.03$, $\lambda = 10, 20$ and $40$ $\upmu$\textup{m} (blue dashed, dot-dashed and dotted lines, respectively). In red, the constant Newtonian period}
\label{fig:YukawaPeriods}
\end{figure*}

In order to decide for a more convenient observable, we analyze the dependence of $T_{\rm BN}^i$ on $(\lambda, \alpha)$. This is shown in  Fig.~\ref{fig:YukawaPeriods}, where the periods for the first 400 revolutions of S around P are drawn for the initial conditions stated above. It can be seen that the larger $\alpha \, \exp({r_0/\lambda})$,  the bigger the relative difference between $T_{\rm BN}^i$ and $T_{\rm N}$. Moreover, a very significant dependence of $T_{\rm BN}^i$ on the revolution number is observed, with the maximal relative difference with $T_{\rm N}$ for the first revolution, when the apoapsis of the Newtonian and Beyond-Newtonian orbits coincide (as we set $\dot r_0 = 0$). For $\lambda = 10$ $\upmu$\textup{m} and $\alpha = 0.50$ (left panel, blue dotted line), $T_{\rm BN}^i - T_{\rm N}$ ranges from $\sim$ 70 s for $i=1$ to $\sim$ 5 s for $i \approx 150$, and 300 revolutions are needed for the two apoapsis to coincide again. This gets much shorter in the case of  $\lambda = 40$ $\upmu$\textup{m} and $\alpha = 0.03$ (right panel, blue dotted line), where the time needed for the apoapsis to perform a $2 \pi$ revolution is approximately of 130 revolutions. 

All this points out that we need an observable sensible to the period variation over several revolutions. We propose the maximum change of the period between any two measures, over a given number of revolutions:
\begin{equation}
\label{eq:TotalPeriodVariation}
\mathrm{\Delta} T_\text{BN}^{\rm max} (\lambda, \alpha)=\max_{1\leq i,j \leq N_\text{rev}}|T_\text{BN}^i (\lambda, \alpha)- T_\text{BN}^j (\lambda, \alpha)| \, .
\end{equation}
To get some intuition, in case the movement is begun at the apoapsis, this would be the difference between the period of the first $2\uppi$-revolution (corresponding to the minima in Fig.~\ref{fig:YukawaPeriods}) and the period of either the revolution in which the periapsis of the BN orbit coincides with the $\theta=0$ line (the maxima of Fig.~\ref{fig:YukawaPeriods}), or of the last measured period if data is not taken for long enough.

In the absence of statistical noise, this observable can easily detect precession, as in the case of Newtonian motion $\mathrm{\Delta} T_\text{BN}^{\rm max} = 0$. 
Once a positive signal were obtained, a more elaborate observable would be needed to try to determine the values of $(\lambda, \alpha)$. Such observable will be later defined in Sect.~\ref{sec:sens}. In the rest of this section we will study the possible outputs and sensitivity of the proposed setup for the chosen initial conditions, using the proposed observable, and working in the noiseless scenario.

\subsection{Collisional region}
\label{sec:YukawaCollision}

\begin{figure*}[h]
    \centering\hspace{-0.4cm}
    \begin{minipage}{0.33\textwidth}
        \centering
        \includegraphics[width=1.\textwidth]{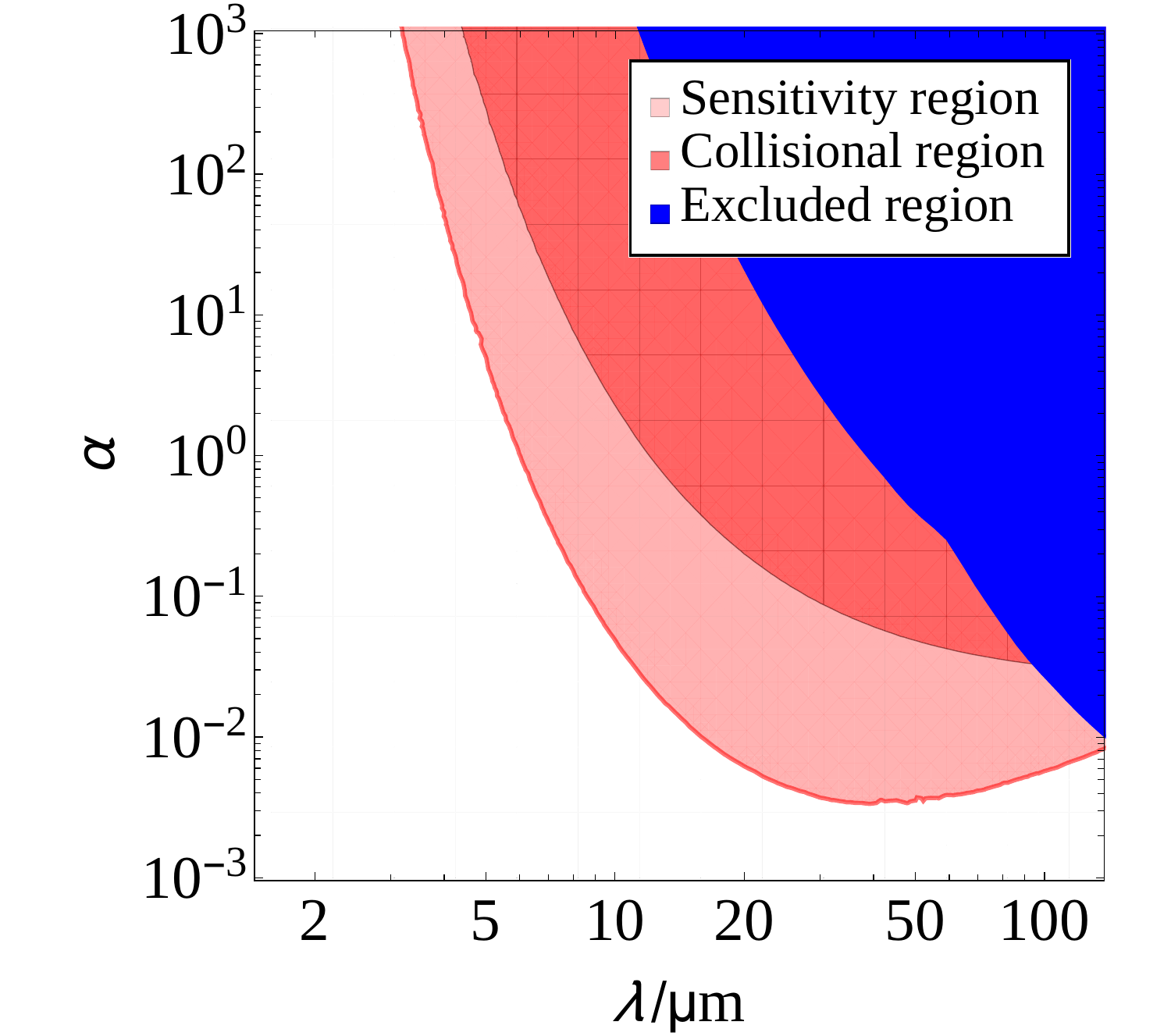}  
    \end{minipage}\hfill
    \begin{minipage}{0.33\textwidth}
        \centering
        \includegraphics[width=1.\textwidth]{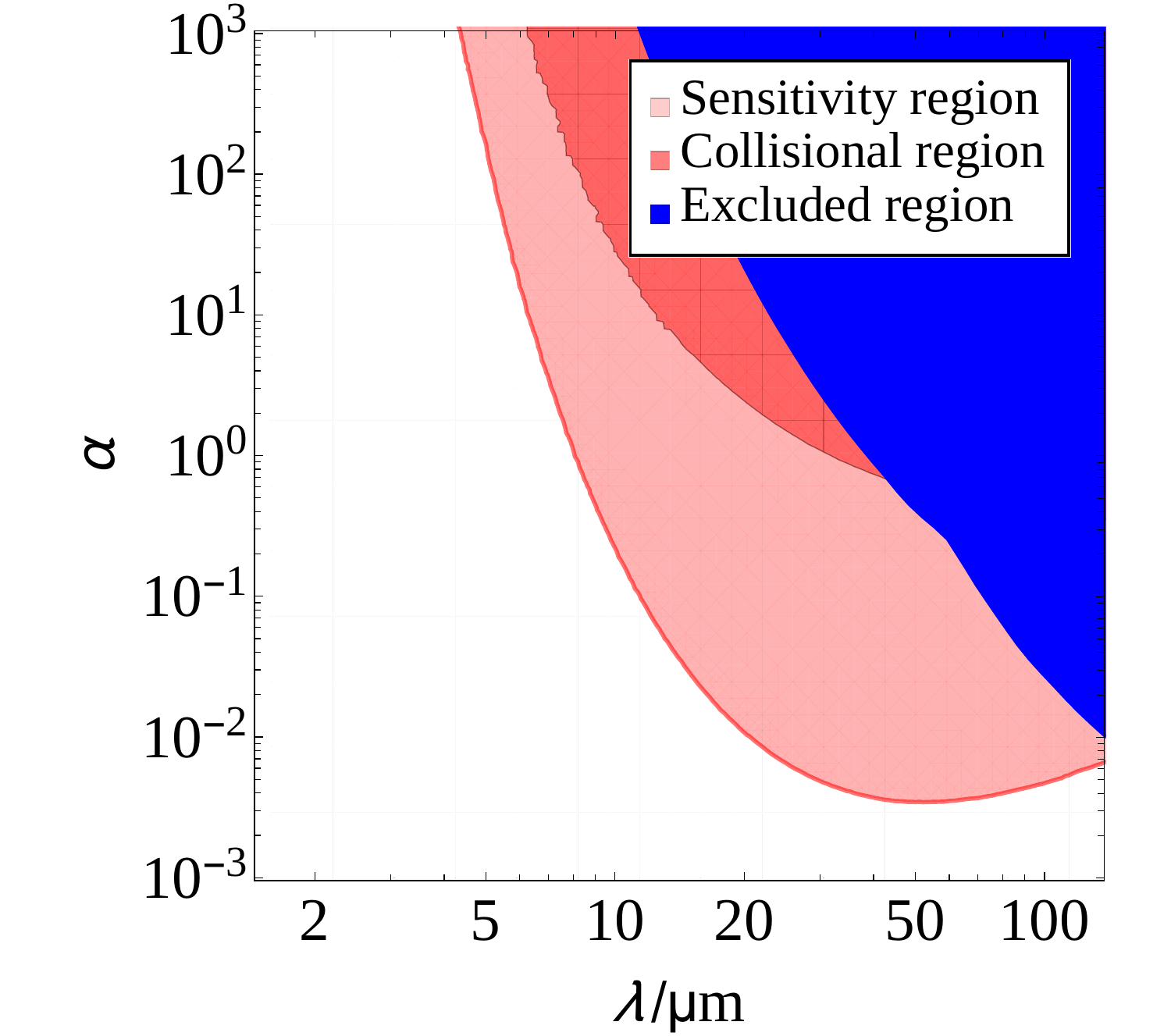} 
    \end{minipage}\hfill
        \begin{minipage}{0.33\textwidth}
        \centering
        \includegraphics[width=1.\textwidth]{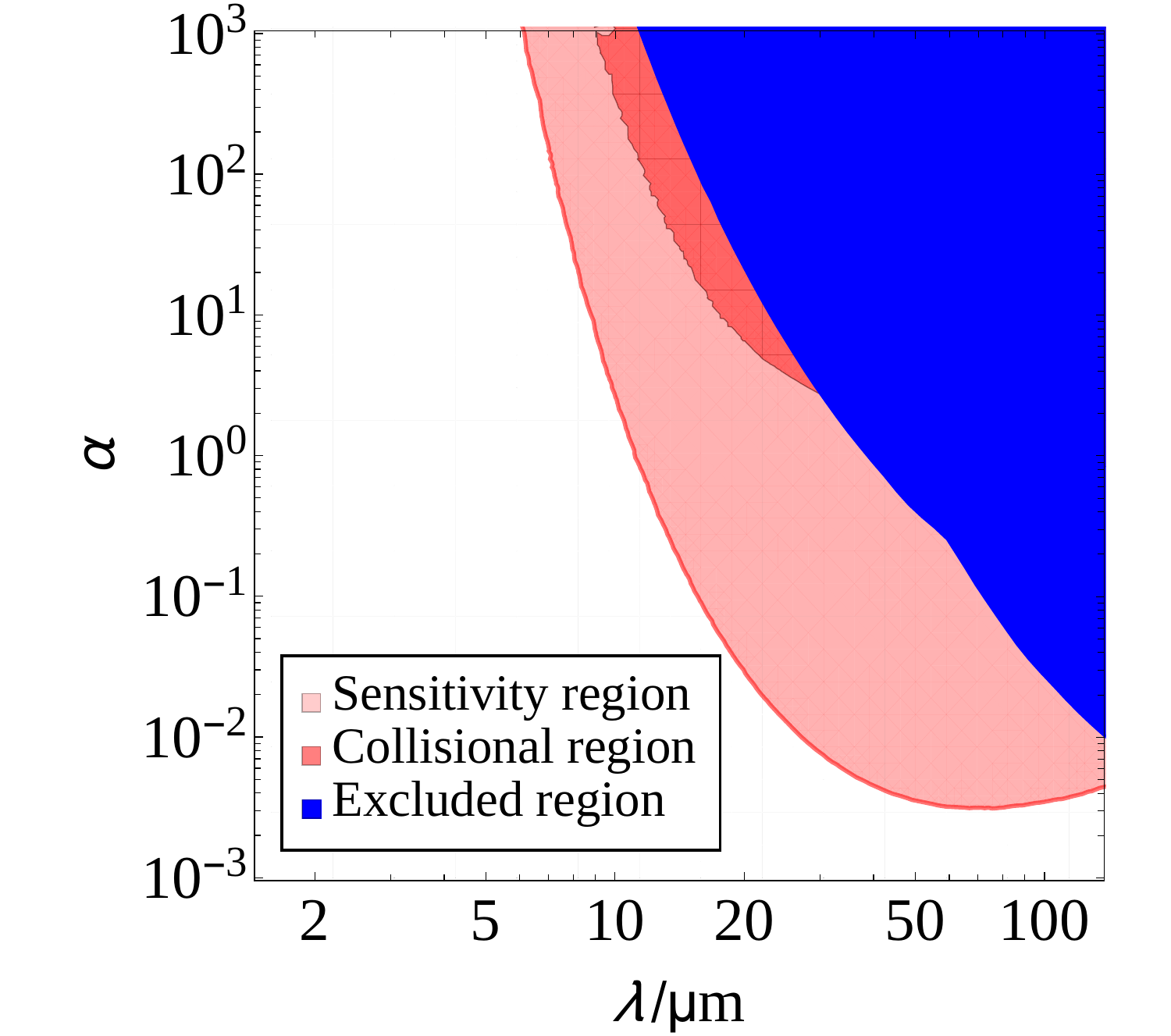} 
    \end{minipage}\hfill
    \caption{\small{\textit{Variation of the size of the collisional region (meshed red) within the region of the parameter space that can be tested with the proposed setup (pink) as a function of the initial conditions.
 Left panel:  $r_0=111.8 \, \upmu$\textup{m}, $\dot{r}_0=30.6$ \textup{nm} \textup{s}$^{-1}$ and $\dot{\theta}_0=491.1 \, \upmu$\textup{rad} \textup{s}$^{-1}$  (corresponding to $r_\text{a} \sim150 \,\upmu$\textup{m}); 
 Middle panel:  $r_0=138.7 \, \upmu$\textup{m}, $\dot{r}_0=13.4$ \textup{nm} \textup{s}$^{-1}$ and $\dot{\theta}_0=363.5 \, \upmu$\textup{rad} \textup{s}$^{-1}$  ($r_\textup{a} \sim160 \,\upmu$\textup{m}); 
 Right panel:  $r_0=177.7 \, \upmu$\textup{m}, $\dot{r}_0=13.4$ \textup{nm} \textup{s}$^{-1}$ and $\dot{\theta}_0=259.3 \, \upmu$\textup{rad} \textup{s}$^{-1}$ ($r_\textup{a}\sim200 \,\upmu$\textup{m} and $r_\textup{p}\sim100 \,\upmu$\textup{m}). 
 The blue region is the area of the parameter space excluded by present experiments. 
 We observe that even in the last case, in which the collisional region is significantly reduced, the sensitivity of the setup is still larger than current bounds.}}}
\label{fig:CollisionRegion}
\end{figure*}

Even if the distance between the surface of the two bodies is larger than 100 $\upmu$m at the apoapsis, it may reduce to less than 
10 $\upmu$m at the periapsis. This is of great importance, as we must take care that the Satellite does not crash onto the Planet 
if we desire to observe one or more complete bounded orbits and take full advantage of precession measurements. As the experimental setup was designed to produce an optimal signal for 
$\lambda = 10$ $\upmu$m and $\alpha =2$, the initial conditions lead to a spatial separation between both spheres at the periapsis that is minimal in such case. 
Should $\lambda$  or $\alpha$ be larger (and thus the potential stronger), we would find a situation for which both bodies collide before a full revolution is completed. A similar situation could occur, even in the Newtonian case, in the presence of external backgrounds.
In that case, it would be impossible to measure the revolution period. We can determine the region of the ($\lambda,\alpha$) plane for which S crashes onto P (``{\em collisional region}'') and, within it, it is still possible to measure differences between the Newtonian and the non-Newtonian behaviour (see Fig.~\ref{fig:linearmotion}). 
In general, we may have three different situations: 
\begin{enumerate}
\item
Both the Newtonian and BN trajectories bring S to collide with P. In this case, we should measure the time required for S to crash onto P and compare it with the expectation 
in case of a Newtonian potential.
\item For some particular choices of $\lambda$ and positive $\alpha$, S may collide with P in the case of the Beyond-Newtonian potential,  while it may not for the Newtonian one; 
or, for negative $\alpha$, the opposite may occur. In these two cases, observation of an unexpected event (either collision when we expect a closed orbit or the opposite) 
would indeed be a striking experimental signal. 
\item Finally, we could have that both the Newtonian and BN trajectories produce bounded orbits. 
\end{enumerate}
We are not going to study in detail any of the two first possibilities, as we want to focus on the region of the parameter space for which a bounded orbit is observed. Nevertheless, when studying the sensitivity
of the setup in the $(\lambda,\alpha)$ plane, we will distinguish the collisional region from the rest of the parameter space. 
In this region we consider collision as a distinctive feature of a Beyond-Newtonian potential (as we mainly focus on the case $\alpha>0$ and choose initial conditions that produce a non-collisional closed orbit in the Newtonian case).
The relevance of the collisional region in the parameter space to be tested can be reduced by an appropriate choice of initial conditions, though. This is shown in Fig.~\ref{fig:CollisionRegion},
where we present the sensitivity of the experiment as will be explained in Sect.~\ref{sec:YukawaLimitsNoBack} (depicted in pink), together with the size of the collisional region (in meshed red), for different choices of the initial conditions.   It gets clear that even with suboptimal choices of $\{r_0, \dot{r}_0, \dot{\theta}_0 \}$ it is possible to reduce
significantly the size of the collisional region with no huge impact on the sensitivity. Note that, as we are using 
a value of $\dot{r}_0 \neq 0$, the initial position of S does not correspond to the apoapsis. The reason why $\dot{r}_0\neq 0$ is used will become clear next.

\subsection{Experimental limits with no backgrounds} 
\label{sec:YukawaLimitsNoBack}

\begin{figure*}[t]
    \centering
    \begin{minipage}{0.5\textwidth}
        \centering
        \includegraphics[width=1\textwidth]{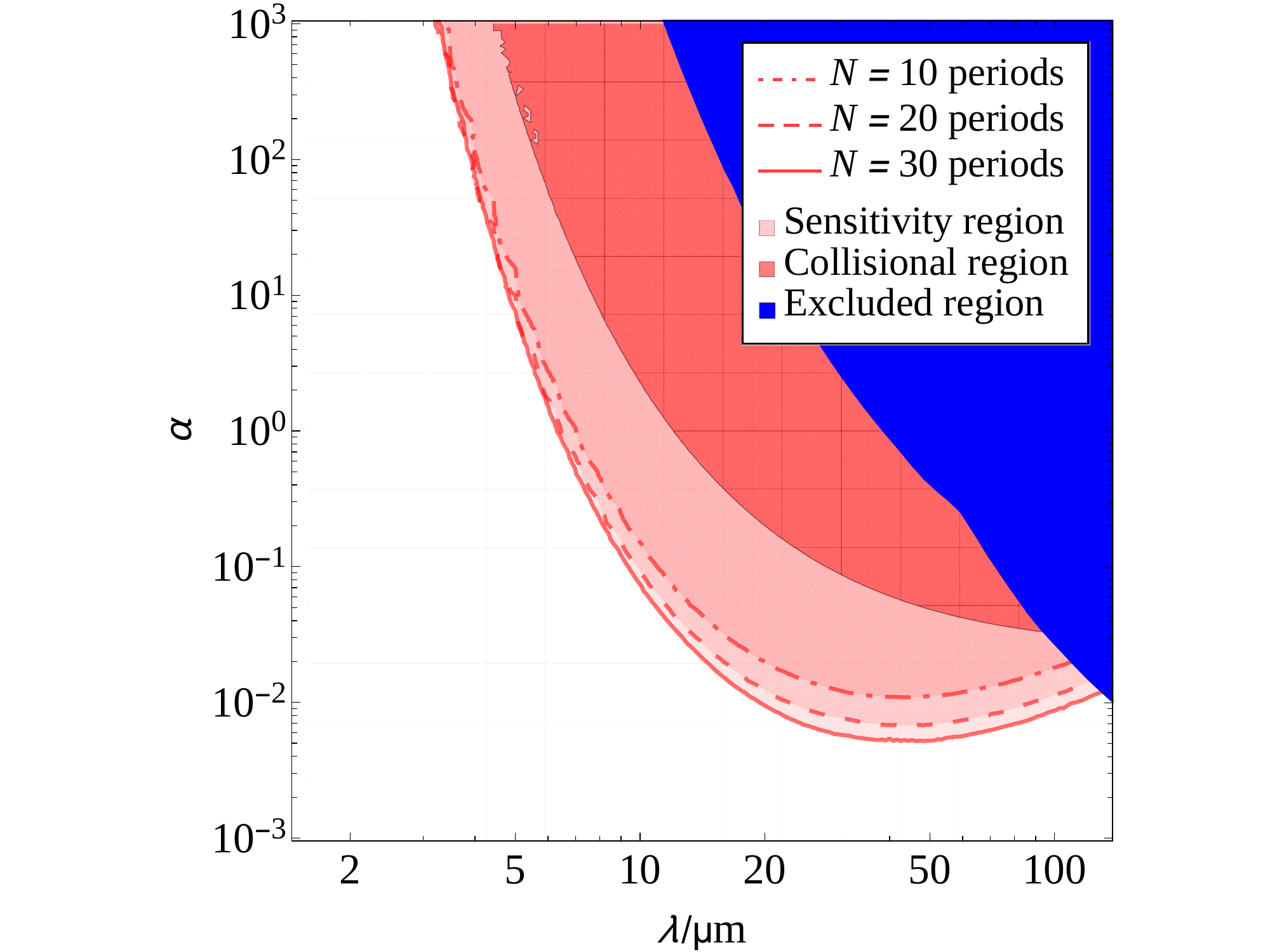} 
    \end{minipage}\hfill
    \begin{minipage}{0.5\textwidth}
        \centering
        \includegraphics[width=1\textwidth]{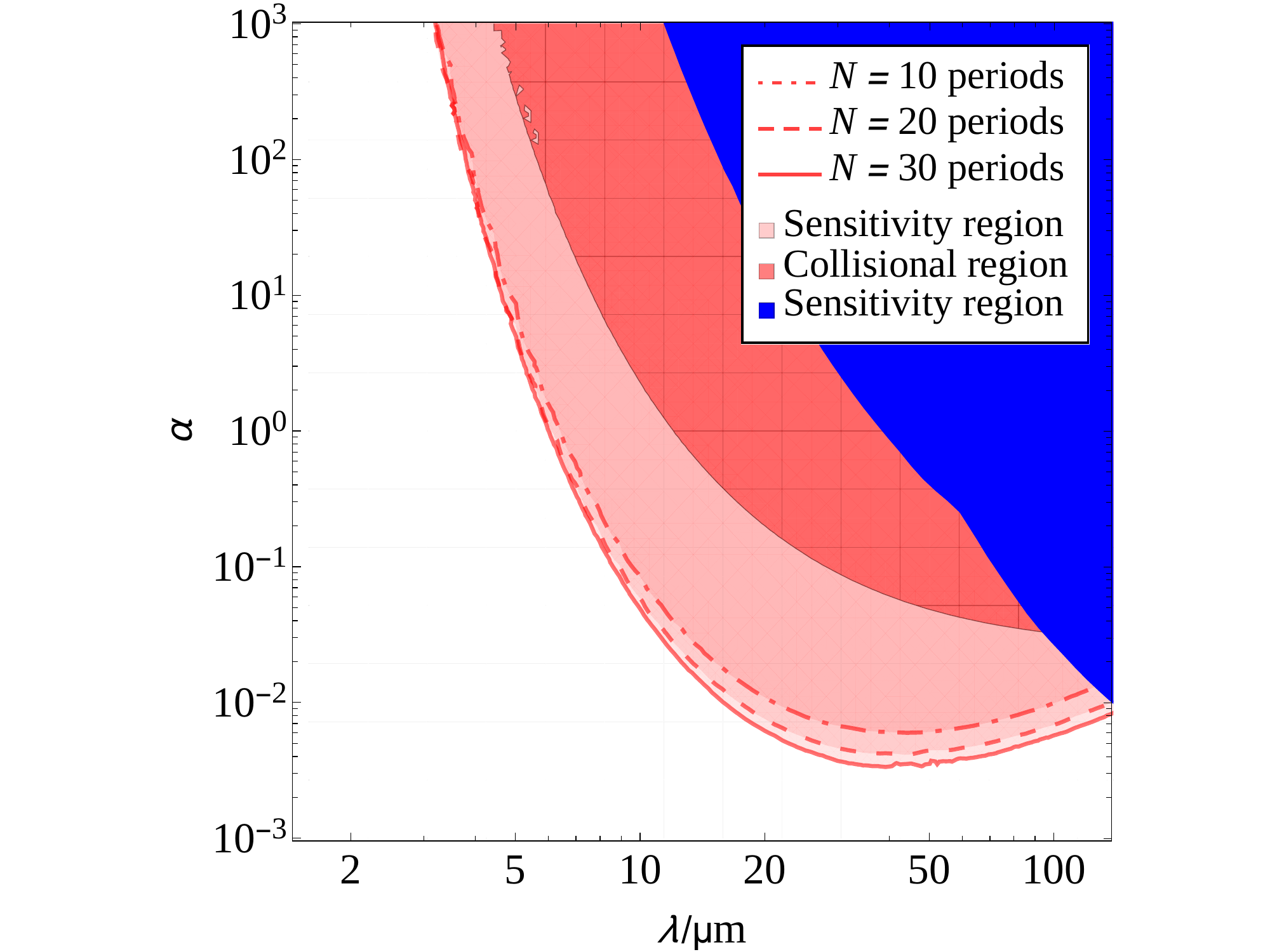} 
    \end{minipage}
    \caption{\small{\textit{
    Sensitivity of the proposed experimental setup in the ($\lambda,\alpha$) plane at the 95\% CL 
    The blue region represents the present experimental bounds (see Fig.~\ref{fig:AlphaLambdaBounds}). 
    The dark red-meshed region is the part of the parameter space for which a Beyond-Newtonian potential induces collision between \textup{S} and \textup{P} (whereas for a Newtonian potential \textup{S} would 
    orbit around \textup{P}). Eventually, the pink-shaded region is the part of the parameter space for which we can detect precession in the presence of New Physics (and that can be thus excluded  in 
    the case no signal is observed). The boundary of the sensitivity region is drawn for values of $\lambda$ and $\alpha$ for which we expect the maximal variation of the period, $\mathrm{\Delta} T_\text{BN}^{\rm max}$, 
    defined in eq.~(\ref{eq:TotalPeriodVariation}), to be as large as 2 \textup{s} (twice the assumed clock precision), 
    for $N_{\rm rev}=10$ (dot-dashed lines), $N_{\rm rev}=20$ (dashed lines), and $N_{\rm rev}=30$ revolutions (solid lines). 
    Left panel: the initial conditions are chosen so that the trajectory of S around P starts at the apoapsis:
    $r_0=r_{\rm apo}=150\,\upmu$\textup{m} and $\dot{\theta}_0=273.0$ $\upmu$\textup{rad} \textup{s}$^{-1}$; 
    Right panel: the initial conditions are chosen so that the trajectory of S around P starts
    at a point between the periapsis and the apoapsis, by choosing a non-vanishing initial radial velocity:
    $r_0=111.8$ $\upmu$\textup{m}, $\dot{r}_0=30.6$ \textup{nm} \textup{s}$^{-1}$ and $\dot{\theta}_0=491.4$ $\upmu$\textup{rad} \textup{s}$^{-1}$. 
    }}}
        \label{fig:Limits1}
\end{figure*}

To understand the impact of backgrounds on the experimental setup, we first present  in Fig.~\ref{fig:Limits1} our sensitivity in their absence using the observable in eq.~(\ref{eq:TotalPeriodVariation}). The blue-shaded area represents the currently excluded region of the parameter space; the red-meshed area is the collisional region 
described in the previous Subsection, and the pink-shaded region represents the sensitivity limit of the experimental setup.
The dot-dashed, dashed and solid  lines stand for $N_{\rm rev} = 10, 20$ and $30$, respectively\footnote{We have not studied larger revolution numbers, as maintaining experimental conditions over several days (30 revolutions take more than 2 days, for the considered initial conditions) may be difficult.}. 
The border of each region represents the values of $(\lambda,\alpha)$ for which the observable $\mathrm{\Delta} T_{\rm BN}^{\rm max}$ is
 twice the expected clock error, $\sigma_{T}=1$ s, {\em i.e.} it delimits the area that can be tested by the proposed setup at 95\% CL.
 
The left panel shows the sensitivity for a Satellite that starts its orbit around the Planet at the apoapsis  with $r_0=r_{\rm a}=150.0\,\upmu$m and $\dot{\theta}_0=273.0$ $\upmu$rad s$^{-1}$. We can see that the more revolutions considered, the larger the sensitivity of the experimental setup, specially for the region of medium-sized $\lambda$ and low $\alpha$ . 
On the other hand, the ultimate sensitivity to $\lambda$ is not significantly dependent on $N_{\rm rev}$.
In the right panel we consider a starting position of S at an intermediate point between the periapsis and the apoapsis, by choosing a non-vanishing radial velocity: 
$r_0=111.8$ $\upmu$m, $\dot{r}_0=30.6$ nm s$^{-1}$ and $\dot{\theta}_0=491.4$ $\upmu$rad s$^{-1}$. 
These initial conditions correspond to a point different from the apoapsis but belonging to the same  BN orbit (with $\lambda=10$ $\upmu$m, $\alpha=2$) as  the previous ones.
We can see that if $\dot{r}_0\neq 0$ is considered, there is some increase in the sensitivity at low $\alpha$ for the considered observable: 
we can reach $\alpha \approx 3 \times 10^{-3}$ in the right panel, compared to $\alpha \approx 5 \times 10^{-3}$ in the left one.
This is because the revolution period $T_\text{BN}^i$ changes slowly when measured near the periapsis or the apoapsis (see Fig.~\ref{fig:YukawaPeriods}), 
whereas it changes more rapidly in the intermediate region between the two extremes. If we can only measure a given number of revolutions (due to problems related to maintaining a stable setup for several days), considering initial conditions 
such that the movement starts at some point different from the apoapsis and periapsis implies a faster variation of $T_{\rm BN}^i$ and, thus, a larger sensitivity. Notice that changing the initial conditions does not make more challenging 
the experimental setup and, thus, is a zero-cost improvement.

\section{Optimization in the presence of backgrounds}
\label{sec:backgrounds}

We are now in position to study the sensitivity of the proposed setup to deviations from Newton's $1/r^2$ law parameterized by the Yukawa potential of eq.~(\ref{eq:YukawaPotential}) in the presence of generic backgrounds. We will study the effect of background sources for two different choices of the initial conditions: 
\begin{itemize}
\item {\bf Case 1}: 
This is the case of maximal sensitivity, that corresponds to an orbit with maximal eccentricity. The initial conditions are $r_0=111.8$ $\upmu$m, $\dot{r}_0=30.6$ nm s$^{-1}$ and $\dot{\theta}_0=491.4$ $\upmu$rad s$^{-1}$, with an expected orbit with $r_{\rm a}\sim 150$ $\upmu$m and $T_{\rm N}\sim2$ h 30 min. 

\item {\bf Case 2}: 
The second case is a more conservative one.  The setup is slightly less sensitive to the signal, but much less affected by possible backgrounds (in particular, the potentially troublesome electric Casimir effect, 
see App.~\ref{app:Casimir}). In this case we use $r_0=177.7$ $\upmu$m, $\dot{r}_0=13.4$ nm s$^{-1}$ and $\dot{\theta}_0=259.2$ $\upmu$rad s$^{-1}$, with $r_{\rm a}\sim 200$ $\upmu$m and $T_{\rm N}\sim4$ h 30 min. 
\end{itemize}
The first case represents the maximum sensitivity  for the considered setup, situation that would only be achieved with a fine control of the possible noise sources, 
whilst the second choice corresponds to a simpler scenario, with a more straightforward calibration phase. In order to set these initial conditions, it was suggested in Ref.~\cite{Donini:2016kgu}  to use a precisely calibrated laser to put the Satellite into motion with the desired radial and angular velocity (see Ref.~\cite{Kobayashi:2012}).

\subsection{Qualitative effects of backgrounds on revolution period}
\label{sec:Qualitative}

So far, we have identified the region of the parameter space for which the proposed experiment will be sensitive to New Physics, and determined within it the part in which collision between S and P may occur. These depend on the chosen initial conditions. Now
we can address the impact of possible backgrounds which, added to the Newtonian potential, may originate a precession signal similar to the BN one, invalidating possible results.
The most important expected backgrounds are the following:
\begin{enumerate}
	\item Electrostatic forces: Coulombian (monopolar) forces go like $1/r^{2} $ and therefore, would not induce precession. Other electric forces, such as multipolar or Van der Waals ones
	have a more exotic behaviour related to the actual shapes of the  bodies, but are expected to be reduced by a proper choice of 
	materials. Recall that the Satellite of the setup proposed here is a diamagnetic sphere a made of pyrolitic graphite, with perpendicular and parallel conductivities     
	$ \sigma_\perp \in [4,6.6]$ S cm$^{-1}$ and $ \sigma_\| \in [2.2,2.8]$ S cm$^{-1}$, respectively, to be compared to that of a conducting material (such as iron, 
	 for which $ \sigma \sim  10^5$ S cm$^{-1}$ \cite{Raheem:19}).
	\item Electric Casimir force: The Casimir force between two parallel conducting planes at distance $d$
	 goes as $1/d^4$ for small separations, whereas for a conducting sphere and a plane (with separation $d$ between surfaces much smaller than the radius of the sphere, $R$) 
	 it goes as $R/d^3$. In the	 case of two conducting spheres at separation $d$ between surfaces much smaller than both radii ($R_1$ and $R_2$), it is proportional to 
	 $ 1/d^3 \times R_1 R_2/(R_1+ R_2)$ \cite{Teo:2011aa}. Due to the choice of materials, being P a conductor and S an insulator, the Casimir force is expected to be repulsive 
	 and reduced compared to the case of two ideal conductors. 
	 
	\item General relativity effects induce a correction to the gravitational force proportional to $1/r^4$. They have been studied in Ref.~\cite{Donini:2016kgu}, 
	finding that they are irrelevant in practice for the considered setup.
	\item Gravitational Casimir forces may also be induced between the two bodies. This force has been computed to go as $1/r^6$ in the plane-plane geometry (see, {\it e.g.,} Ref.~\cite{Hu:2016lev}), so we expect that for the proposed setup they would also be strongly marginal.
\end{enumerate}
In App.~\ref{app:Casimir}  we discuss the most relevant among these backgrounds, focusing mainly on the electric Casimir effect.

As we have previously commented, dominant backgrounds of electrical origin have the same $r$-dependence as the Newtonian gravitational potential. However, other backgrounds would have a different $r$-dependence, 
as it is the case of General Relativity corrections or of the electric Casimir force. In order to study the impact of generic background sources, we can define a {\em modified-Newtonian potential} (mN) 
assuming the following functional form:
\begin{equation}
\label{eq:mNpotential1}
V_{\rm mN}(r)=-\frac{G_{\rm N} \, m_{\rm P}}{r}\left(1+Q_2+\frac{Q_3}{r}+\frac{Q_4}{r^2} + \dots \right),
\end{equation}
where the names $Q_k$ refer to the $1/r^k$-dependence of the corresponding terms in the modified gravitational force. For simplicity, we have not considered terms with $k > 4$ that should be sub-dominant (we will estimate their impact on the sensitivity in Sect.~\ref{sec:sens}, though). The motivation behind this particular choice of modified potential is the following:  all the foreseeable backgrounds listed above can be written as central potentials
(as is the case of most electrically-induced corrections, General Relativity correction, and terms related to Casimir force, either electrical or gravitational). 
Deviations from homogeneity or isotropicity\footnote{
One possible background that we are not considering here and that could fall into this category is that originating from inhomogeneities
in the magnetic field used to levitate S, assuming it negligible with respect to other background sources. A careful calibration of the setup used to levitate the Satellite
should be performed before the start of the experiment, though.  We refer the reader to App.~\ref{app:MC} for details on the magnets setup.
}
 could also be parameterized in a polynomial expansion as long as the typical length scale of the deviations is much smaller than the distance between S and P. 

In the rest of the section, we will consider attractive backgrounds, only. This means that $Q_2$, $Q_3$ and $Q_4$ will all be taken to be positive. There are several reasons why this is done. 
First, many of the possible backgrounds are expected to be attractive, and so the presence of additional repulsive effects would only counter them. 
Second, for a given absolute value of the background the impact on the experiment would be higher should it be attractive. 
Finally, as we are focusing on $\alpha>0$, precession induced by a dominant repulsive background would be clearly distinguishable from the desired signal, since it would proceed in the opposite sense as the one expected from BN effects. Therefore, this choice puts us in the worst case scenario. 
Knowledge of the sign of the background sources from a calibration phase will, thus, tell us in advance if the sensitivity of the setup is maximal for positive or negative $\alpha$. The same analysis could easily be repeated for repulsive backgrounds.

\begin{figure}[t]
    \centering
    \includegraphics[width=0.43\textwidth]{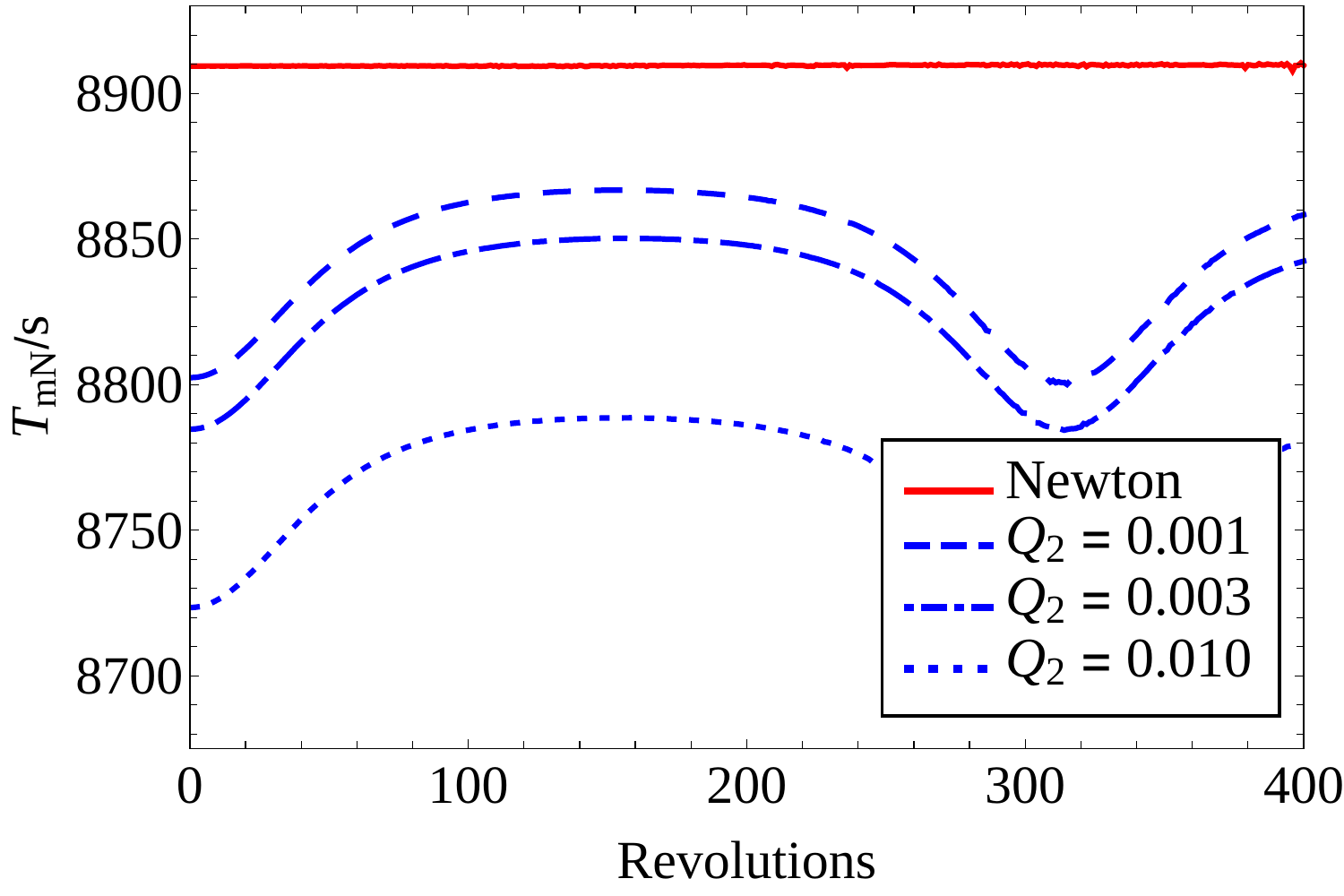} 
    \caption{\small{\textit{Evolution of the orbital period of \textup{S} around \textup{P} (in seconds) during the first 400 revolutions for different values of $Q_2$ in the mN-potential,  and constant $Q_3=0.2$ $\upmu$\textup{m} and $Q_4=1$ $\upmu$\textup{m}$^2$. 
    We observe the $Q_2$ term has no measurable effect in the precession, that is due to non-vanishing $Q_3$ and  $Q_4$. 
    However, it reduces the period of S, as it acts as an effective increase of the gravitational constant, $G_{\rm N}\rightarrow G_{\rm N} (1 + Q_2)$. In solid red, we depict the constant Newtonian period.}}}
        \label{fig:BackgroundPeriodQ2}
\end{figure}

We can now qualitatively study the movement of S under the action of a mN-potential in the absence of BN effects. We observe in eq.~\ref{eq:mNpotential1} that $Q_2$ is a background which can be added to the strength of the $1/r$ Newtonian term 
in the potential with the effect of increasing the effective Newton's constant, $G_{\rm N} \to G_{\rm N} (1 + Q_2)$.  Therefore, as observed in Fig.~\ref{fig:BackgroundPeriodQ2}, it only produces a reduction of the period, which is the expected effect
for all of the dominant electrically-induced attractive backgrounds. As previously stated, such backgrounds would not affect the sensitivity of the experiment, although they must be controlled 
since they could lead to an undesired collision between S and P.
The variation of the period between consecutive revolutions, observable in the plot, signals a precession of the orbit. It is, however, induced non-vanishing values of $Q_3$ and $Q_4$ ($Q_3 = 0.2$ $\upmu$m and $Q_4 = 1$ $\upmu$m$^2$ in Fig.~\ref{fig:BackgroundPeriodQ2}).

\begin{figure*}[t]
    \centering
    \begin{minipage}{0.5\textwidth}
        \centering
        \includegraphics[width=0.95\textwidth]{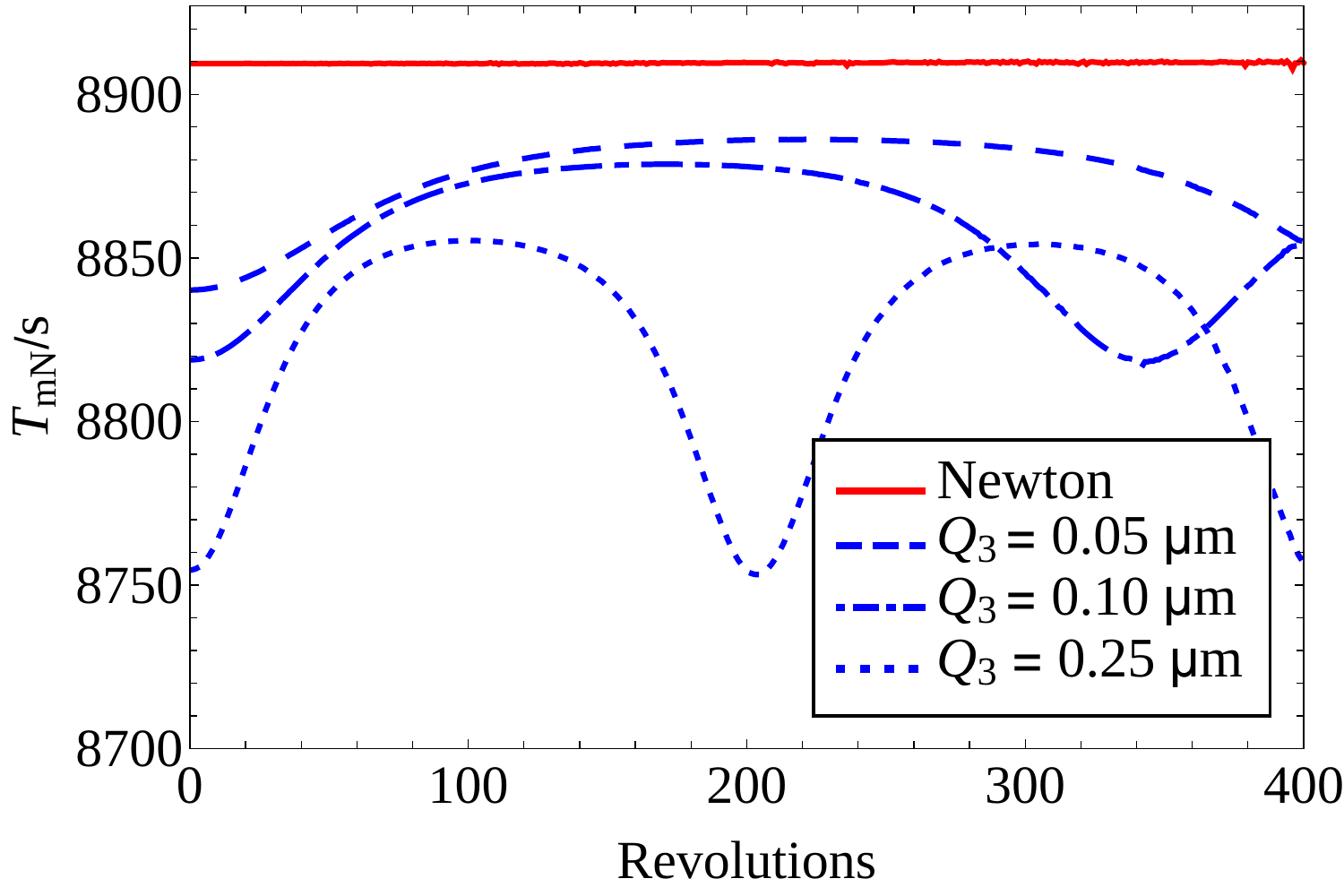} 
    \end{minipage}\hfill
    \begin{minipage}{0.5\textwidth}
        \centering
        \includegraphics[width=0.95\textwidth]{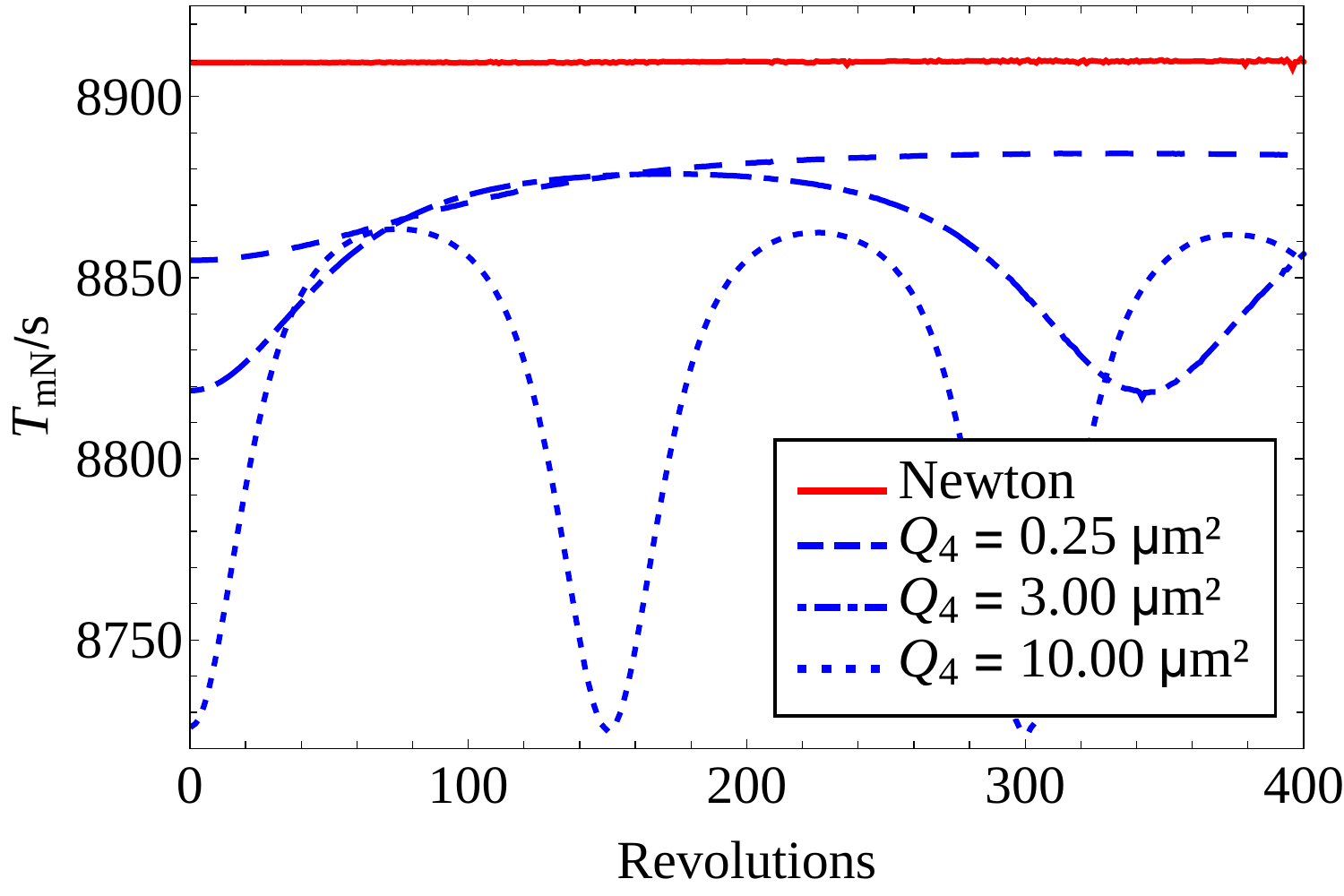} 
    \end{minipage}
    \caption{\small{\textit{Evolution of the orbital period of \textup{S} around \textup{P} (in seconds) during the first 400 revolutions for different values of the parameters in the mN-potential. In each panel we keep constant two parameters and vary the last one, being 
    it $Q_3$ on the left (for $Q_2=0.001$, $Q_4=3$ $\upmu$\textup{m}$^2$) and $Q_4$ on the right (for $Q_2=0.001$, $Q_3=0.1$ $\upmu$\textup{m}). We observe that both $Q_3$ and $Q_4$ produce similar effects, 
    albeit the former has a bigger impact.}}}
        \label{fig:BackgroundPeriodQ3Q4}
\end{figure*}

The two background sources $Q_3$ and $Q_4$ have similar implications on the revolution time, as is depicted in Fig.~\ref{fig:BackgroundPeriodQ3Q4}. In the left panel, we show the $Q_3$-dependence
of the revolution time for fixed $Q_2 = 0.001$ and $Q_4 = 3$ $\upmu$m$^2$. In the right panel, we show the same for $Q_4$, with $Q_2=0.001$, $Q_3=0.1$ $\upmu$m. Both background sources
induce a reduction of the average period and a separation between the maximum BN period values and the Newtonian one, together with a notorious precession similar to that observed in Fig.~\ref{fig:YukawaPeriods}. 
Due to the softer $r$-dependence, the impact of $Q_3$ is, however, larger than that of $Q_4$ (notice the vertical scale). 

In conclusion, both mN- and BN-potentials produce qualitatively similar signals.
Hence the question is how much should backgrounds be controlled so that any positive signal can be correctly attributed to New Physics. To this end, we will compare the possible outputs of a purely BN-potential to the mN-case to see for which values of the background parameters the signal produced by the first can be statistically distinguishable from one generated by the second\footnote{In case some background source is accurately known from a calibration phase, this same analysis could be repeated including it to the BN case.}. All statements will be made for a 95\% CL, meaning the signals produced by the mN-potential would be at least 2 s smaller than those produced in the BN case.
Once a measurement with some clear BN signature is made, it would be necessary to combine New Physics and background effects in a single potential, and to use a more elaborated observable to be able to precisely determine a point in the $( \lambda, \alpha)$ plane. A study of the positive signal scenario is made in Sect.~\ref{sec:positive}.

 
\subsection{Impact of backgrounds on the experimental limits}
\label{sec:backlimits}

We will now study whether the proposed setup is
capable of producing a signal statistically different from one generated by the mN-potential. First, we will study the amount of noise that could be allowed if we want to avoid collision. Second, we will determine which levels would still allow to differentiate some BN signal from the background effect in most of the sensitivity region of the experiment.


\begin{figure*}[h]
    \centering\hspace{-1.1cm}
    \begin{minipage}{0.333\textwidth}
        \centering
        \includegraphics[width=1.2\textwidth]{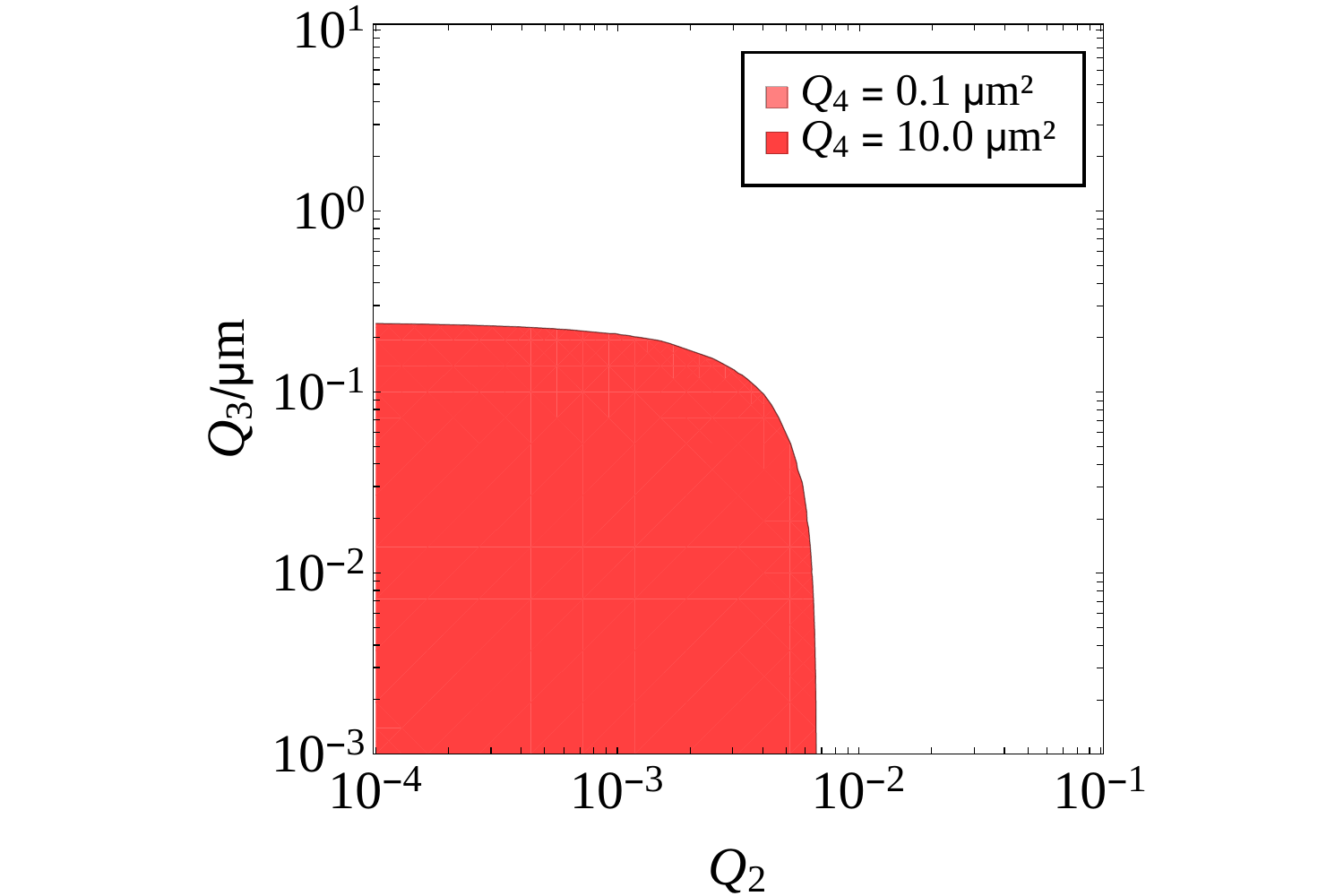} 
    \end{minipage}\hfill
    \begin{minipage}{0.333\textwidth}
        \centering
        \includegraphics[width=1.2\textwidth]{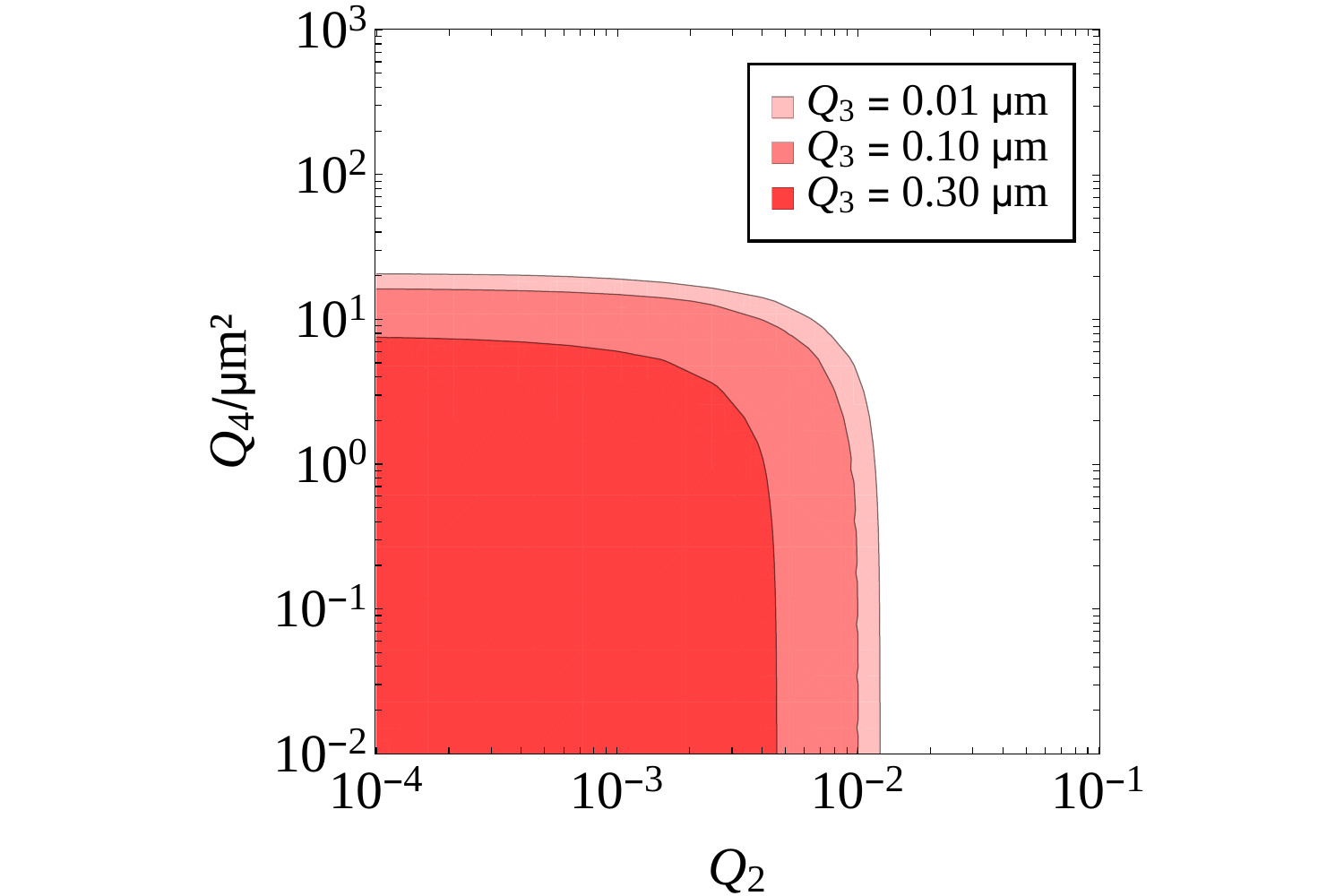} 
    \end{minipage}\hfill
        \begin{minipage}{0.333\textwidth}
        \centering
        \includegraphics[width=1.2\textwidth]{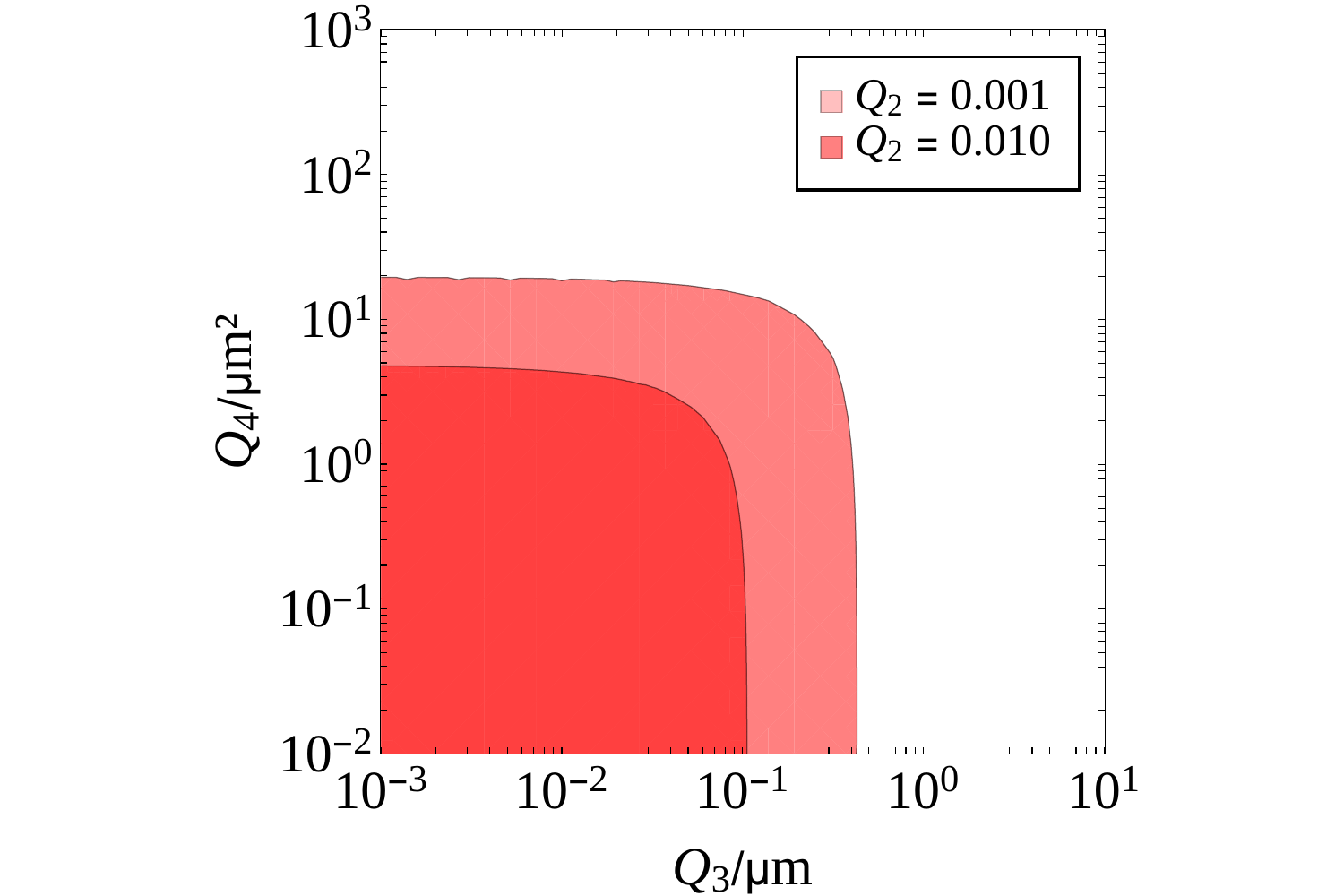} 
    \end{minipage}\hfill
    
	\centering\hspace{-1.1cm}       
    \begin{minipage}{0.333\textwidth}
        \centering
        \includegraphics[width=1.2\textwidth]{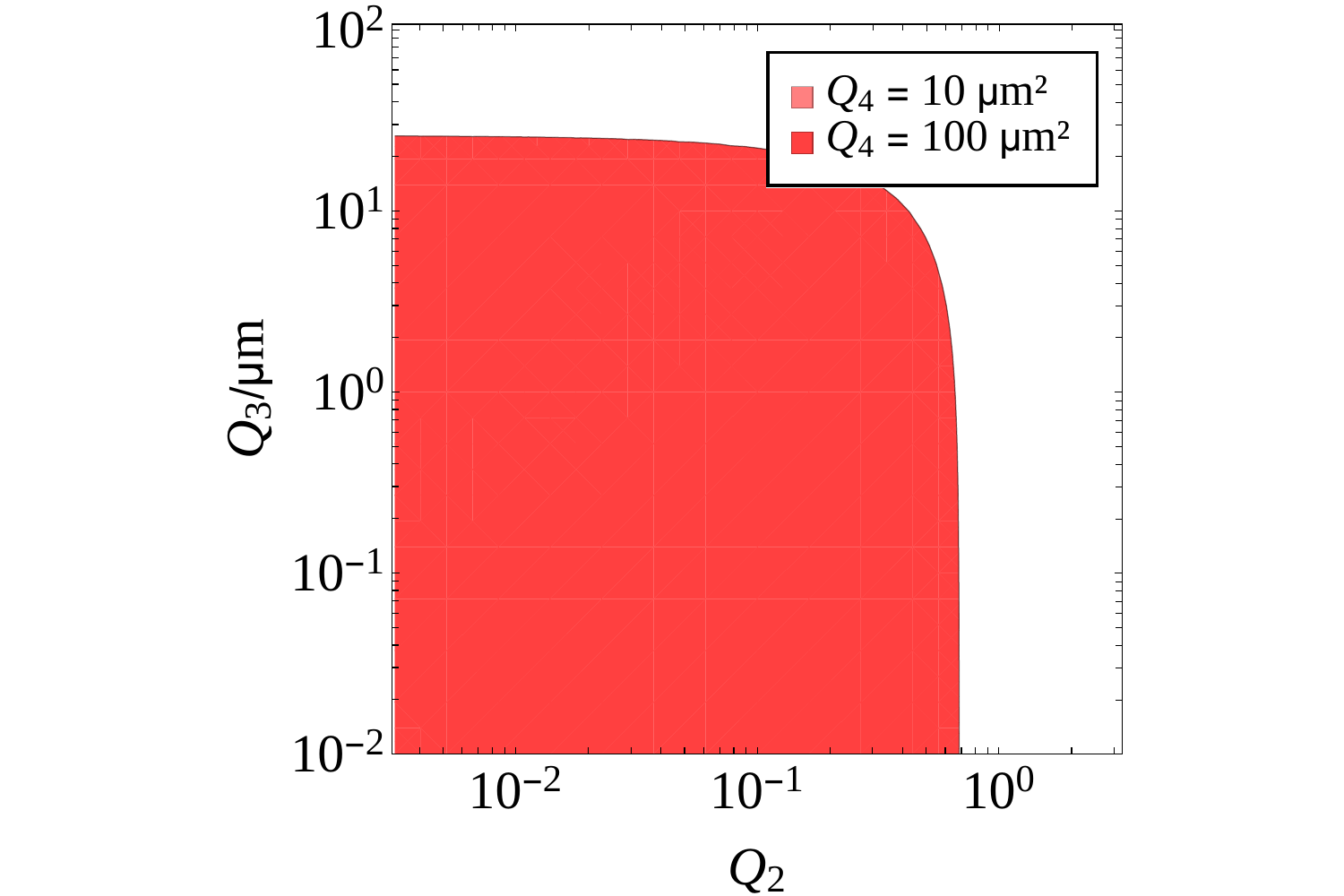} 
    \end{minipage}\hfill
    \begin{minipage}{0.333\textwidth}
        \centering
        \includegraphics[width=1.2\textwidth]{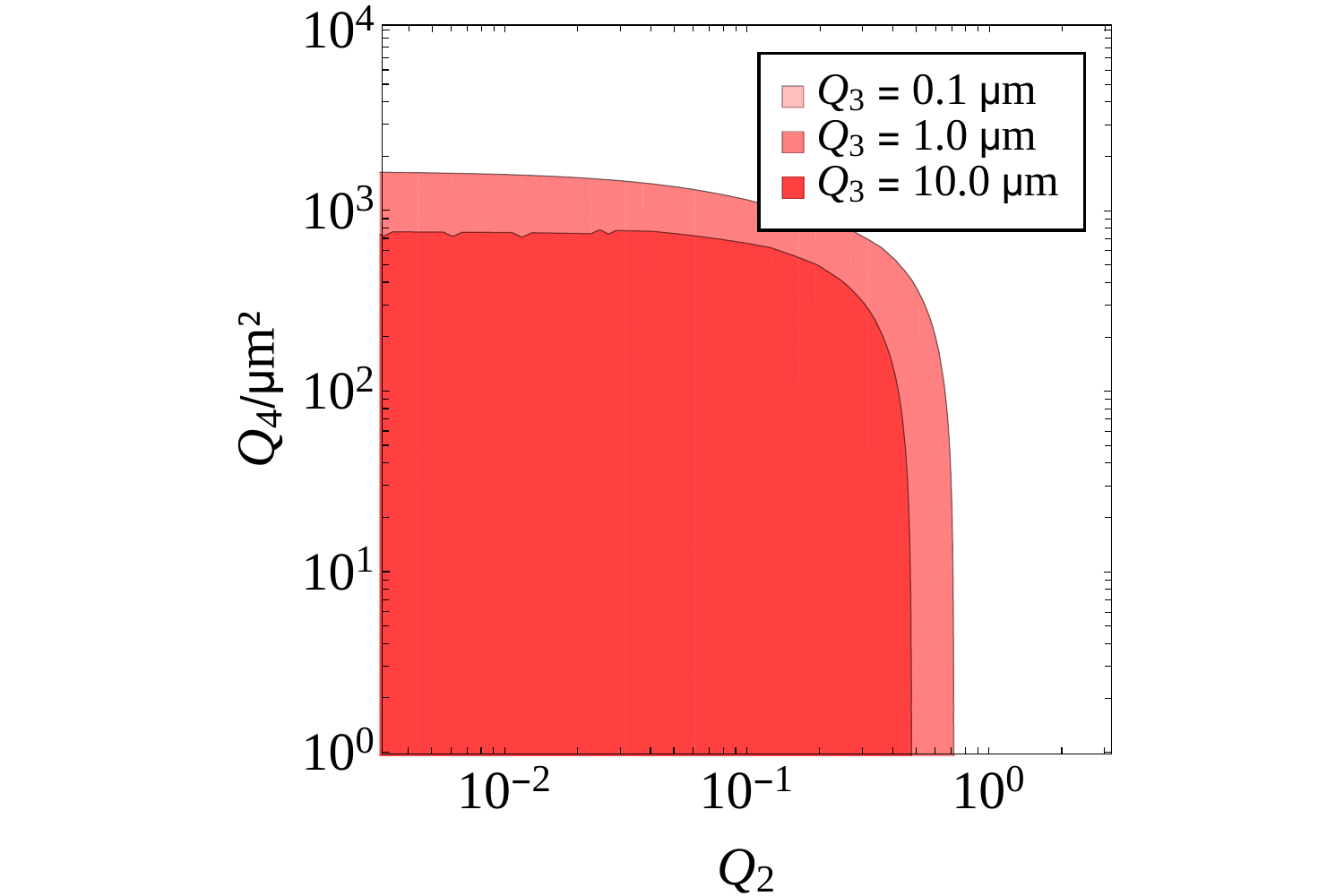} 
    \end{minipage}\hfill
        \begin{minipage}{0.333\textwidth}
        \centering
        \includegraphics[width=1.2\textwidth]{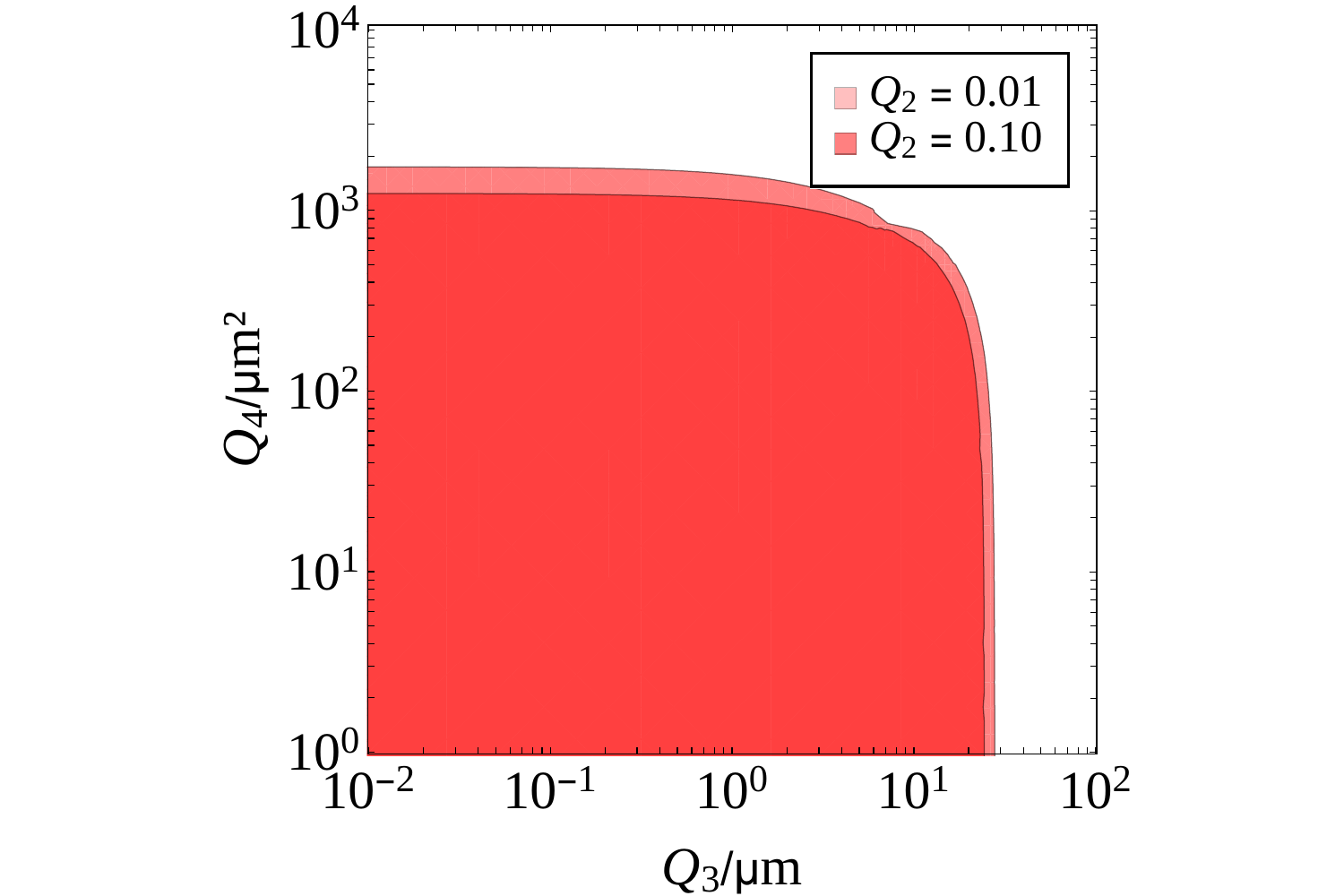} 
    \end{minipage}\hfill
    \caption{\small{\textit{Values of the background sources $Q_2$, $Q_3$ and $Q_4$ (red shaded regions) for which no collision between \textup{S} and \textup{P} is observed when using the mN-potential. Within these regions neither the Newtonian
    nor the modified-Newtonian potential induce collision. From left to right, correlations between the different background sources in the planes ($Q_2,Q_3$), ($Q_2,Q_4$) and ($Q_3,Q_4$) are shown. Top row: {\bf Case 1}; bottom row: {\bf Case2}. 
    Different values of the parameters not represented in each panel are shown, as  explained in the respective legends. 
    }}
    }
    \label{fig:CollisionBackground}
\end{figure*}

\begin{figure*}[h!]
    \centering\hspace{-1.1cm}
    \begin{minipage}{0.333\textwidth}
        \centering
        \includegraphics[width=1.2\textwidth]{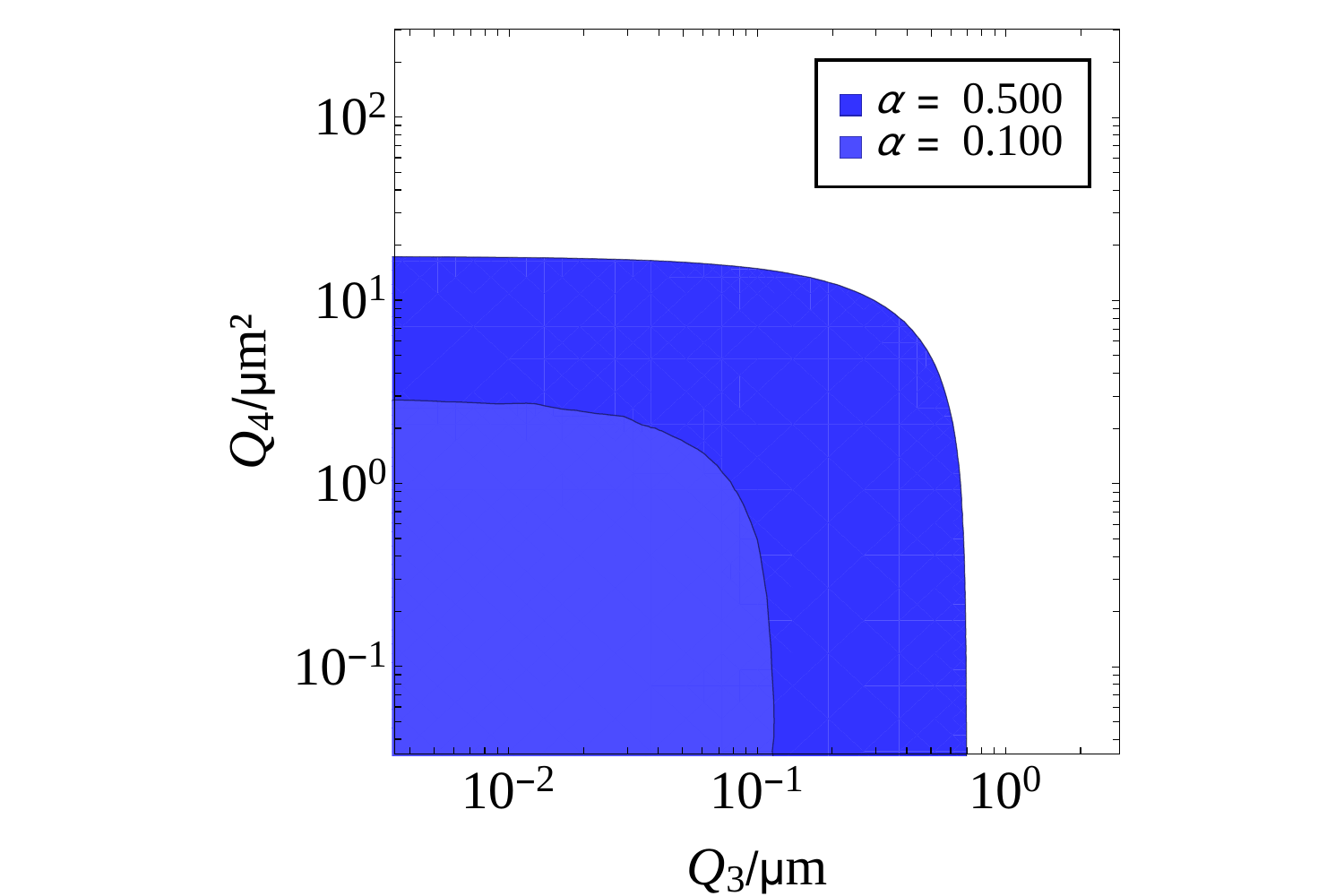} 
    \end{minipage}\hfill
    \begin{minipage}{0.333\textwidth}
        \centering
        \includegraphics[width=1.2\textwidth]{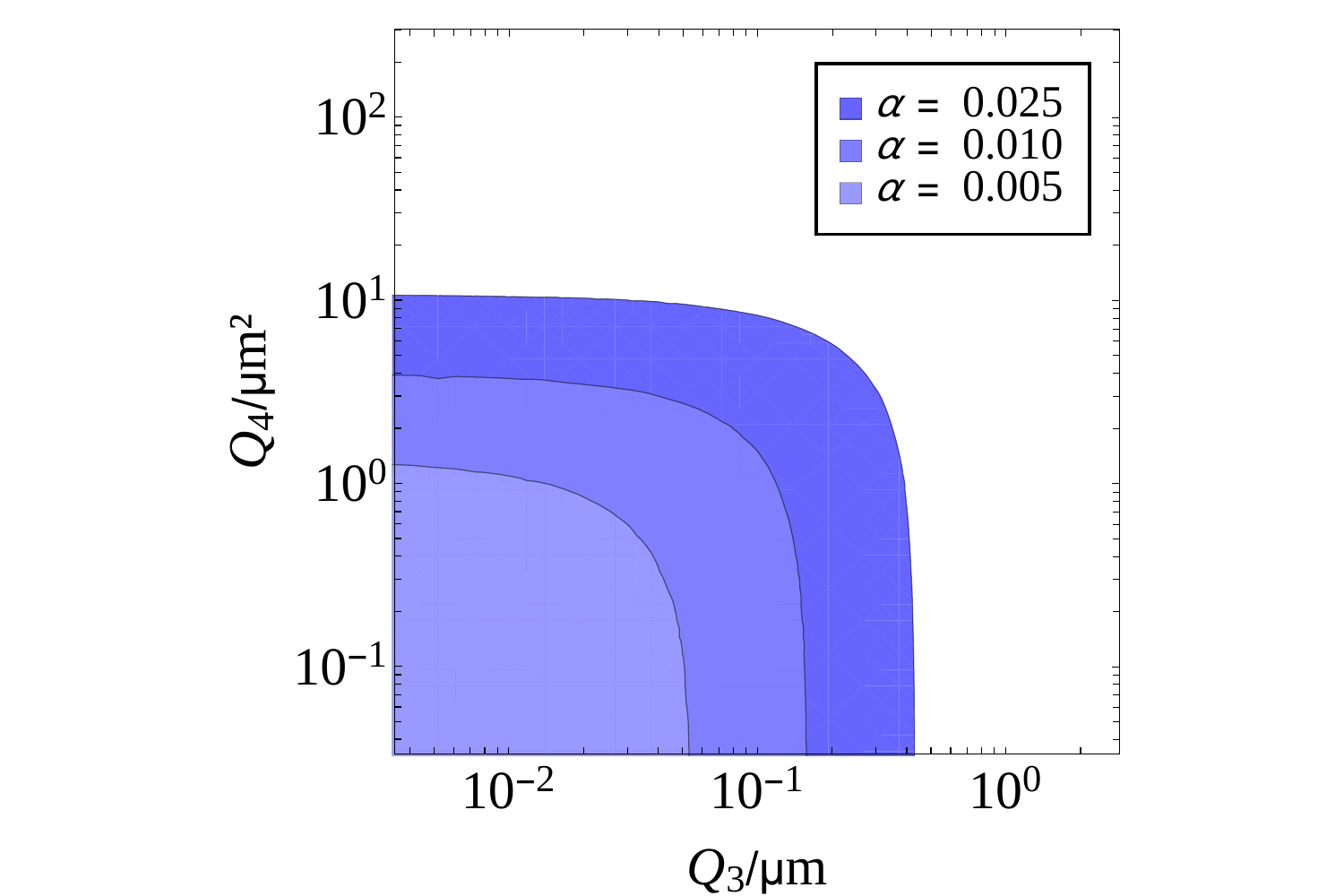} 
    \end{minipage}\hfill
        \begin{minipage}{0.333\textwidth}
        \centering
        \includegraphics[width=1.2\textwidth]{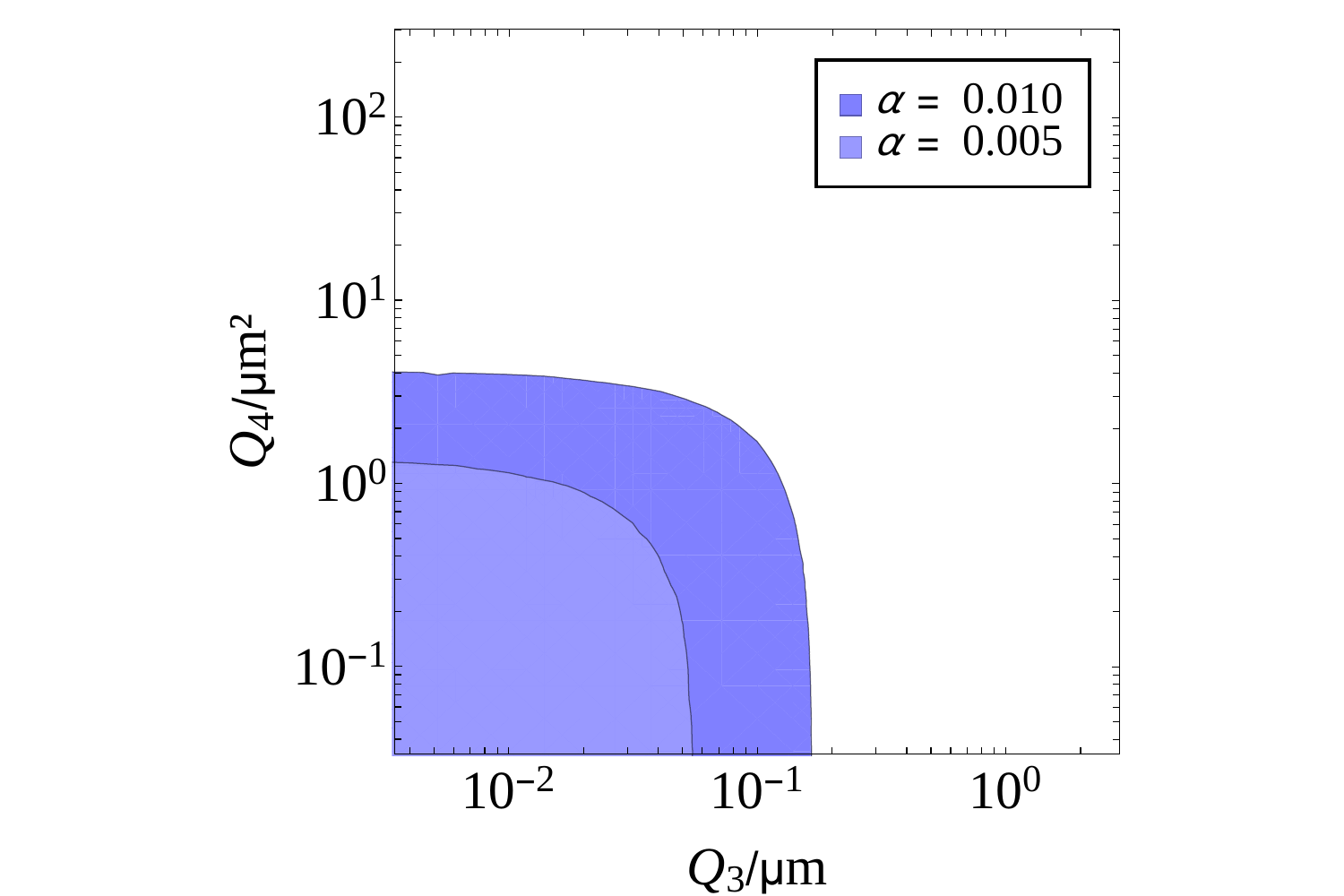} 
    \end{minipage}\hfill
    
    \centering\hspace{-1.1cm}
    \begin{minipage}{0.333\textwidth}
        \centering
        \includegraphics[width=1.2\textwidth]{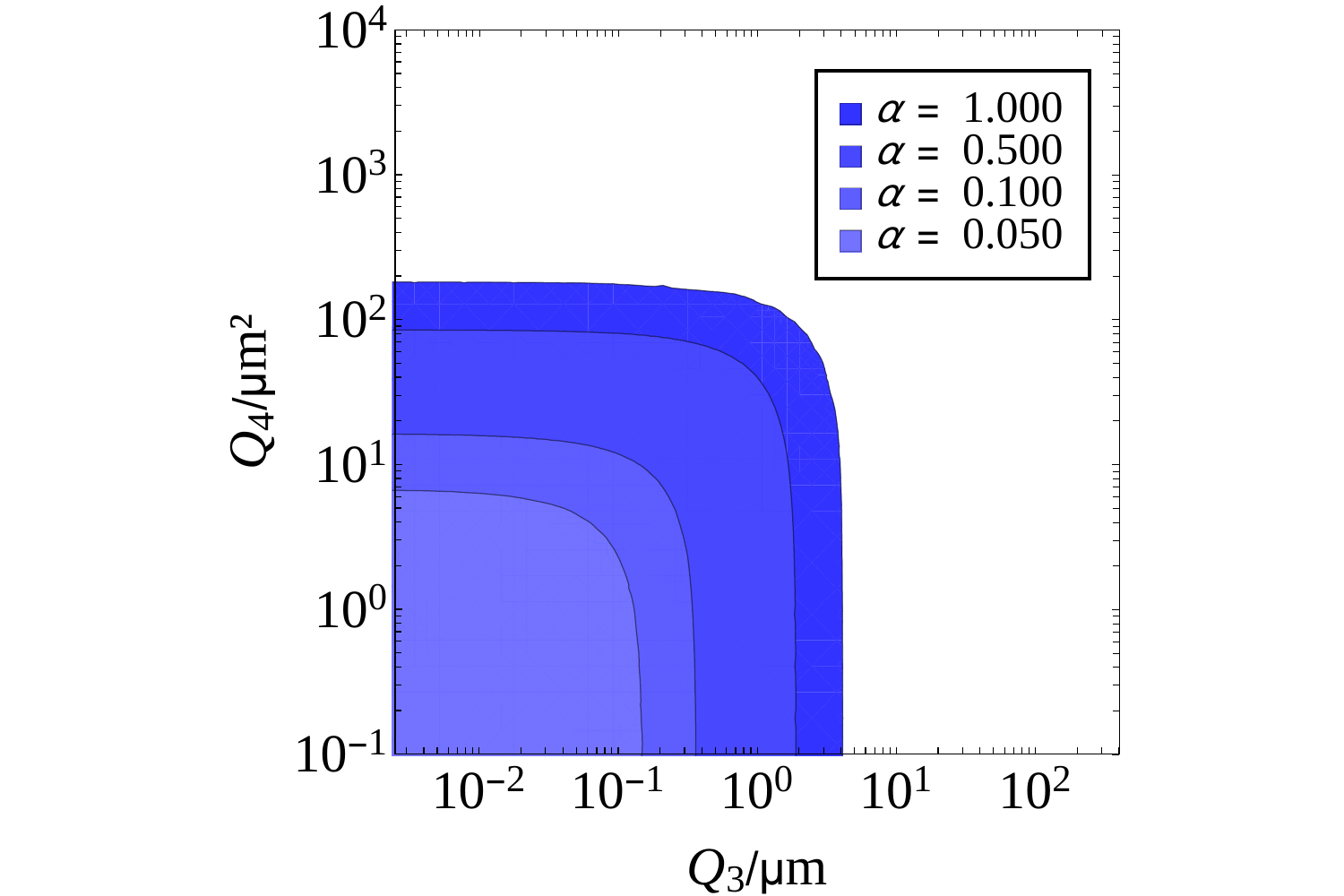} 
    \end{minipage}\hfill
    \begin{minipage}{0.333\textwidth}
        \centering
        \includegraphics[width=1.2\textwidth]{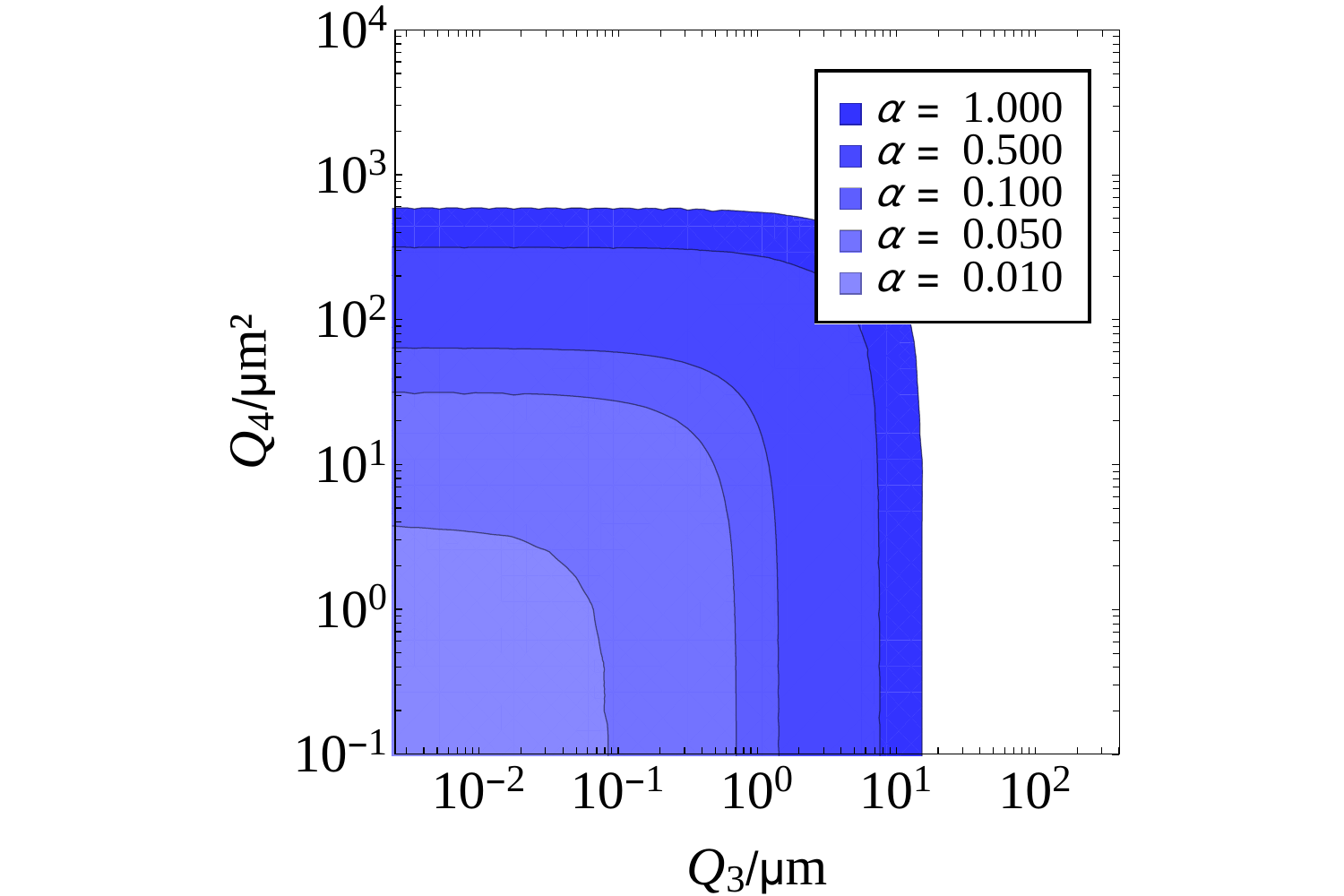} 
    \end{minipage}\hfill
        \begin{minipage}{0.333\textwidth}
        \centering
        \includegraphics[width=1.2\textwidth]{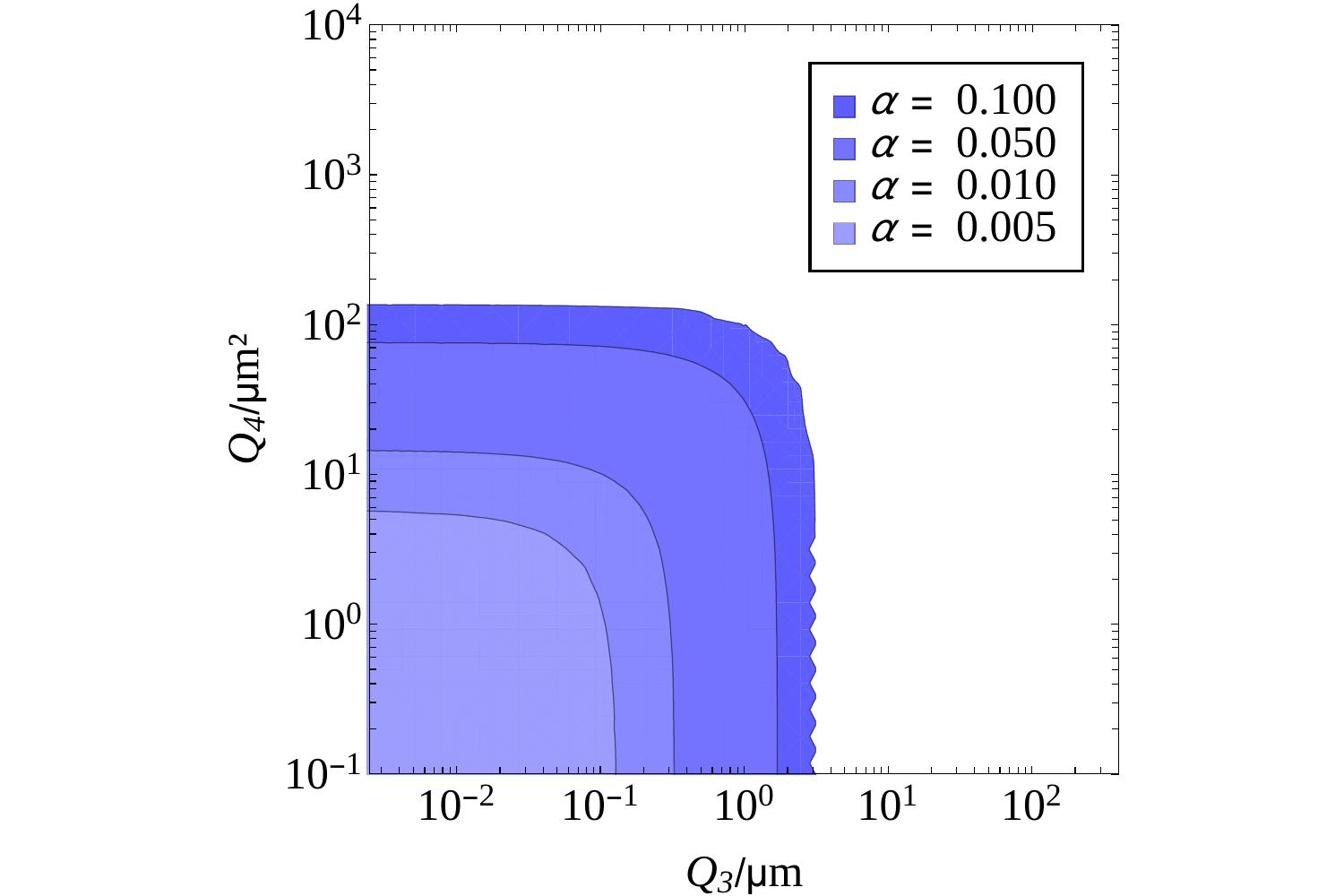} 
    \end{minipage}\hfill

    \caption{\small{\textit{
	Values of background sources $Q_3$ and $Q_4$ (blue shaded regions)     
    for which the precession induced by the mN-potential differs from that induced by
    the BN-potential, for specific points in the ($\lambda,\alpha$) plane, by a value larger than twice the clock precision (2 \textup{s}). Within this region we could claim that any observed 
    precession would be due to New Physics and not to background sources. 
    From left to right: Top row ({\bf Case 1}): $\lambda = 10$ $\upmu$\textup{m}, $30$ $\upmu$\textup{m} and $60$ $\upmu$\textup{m}; Bottom row ({\bf Case 2}): $\lambda = 20$ $\upmu$\textup{m}, $30$ $\upmu$\textup{m} and $60$ $\upmu$\textup{m}. 
    In each panel, several values of $\alpha$ are shown, as explained in the legends.
    }}
    }
     \label{fig:BackgroundAlpha}
\end{figure*}

The first aspect is studied in Fig.~\ref{fig:CollisionBackground}.
 We depict the correlation between two background sources with the third one fixed to some specific value  as shown in the plot legend. 
From left to right, we present the ($Q_2,Q_3$), ($Q_2,Q_4$) and ($Q_3,Q_4$) planes, respectively. The red-shaded areas represent the regions of the parameter space for which the background sources are small enough so that with the chosen initial conditions
that avoid collision between S and P in the Newtonian case, the mN-potential also implies non-collisional trajectories\footnote{Remind that we are considering here only positive values for all $Q_i$. 
The same analysis could be repeated taking into account negative ones, although in that case there would be a cancellation effect that would prevent collision.}. The color-coding corresponds to different values of the third background source in each panel.  

\begin{figure*}[h!]
    \centering\hspace{-1.1cm}
    \begin{minipage}{0.333\textwidth}
        \centering
        \includegraphics[width=1.2\textwidth]{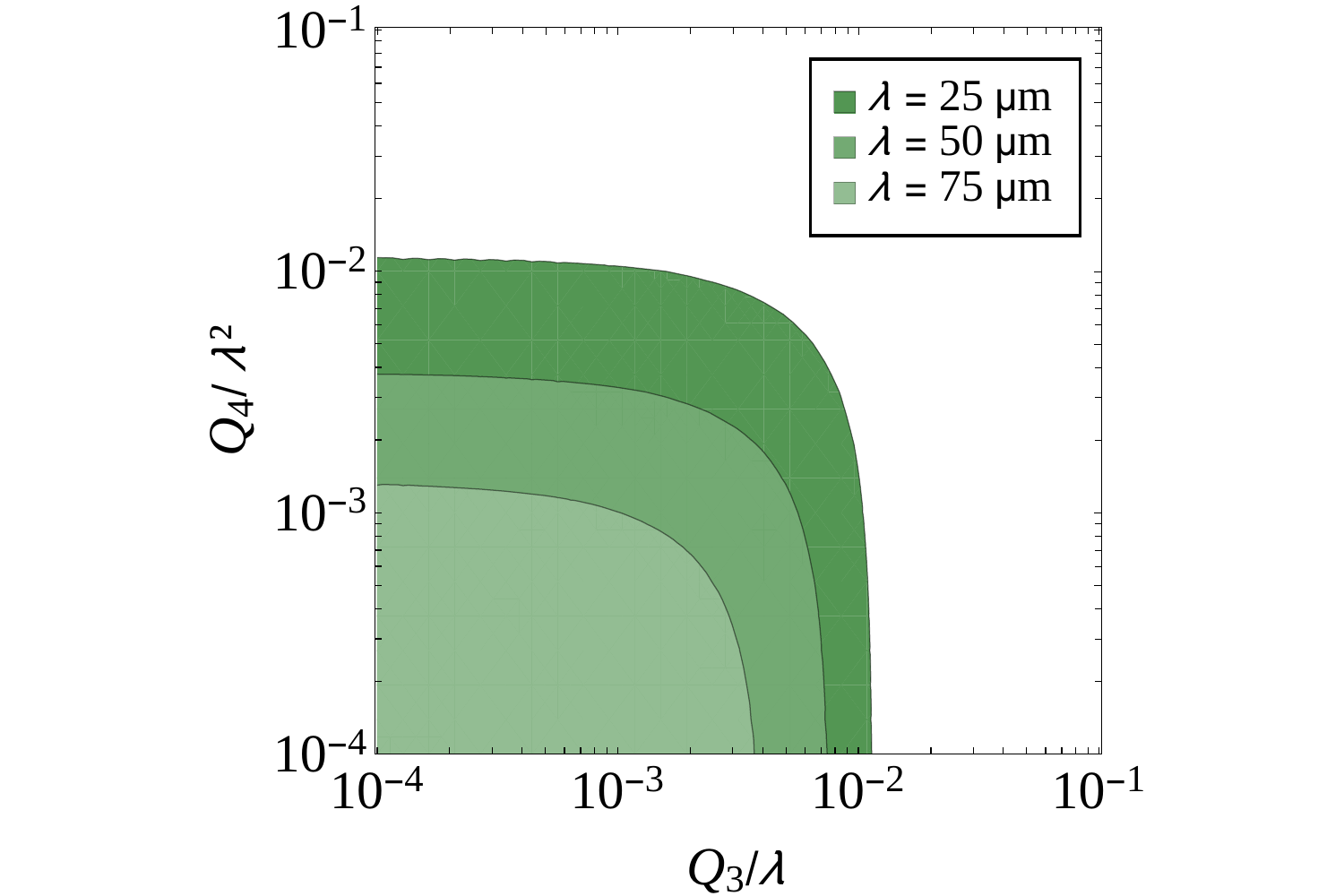} 
    \end{minipage}\hfill
    \begin{minipage}{0.333\textwidth}
        \centering
        \includegraphics[width=1.2\textwidth]{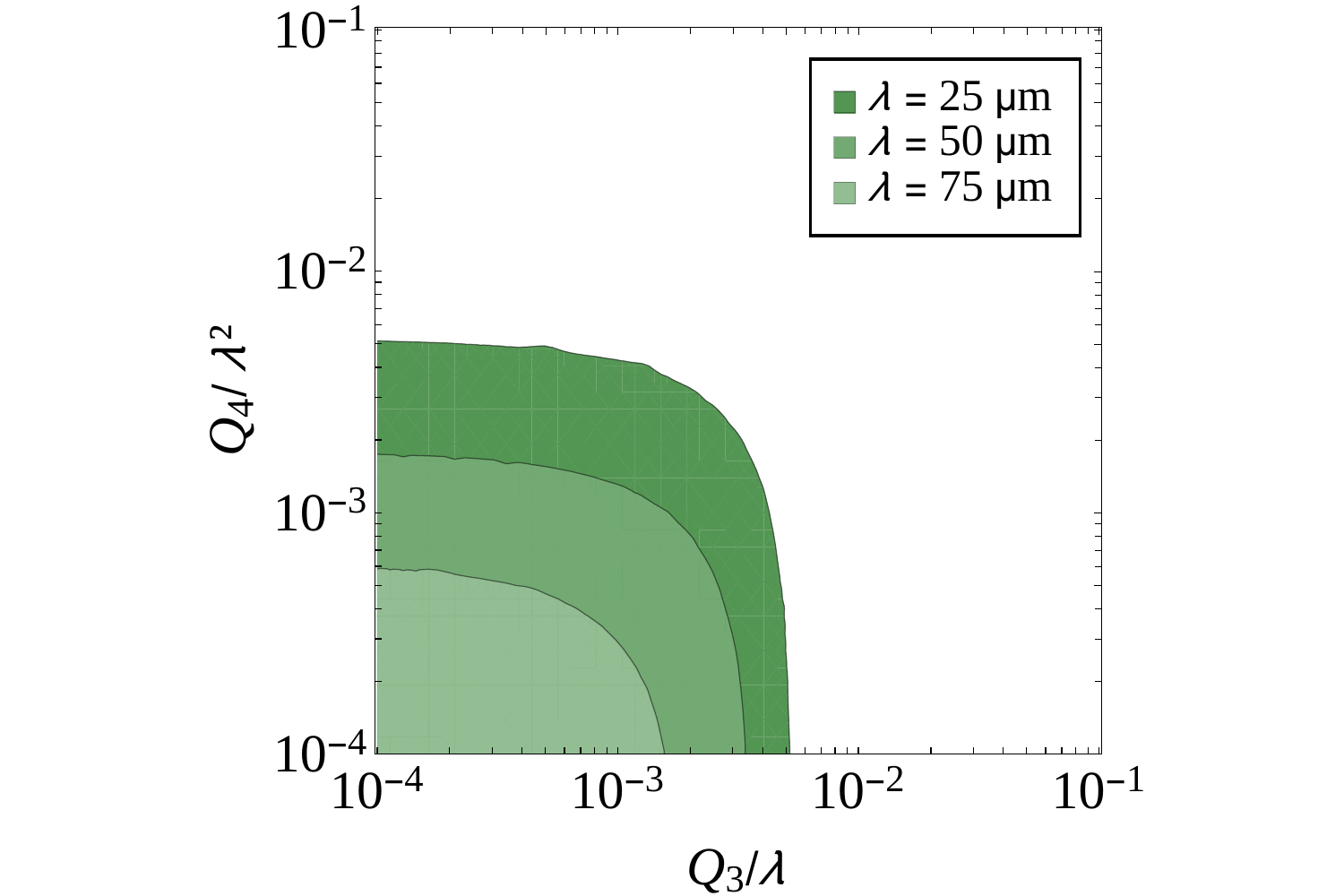} 
    \end{minipage}\hfill
        \begin{minipage}{0.333\textwidth}
        \centering
        \includegraphics[width=1.2\textwidth]{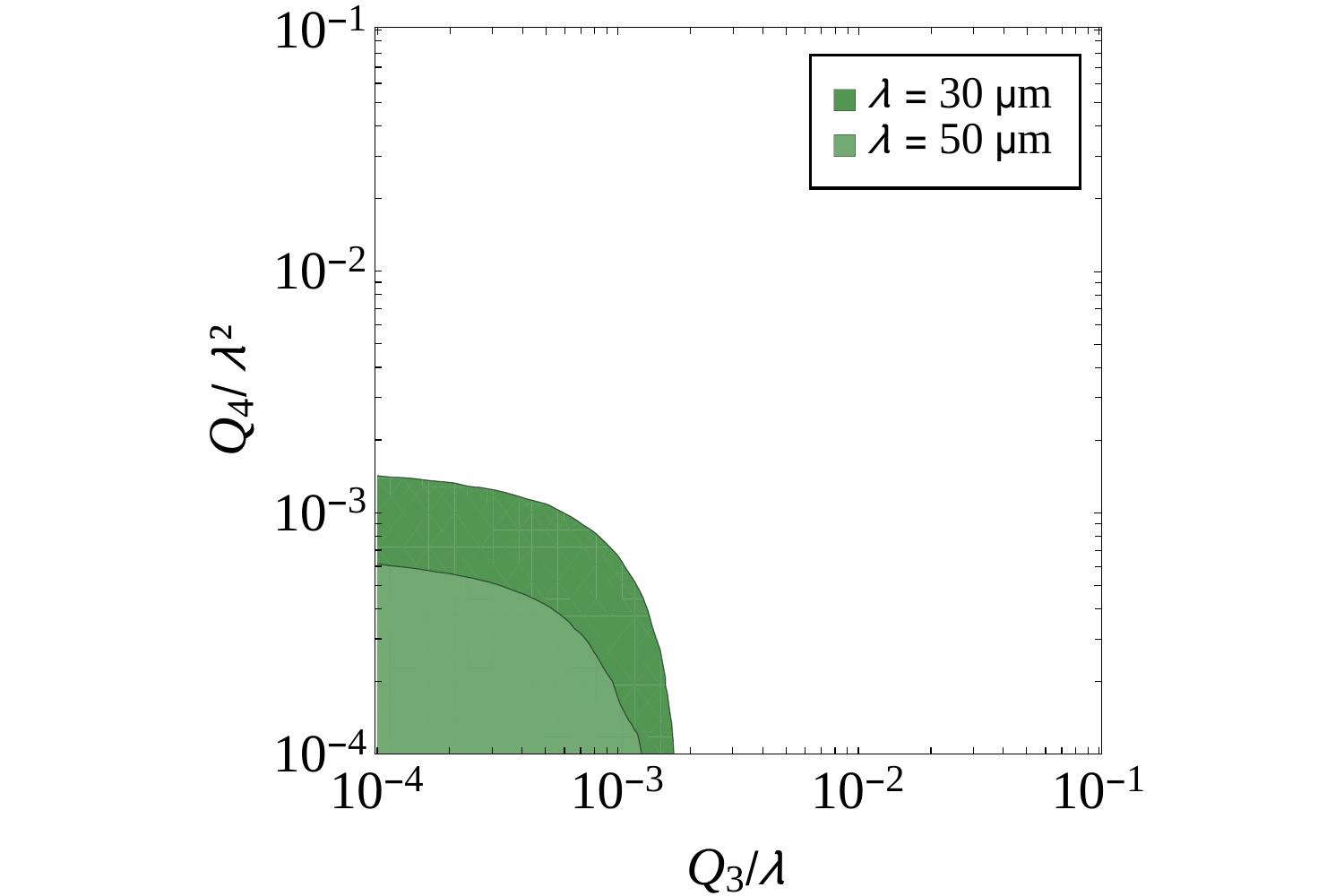} 
    \end{minipage}\hfill
    
    \hspace{-1.1cm}
    \begin{minipage}{0.333\textwidth}
        \centering
        \includegraphics[width=1.2\textwidth]{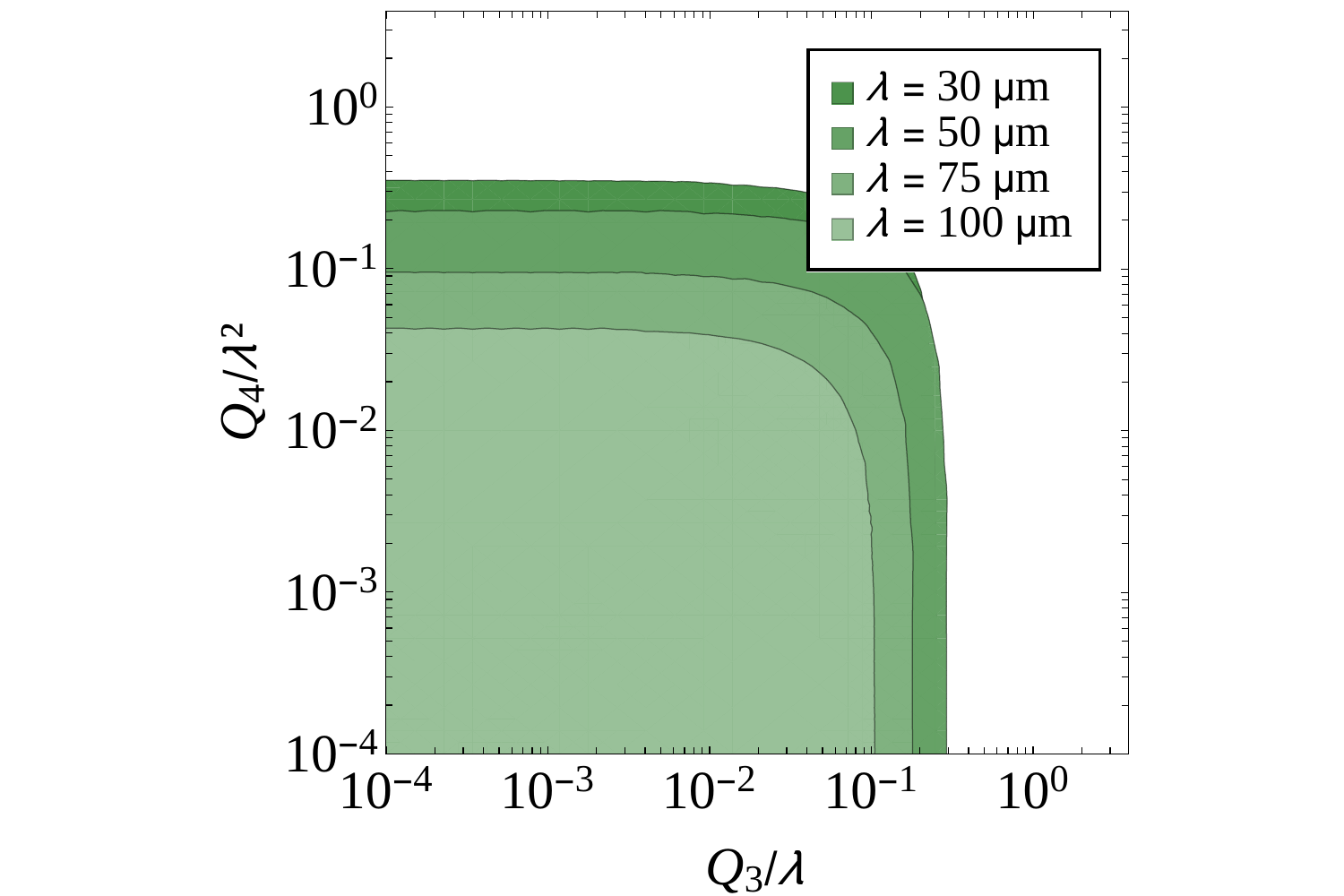} 
    \end{minipage}\hfill
    \begin{minipage}{0.333\textwidth}
        \centering
        \includegraphics[width=1.2\textwidth]{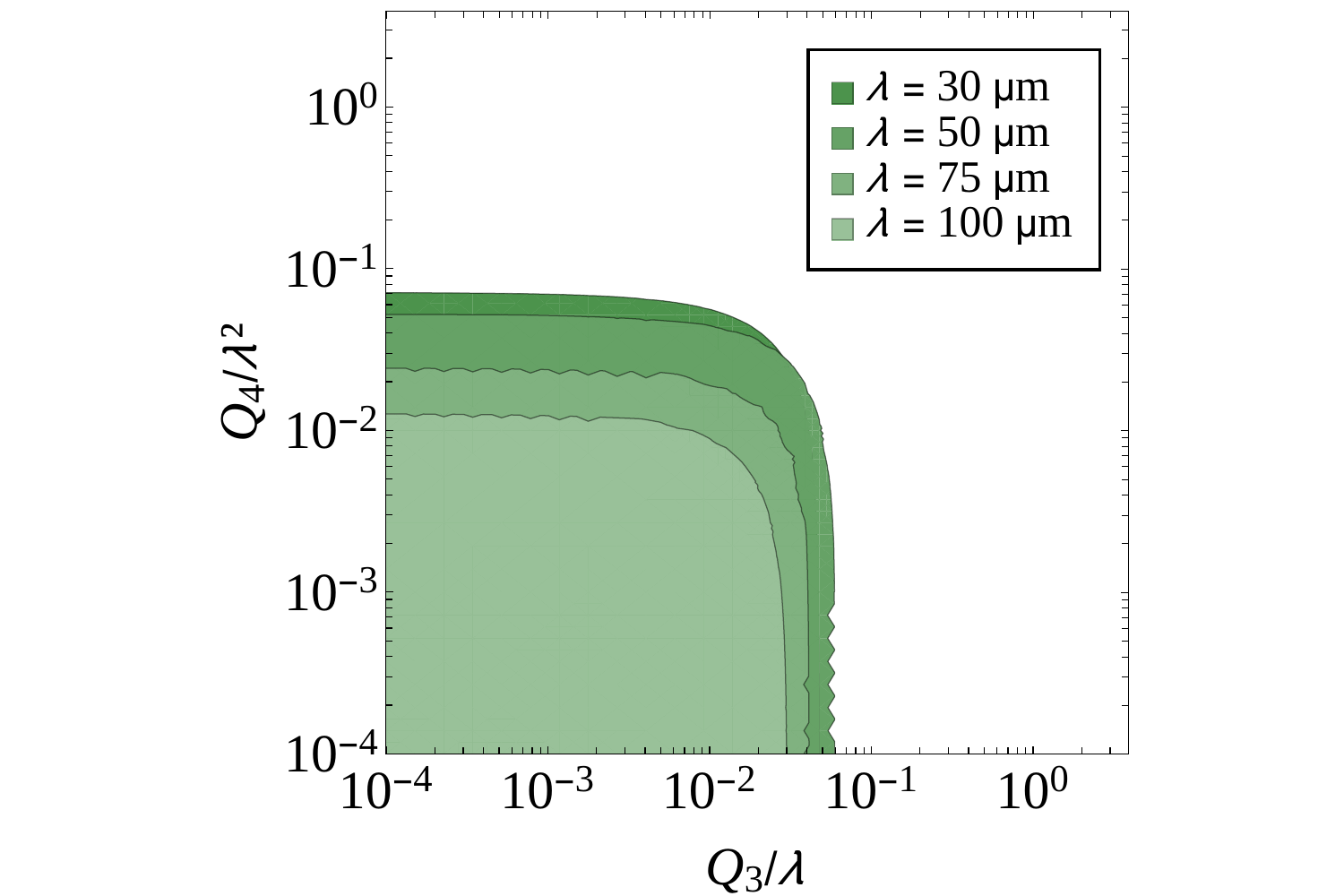} 
    \end{minipage}\hfill
        \begin{minipage}{0.333\textwidth}
        \centering
        \includegraphics[width=1.2\textwidth]{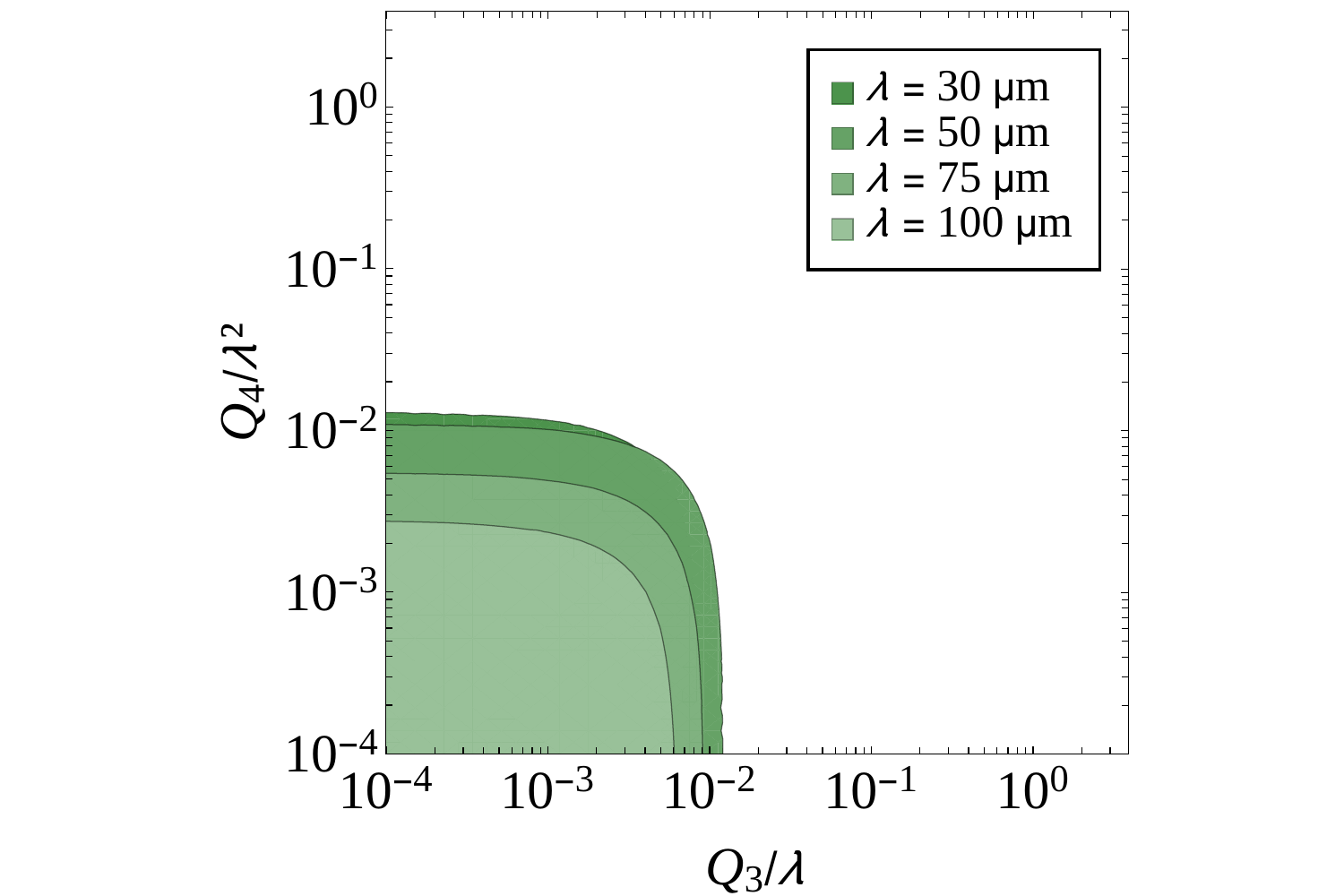} 
    \end{minipage}\hfill
    
    \caption{\small{\textit{
     Values of background sources $Q_3$ and $Q_4$ (green shaded regions)
     for which the precession induced by the mN-potential differs from that induced by
    the BN-potential, for specific points of the ($\lambda,\alpha$) plane, by a value larger than twice the clock precision (2 \textup{s}). Within this region we could claim that any observation of 
    precession would be due to New Physics and not to large background sources. From left to right: Top row ({\bf Case 1}): $\alpha = 0.020, 0.010$, and $0.005$; Bottom Row ({\bf Case 2}): $\alpha = 0.50, 0.10$, and $0.02$. 
    In each panel, several values of $\lambda$ are shown, as explained in the legends. The background sources have been normalized to the correct power of $\lambda$ in order to have comparable results
    using adimensional coefficients.}}
    }
        \label{fig:BackgroundLambda}
\end{figure*}

 The top row corresponds to {\bf Case 1}, while the bottom one represents 
{\bf Case 2}. 
By looking at our results for the former, we can make a few observations: the ultimate values for which the behaviour of the setup is unspoiled by backgrounds are $Q_ 2 \leq 10^{-2}$ and $ Q_3 \leq 3 \times 10^{-1}$ $\upmu$m (top row, left panel). Note that the two shaded regions corresponding to different values of $Q_4$ overlap almost perfectly, implying that these bounds are mostly independent on the value of the remaining background. Moreover, these two background sources have very small correlations (as the red-shaded areas are approximately rectangular-shaped). Nevertheless, some correlations can be appreciated in the other two panels of the top row, where we can see that the upper bound for which the behaviour is unspoiled in the ($Q_2,Q_4$) and ($Q_3,Q_4$) planes 
changes for varying $Q_3$ (middle panel) and $Q_2$ (right panel). The upper bound on $Q_4$ depends, indeed, on the values of $Q_2, Q_3$, with $Q_4^{\rm max} \in [10^1,10^2]$ $\upmu$m$^2$.
 We conclude that $1/r$ and $1/r^2$ backgrounds affect the most the qualitative behaviour of the orbit of S around P. This effect should be more important 
the smaller the initial distance between S and P is, what can indeed be seen in the bottom row, corresponding to {\bf Case 2} (with a larger $r_0$ than {\bf Case 1}). 
We get in this case:  $Q_ 2 \leq 5 \times 10^{-1}, Q_3 \leq 10^{1}$ $\upmu$m and $Q_4 \leq 10^3$ $\upmu$m$^2$, approximately. For this choice of initial conditions,
we can see that correlations between the three background sources are also small. 

After determining the maximum values of $Q_i$ for which the sole effect of backgrounds would not induce collision, 
we study which levels of noise are small enough to preserve the sensitivity of the experimental setup, and how the ultimate sensitivity in the ($\lambda,\alpha$) plane 
is affected by noises. In order to study the impact of the different background sources, 
we fix  $Q_2 = 0.001$ in {\bf Case 1} and $Q_2 = 0.01$ in {\bf Case 2} (since this parameter does not induce precession of the orbit), and compare the signal induced by a mN-potential for different values of $Q_3$ and $Q_4$ with that expected from a BN-potential for different values of $(\lambda,\alpha)$ inside the sensitivity region. This indicates which the maximum backgrounds allowed to a have for a clear New Physics signal are.

 Our results for different values of $\alpha$ at fixed $\lambda$ are given in Fig.~\ref{fig:BackgroundAlpha}, again for two possible choices of initial conditions: {\bf Case 1} (top row) and {\bf Case 2} (bottom row). 
 In each panel we compute $\mathrm{\Delta} T_{\rm BN}^{\rm max}$ over $N_\text{rev}=30$ revolutions
 using the BN-potential with a fixed value of $\lambda$ ($\lambda = 10$ $\upmu$m, $30$ $\upmu$m and $60$ $\upmu$m, from left to right) for several values of $\alpha$, 
different in each panel. We then compute the signal induced using the mN-potential and find the values of the background parameters that give a signal that differs from the BN one by less than 2 s (twice our choice for the clock precision). The regions in the ($Q_3,Q_4$) plane for which we can distinguish the two models are depicted by blue-shaded areas, whose hue depend on the particular value of $\alpha$. 
For larger values of $Q_3$ and $Q_4$, the two potentials give results that are too similar to be distinguished and therefore, it would not be possible to claim that observing precession is a distinctive signature of a BN-gravitational potential. 
A general comment that can be drawn is that the effect of
backgrounds is much more relevant for those points of the parameter space that are nearer to the ultimate sensitivity of the experimental setup (see Fig.~\ref{fig:Limits1}).

In Fig.~\ref{fig:BackgroundLambda} we show the same kind of analysis, albeit for fixed values of $\alpha$ and  several different values of $\lambda$ (as shown in the legend of each panel). As before, top and bottom rows  refer
to different choices of the initial conditions: top row stands for {\bf Case 1}, and bottom row for {\bf Case 2}. In order to compare in the same plot values for dimensionful background sources ($Q_3$ and $Q_4$) at different values of $\lambda$, 
we have normalized them by powers of $\lambda$: $\hat Q_3 = Q_3/\lambda$; $\hat Q_4 = Q_4/\lambda^2$. This way we have adimensional quantities easier to compare, as it is the case of $Q_2$. 
As before, in each panel we first compute the expected maximum period variation over 30 revolutions defined in eq.~(\ref{eq:TotalPeriodVariation}), $\mathrm{\Delta} T_\text{BN}^{\rm max}$, using the BN-potential for a particular point 
in the ($\lambda,\alpha$) plane. Second, we compute the signal induced using the mN-potential by increasing values of $Q_3$ and $Q_4$. 
Eventually we find the value for which this background-induced signal differs from the BN one by less than 2 s. 
The regions in the ($\hat Q_3,\hat Q_4$) plane for which we can distinguish the two models are depicted by green-shaded areas, the hue of which depends on the particular value of $\lambda$. 
As before, we observe how the effect of backgrounds is more relevant on the border of the ultimate sensitivity of the experimental setup (see, again, Fig.~\ref{fig:Limits1}). For example, looking at the top left panel in Fig.~\ref{fig:BackgroundLambda}, 
we can see that, for $\lambda = 75$ $\upmu$m and $\alpha = 0.004$, we require $\hat Q_3 \lesssim 0.025$, 
whereas for $\lambda = 25$ $\upmu$m, $\hat Q_3 \lesssim 0.010$ is enough.

To sum up, we have found that in order to preserve most of the ultimate sensitivity of the experimental setup as depicted in Fig.~\ref{fig:Limits1}, some limits on the strength of the backgrounds should be established. For the choice of initial conditions corresponding to {\bf Case 1}
we must keep $Q_3 \lesssim 0.05$ $\upmu$m and $Q_4 \lesssim 1$ $\upmu$m$^2$. On the other hand, for {\bf Case 2} it would be enough to have $Q_3 \lesssim 0.1$ $\upmu$m and $Q_4 \lesssim 5$ $\upmu$m$^2$. 
In both cases, however, larger values of the background sources can still be allowed if we are focusing on points far from the edge of the sensitivity region. 
Armed with these results, we can now move to show the expected performances of the experimental setup in the presence of backgrounds.

\section{Sensitivity and measurements}
\label{sec:sens}

After optimizing the proposed experimental setup and studying the effect of possible background sources in detail, we sum up our final results in this section. 
First, we show in Sect.~\ref{sec:limitswithbackground} our updated sensitivity limits when background sources are added to the 
Newtonian potential, as done in eq. \ref{eq:mNpotential1}. These results should be compared with those presented in Ref.~\cite{Donini:2016kgu} and in Fig.~\ref{fig:Limits1} of the present work. 
Then in Sect.~\ref{sec:positive}, we consider the case in which a positive signal is detected. We assume that the signal is produced by some benchmark points in the $(\lambda, \alpha)$ plane with a given value of the polynomial backgrounds, for which we expect a sizeable precession of the orbit of S around P. Then, we check the precision 
 that could be attained on $\lambda$ and $\alpha$ by the proposed experimental setup and the impact of changing the initial conditions.


\subsection{Sensitivity limits with backgrounds}
\label{sec:limitswithbackground}

\begin{figure*}[h]
    \centering
    \begin{minipage}{0.5\textwidth}
        \centering
        \includegraphics[width=\textwidth]{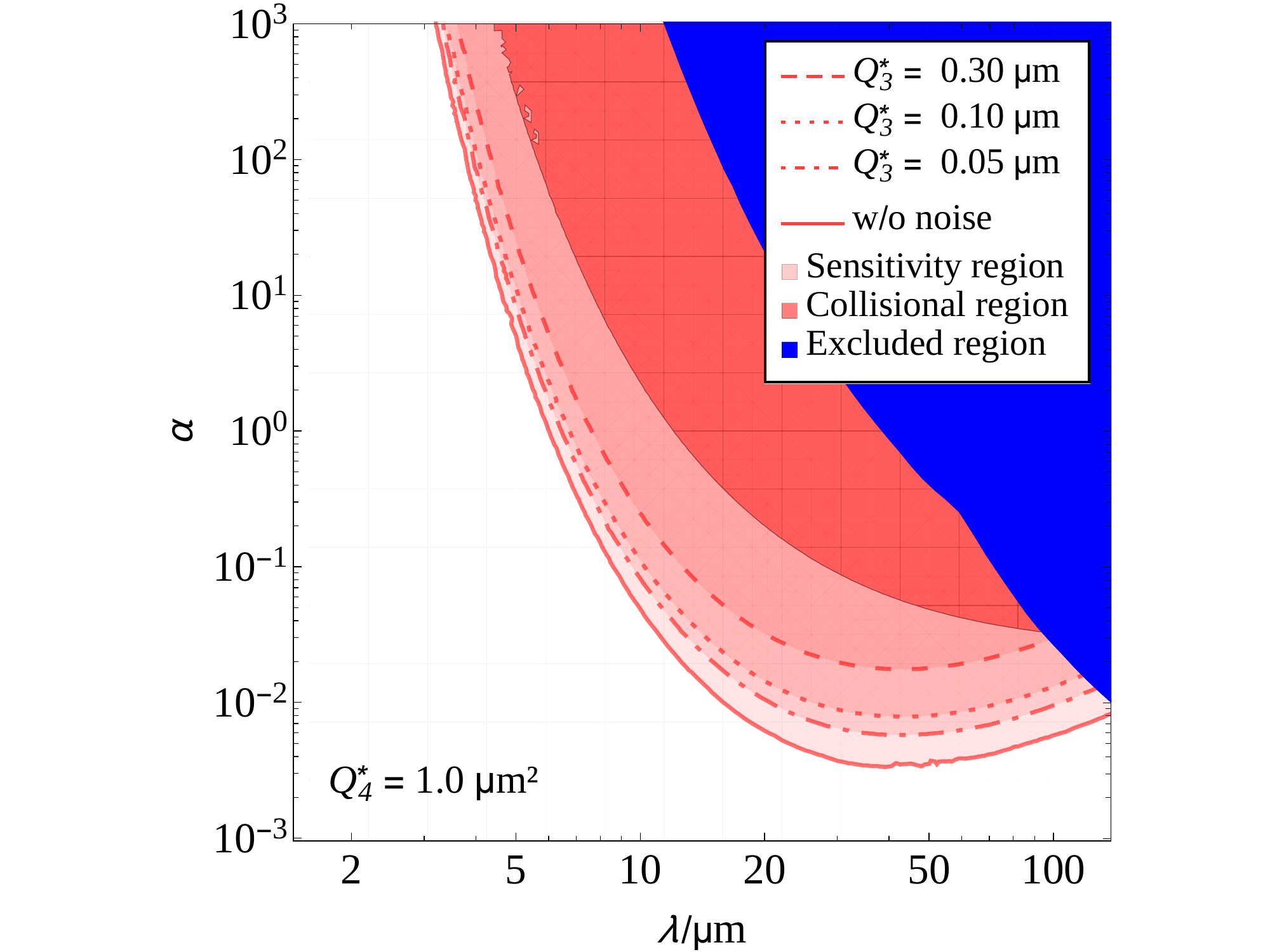} 
    \end{minipage}\hfill
    \begin{minipage}{0.5\textwidth}
        \centering
        \includegraphics[width=\textwidth]{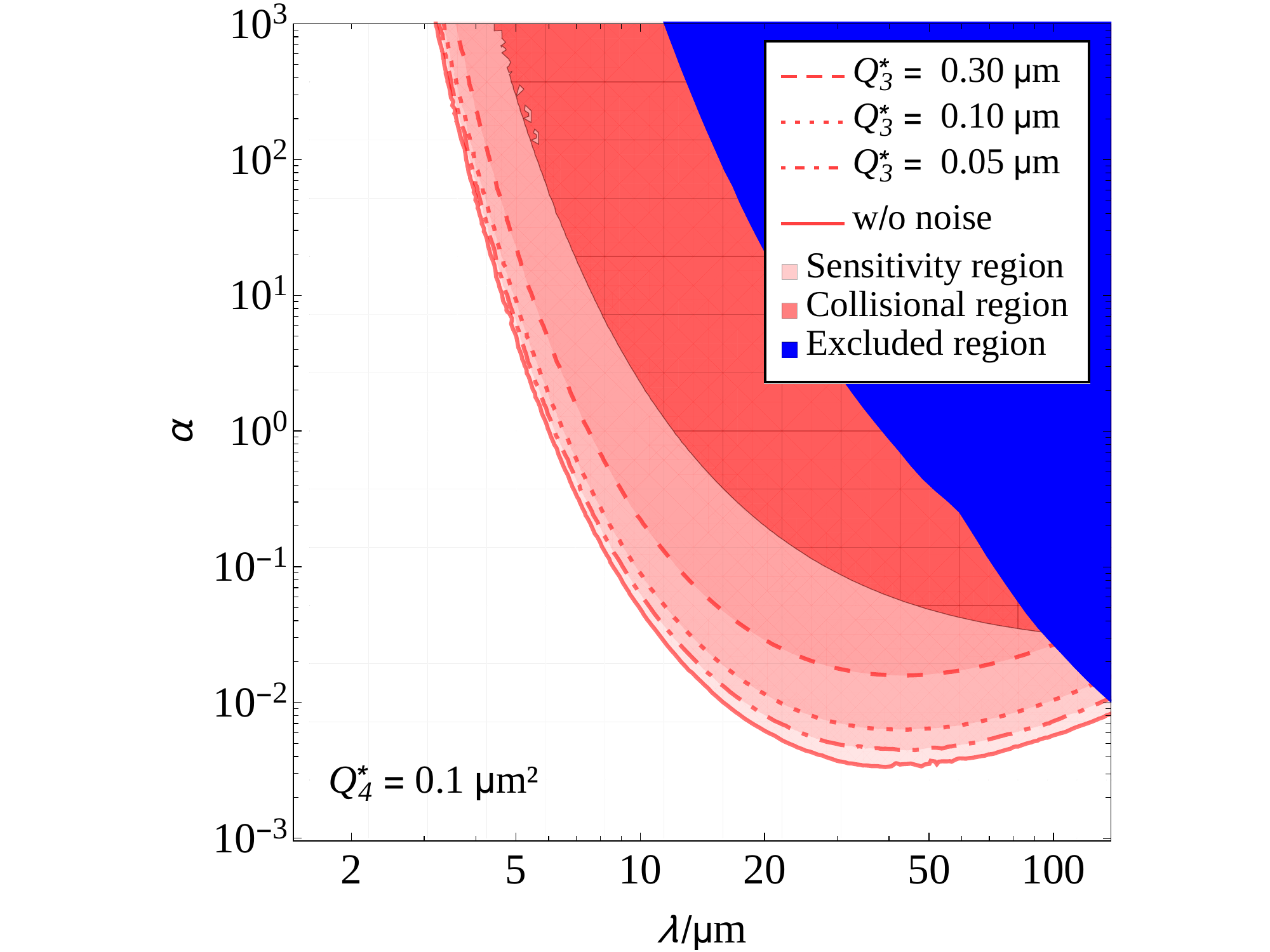} 
    \end{minipage}
    \vspace{0.2 cm}
    
        \centering
    \begin{minipage}{0.5\textwidth}
        \centering
        \includegraphics[width=\textwidth]{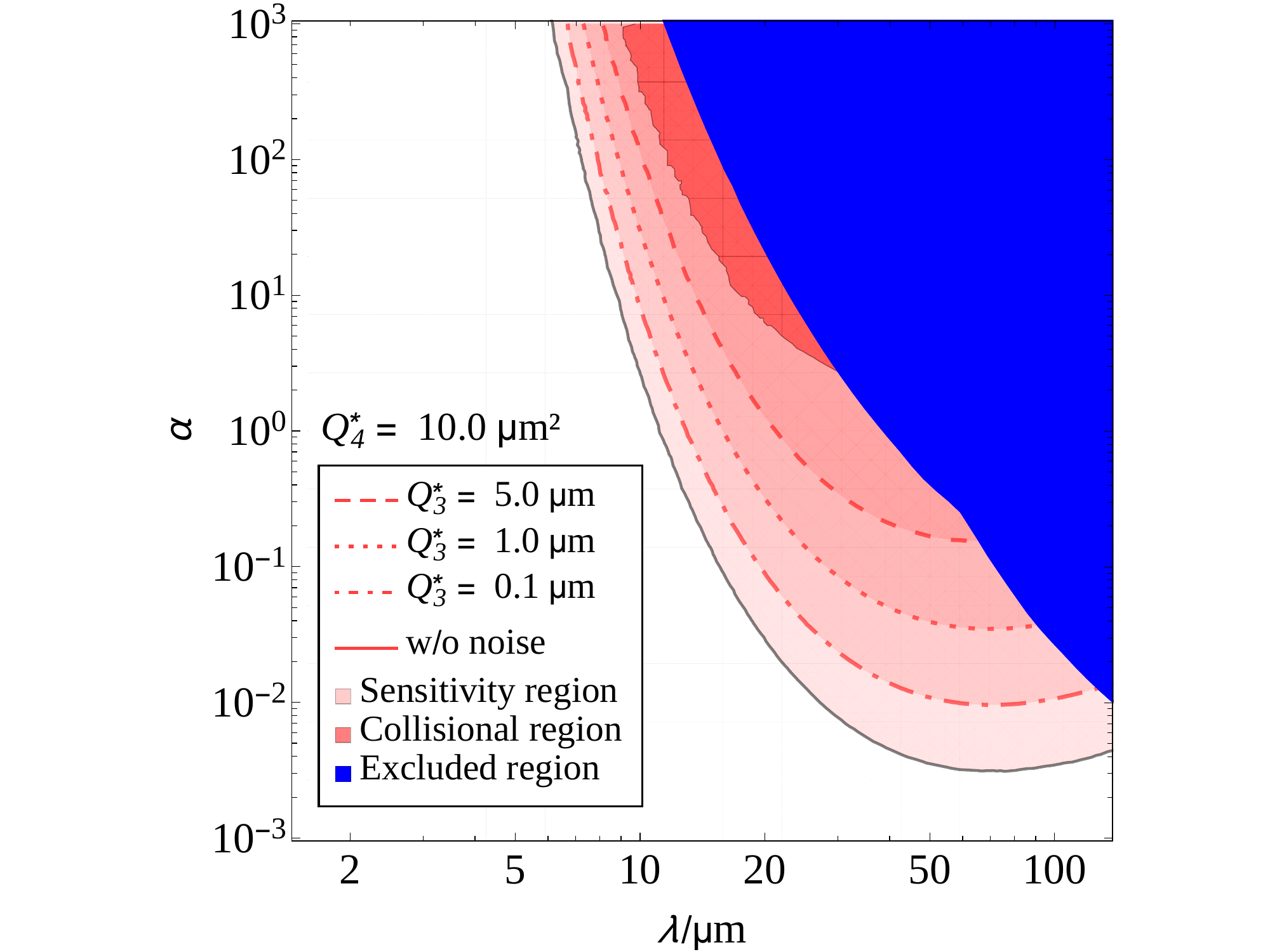} 
    \end{minipage}\hfill
    \begin{minipage}{0.5\textwidth}
        \centering
        \includegraphics[width=\textwidth]{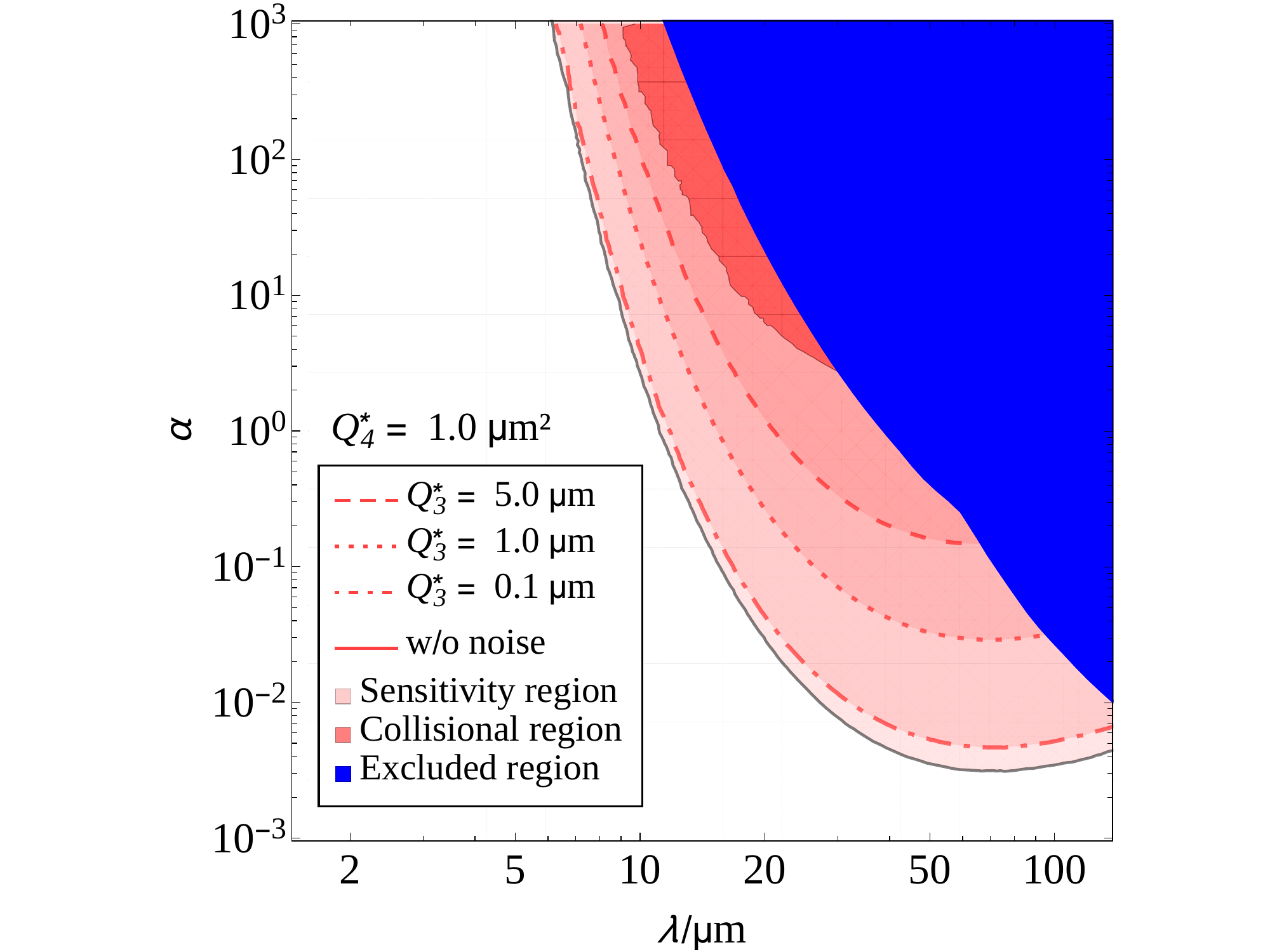} 
    \end{minipage}

    \caption{\small{\textit{
    Sensitivity of the proposed experimental setup in the ($\lambda,\alpha$) plane in the presence of $Q_i$ attractive backgrounds with $1/r$, $1/r^2$ and $1/r^3$ dependence which are only known to be smaller than some upper limits ${\cal Q}_i$. 
    The blue region represents the present experimental bounds (see Fig.~\ref{fig:AlphaLambdaBounds}). 
    The red-meshed region is the part of the parameter space for which a Beyond-Newtonian potential will induce collision between \textup{S} and \textup{P}. 
    The pink-shaded region is the part of the parameter space for which we would detect precession in the presence of New Physics, expected to be large enough to be distinguishable from the signal induced by the backgrounds (and that can be thus excluded  at 95\% CL in the case no signal is observed). 
    The boundary of the sensitivity region is drawn for values of $\lambda$ and $\alpha$ for which we expect the difference between $\mathrm{\Delta} T_{\rm BN}^{\rm max}$ and $\mathrm{\Delta} T_{\rm mN}^{\rm max}$ 
    to be equal to twice the expected clock error, $\sigma_{T}=1$ \textup{s}. Top row: {\bf Case 1}. Both panels assume $Q_2 = 10^{-3}$ and fixed ${\cal Q}_4$ 
    (left panel: ${\cal Q}_4 = 1$ $\upmu$\textup{m}$^2$; right panel: ${\cal Q}_4 = 0.1$ $\upmu$\textup{m}$^2$). 
    In each panel, several values of ${\cal Q}_3$ are shown: ${\cal Q}_3 = 0.05$ $\upmu$\textup{m} (dot-dashed line), ${\cal Q}_3 = 0.1$ $\upmu$\textup{m} (dotted line), 
    and ${\cal Q}_3 = 0.5$ $\upmu$\textup{m} (dashed line), as reported in each plot legend.
    Bottom row: {\bf Case 2}. Both panels assume $Q_2 = 10^{-2}$ and fixed $Q_4$  (left panel: ${\cal Q}_4 = 10$ $\upmu$\textup{m}$^2$; right panel: ${\cal Q}_4 = 1.0$ $\upmu$\textup{m}$^2$). 
    In each panel, we have: ${\cal Q}_3 = 0.1$ $\upmu$\textup{m} (dot-dashed line), ${\cal Q}_3 = 1.0$ $\upmu$\textup{m} (dotted line), and ${\cal Q}_3 = 5.0$ $\upmu$\textup{m} (dashed line). In all cases the solid line corresponds to the noiseless scenario as presented in Fig.~\ref{fig:Limits1}.
    }}}
        \label{fig:Limits2}
\end{figure*}

Our final results for the expected sensitivity of the optimized setup in the ($\lambda,\alpha$) plane, for positive $\alpha$, are given in  Fig.~\ref{fig:Limits2}. These results are presented in presence of attractive backgrounds, whose exact values are unknown, but for which some upper limits have been set in a previous calibration phase, denoted as ${\cal Q}_i$ (so $0\leq Q_i\leq {\cal Q}_i$). As in the previous Section, measures are done over 30 revolutions and we consider two possible choices
of the initial conditions: 
\begin{itemize}
\item {\bf Case 1} (top row): $r_0=111.8$ $\upmu$m, $\dot{r}_0=30.6$ nm s$^{-1}$ and $\dot{\theta}_0=491.1$ $\upmu$rad s$^{-1}$; 
          for which we have $r_{\rm a}\sim 150$ $\upmu$m, $T_{\rm N}\sim2$  h 30 min.
\item {\bf Case 2} (bottom row): $r_0=177.7$ $\upmu$m, $\dot{r}_0=13.4$ nm s$^{-1}$ and $\dot{\theta}_0=259.2$ $\upmu$rad s$^{-1}$; 
         for which we have $r_{\rm a}\sim 200$ $\upmu$m, $T_{\rm N}\sim4$ h 30 min.
\end{itemize}
The main difference between the two cases is that the second choice is more conservative: as the distance between both bodies is larger, the impact of $Q_3$ and $Q_4$-like backgrounds is suppressed. 
This comes however with some loss in the ultimate sensitivity of the experimental setup, mainly for small $\lambda$. 

Both cases give rise to a bounded orbit when using the mN-potential with attractive backgrounds chosen within certain limits to avoid collision between the spheres (as discussed at the beginning of Sect.~\ref{sec:backlimits}). These backgrounds, however, lead to a loss of sensitivity.
On the other hand, in the case of repulsive backgrounds, as is expected to be the case of the electric Casimir effect, we expect them to prevent collision from occurring and to induce precession in the opposite sense to a BN-potential with positive $\alpha$, meaning the impact on the setup would be smaller. However, unless kept small, they may lead to open trajectories. 
In this scenario, a re-optimization of the setup and a new analysis analogous to that of Sect.~\ref{sec:backgrounds} for repulsive background sources would probably allow to deal with them with just a minor loss in the ultimate sensitivity.
 

We now discuss Fig.~\ref{fig:Limits2}. We have shown in previous Sections that $Q_2$ does not affect our sensitivity to New Physics as it does not induce precession. Thus in all panels $Q_2$ is assumed smaller than some fixed value. 
In particular, for {\bf Case 1} we choose ${\cal Q}_2= 10^{-3}$, and for {\bf Case 2}, ${\cal Q}_2 = 10^{-2}$. 
Within each row, left and right panels show the sensitivity for two different bounds on the sub-dominant background source, $Q_4$. For {\bf Case 1} we set ${\cal Q}_4 = 1$ $\upmu$m$^2$ (left panel) and ${\cal Q}_4 = 0.1$ $\upmu$m$^2$ (right panel). 
For {\bf Case 2} we use ${\cal Q}_4 = 10$ $\upmu$m$^2$ (left panel) and ${\cal Q}_4 = 1$ $\upmu$m$^2$ (right panel). The dependence of the sensitivity on the dominant background source, 
$Q_3$ is shown by drawing different contours in each panel corresponding to different values of ${\cal Q}_3$. Different dashing types (defined in the plot legend of each panel) 
refer to ${\cal Q}_3 = 0.05$ $\upmu$m, $0.1$ $\upmu$m and $0.3$ $\upmu$m in the top row ({\bf Case 1}), and  ${\cal Q}_3 = 0.5$ $\upmu$m, $1.0$ $\upmu$m and $5.0$ $\upmu$m in the bottom row ({\bf Case 2}).
Notice that in {\bf Case 2} we may allow for generically larger background sources compared to {\bf Case 1}, as the choice of initial conditions is more conservative.

As always, the region excluded by present experiments \cite{Lee:2020zjt} is depicted in blue and the collisional region is represented in meshed red. Notice that for {\bf Case 1} the collisional region is significantly larger than for {\bf Case 2}\footnote{Note that the depicted collisional regions are the same than in the noiseless scenario. The reason behind this is that we focus on bounded non-collisional trajectories and only assume upper bounds for attractive background sources. In order to deal with the collisional case, a new analysis would be required.}. In the former case, it is thus more likely that we observe crashing of S onto P, while in the latter, most of the parameter space can  be tested by looking for precession in bounded non-collisional orbits of S around P. 
 Non-observation of crashing will exclude region of the parameter space corresponding to the collisional region. 
The ultimate sensitivity of the setup using a precession measurement  is presented in pink for increasing values of ${\cal Q}_3$, while the white region on the left corresponds to the region of the parameter space where we are not able to distinguish a New Physics
signal irrespectively of the backgrounds.
The borders of each region represent the value of ($\lambda,\alpha$) for which the difference between the observable $\mathrm{\Delta} T_{\rm BN}^{\rm max}(\lambda,\alpha)$ and  $\mathrm{\Delta} T_{\rm mN}^{\rm max}({\cal Q}_2,{\cal Q}_3,{\cal Q}_4)$  
is equal to twice the expected clock error, $\sigma_{T}=1$ s, {\it i.e.,} 
\begin{equation}
\label{eq:chi2sens}
\chi^2 = \frac{[\mathrm{\Delta} T_{\rm BN}^{\rm max}(\lambda, \alpha) - \mathrm{\Delta} T_{\rm mN}^{\rm max}({\cal Q}_2,{\cal Q}_3,{\cal Q}_4)]^2}{\sigma_T^2} = 4 \, . 
\end{equation}
where $\mathrm{\Delta} T_{\rm mN}^{\rm max}$ is defined in an analogous way to $\mathrm{\Delta} T_{\rm BN}^{\rm max}$. Therefore,  contours delimit the area of the parameter space that could be excluded by the experimental setup at 95\% CL.
The best sensitivity is, of course, obtained without backgrounds (solid line). 
As expected, we observe that the sensitivity of the setup decreases for increasing $Q_3$. 

Notice that the impact of $Q_3$ is much larger than that of $Q_4$, as expected. Using as a benchmark $\alpha \sim 1$ (corresponding to the generic size of $\alpha$ for LED models \cite{Antoniadis:1990ew,Antoniadis:1998ig}), we can see that the ultimate sensitivity bounds of the setup in {\bf Case 1} in the absence of backgrounds is, approximately, $\lambda \lesssim 5$ $\upmu$m. On the other hand, when a hindering
combination of background sources is considered ({\it e.g.}, ${\cal Q}_2 = 10^{-3}$, ${\cal Q}_3 = 0.3$ $\upmu$m and ${\cal Q}_4 = 1.0$ $\upmu$m$^2$), we can reach $\lambda \lesssim 7$ $\upmu$m. This should be compared with the present sensitivity,  $\lambda \leq 40$ $\upmu$m. 
For {\bf Case 2}, the sensitivity for $\alpha \sim 1$ is a bit worse: it ranges from  $\lambda \lesssim 10$ $\upmu$m with no backgrounds to  $\lambda \lesssim 20$ $\upmu$m for the worst considered case 
(${\cal Q}_2 = 10^{-2}$, ${\cal Q}_3 = 5.0$ $\upmu$m and ${\cal Q}_4 = 10.0$ $\upmu$m$^2$). Still, it is a factor two better than present bounds. 

A significant difference between {\bf Case 1} and {\bf Case 2} is the fact that, for the worst possible choice
of background sources, in the former case the region of the parameter space where we expect precession is rather small compared to the region for which the bodies would collide. This is a severe limitation, since it require a much more precise calibration of initial conditions and because the experiment has been designed to take full advantage of measurement repetition of the revolution time. On the other hand, for {\bf Case 2} the collisional region is 
very small, and the experimental setup can exploit repeated measurements at its best. It is thus clear that for {\bf Case 1} it is much more important to keep the background level under control; for example, 
compared to ${\cal Q}_3 = 0.3$ $\upmu$m, a significant increase in the sensitivity in presence of backgrounds is obtained if we manage to keep $Q_3 \leq 0.1$ $\upmu$m, irrespectively of $Q_4$. 
For $Q_3 \leq 0.05$ $\upmu$m the impact of $Q_4$ becomes more relevant, as it can be seen comparing the difference between the line of ultimate sensivity for no backgrounds (solid line) and that for ${\cal Q}_3 = 0.04$ $\upmu$m (dot-dashed line) in both figures. 

Finally, we have quantified the impact of possible backgrounds that could appear as terms with an $r$-dependence proportional to $1/r^k$ with $k \geq 5$ in eq.~(\ref{eq:mNpotential1}). 
These backgrounds include, for example, the gravitational Casimir effect (whose contribution to the force goes as $1/r^6$). 
We have found that the results shown in Fig.~\ref{fig:Limits2} do not significantly change if we include a $Q_5/r^3$ term  with $Q_5 \leq 10$ $\upmu$m$^3$ in {\bf Case 1} or $Q_5 \leq 25$ $\upmu$m$^3$ in {\bf Case 2}. Therefore, our results are robust under the hypothesis of neglecting higher-order corrections to Newton's potential, as long as such corrections are smaller than the considered backgrounds and the quoted bounds, something that should be checked in a calibration phase once materials and environmental conditions under which the experiment is carried 
on are defined.

\subsection{Performance of the experimental setup in case of a positive signal}
\label{sec:positive}

We will now study a different possibility: that the experimental setup measures an unambiguously positive signal of deviation from  Newton's $1/r^2$ law induced by New Physics. In this case, our main interest shifts to understanding the attainable
precision with which the experimental setup is able to determine the two parameters of the BN-potential, $\lambda$ and $\alpha$. There are two possible positive signals depending on the region of the parameter space where $(\lambda, \alpha)$ lie: 
\begin{enumerate}
\item {\bf Within the collisional region:} 
We observe a collision between S and P for a choice of the initial conditions for which the Newtonian potential (and the modified Newtonian potential, once $Q_2, Q_3$ and $Q_4$ are known and controlled) would give a bounded non-collisional trajectory.
\item {\bf Outside of the collisional region:} 
We are able to detect precession statistically different from that expected for the modified-Newtonian potential with a given set of $Q_2, Q_3$ and $Q_4$. 
\end{enumerate}

To properly determine $\lambda$ and $\alpha$, it is necessary to explicitly include the effect of background sources (which are controlled from a calibration phase) into the BN-potential. To this end we would use a {\em modified} Beyond-Newtonian potential
\begin{equation}
\label{eq:mBNpotential}
V_{\rm mBN} (r) = -\frac{G_\text{N}m_\text{P}}{r} \left ( 1 + Q_2 + \frac{Q_3}{r} + \frac{Q_4}{r^2} + \alpha\text{e}^{-r/\lambda} \right ).
\end{equation}
As before, it is probable that only some bounds for the $Q_i$-background are known. This would introduce an additional uncertainty in the determination of the BN parameters.

In the first case, we could measure the time  that takes S to fall onto P, $T_{\rm obs}^{\rm crash}$. 
Using the modified Beyond-Newtonian potential we could compute which values of $\lambda$ and $\alpha$ will give a trajectory that last for the same time (within errors). From this measurement, we could eventually 
draw contours  in the ($\lambda,\alpha$) plane corresponding to any number of standard deviations. To do so, we would compute

\begin{equation}
\label{eq:chi2collision}
\chi^2 = \max_{Q_i}\sum_{j = 1}^{N_{\rm rep}} \frac{\left [ \left .T_{\rm mBN}^{\rm crash} (\lambda,\alpha) \right |_{Q_2,Q_3,Q_4} - T_{\rm obs}^{\rm crash} \right ]_j^2}{\sigma_T^2} \, ,
\end{equation}
where, since we do not know the particular values of $Q_i$, we use the ones, within the bounds determined in a calibration phase, which maximize the $\chi^2$ at each point in the ($\lambda$, $\alpha$) plane. Also $N_{\rm rep}$ is the number of repetitions (each one labelled with  $j$) of the measurement that we perform. The procedure of repeating is straightforward, and it should allow to take into account errors in the choice of the initial conditions. However, repetitions could also include new measurement with different initial conditions that could allow for a better determination of $(\lambda, \alpha)$. 


Recall, however, that our experimental setup has not been optimized to exploit collisional trajectories. In contrast, its optimal performance
should be obtained in the case of a positive signal in the non-collisional region, where a large  amount of information on repeated orbits of S around P can be stored\footnote{We are using in this paper just a simple measurement
of the time needed to perform a $2 \pi$-revolution of S around P, but of course many more details of the orbits could be studied by means of a micro-camera, for example.}. In this second region, 
the statistical distribution to be studied is the following:
\begin{equation}
\label{eq:chi2precession}
\chi^2 = \max_{Q_i} \sum_{j = 1}^{N_{\rm rep}} \sum_{i = 1}^{N_{\rm rev}} \frac{\left [ \left . T^i_{\rm mBN} (\lambda,\alpha) \right |_{Q_2,Q_3,Q_4} - T^i_{\rm obs} \right ]_j^2}{\sigma_T^2} \, ,
\end{equation}
where $T^i_{\rm obs}$ is the time it takes to S to perform the $i$-th  $2\pi$-revolution around P, and $\left . T^i_{\rm mBN} (\lambda,\alpha) \right |_{Q_2,Q_3,Q_4}$ is the expected 
$i$-th revolution time for a given choice of $\lambda$ and $\alpha$, and fixed values of $Q_2, Q_3$ and $Q_4$, computed using eq.~(\ref{eq:mBNpotential}). 
As before, we explicitly indicate the fact that the measurement could be repeated $N_{\rm rep}$ times. 

\begin{figure*}[h]
    \centering
    \begin{minipage}{0.5\textwidth}
        \centering
        \includegraphics[width=\textwidth]{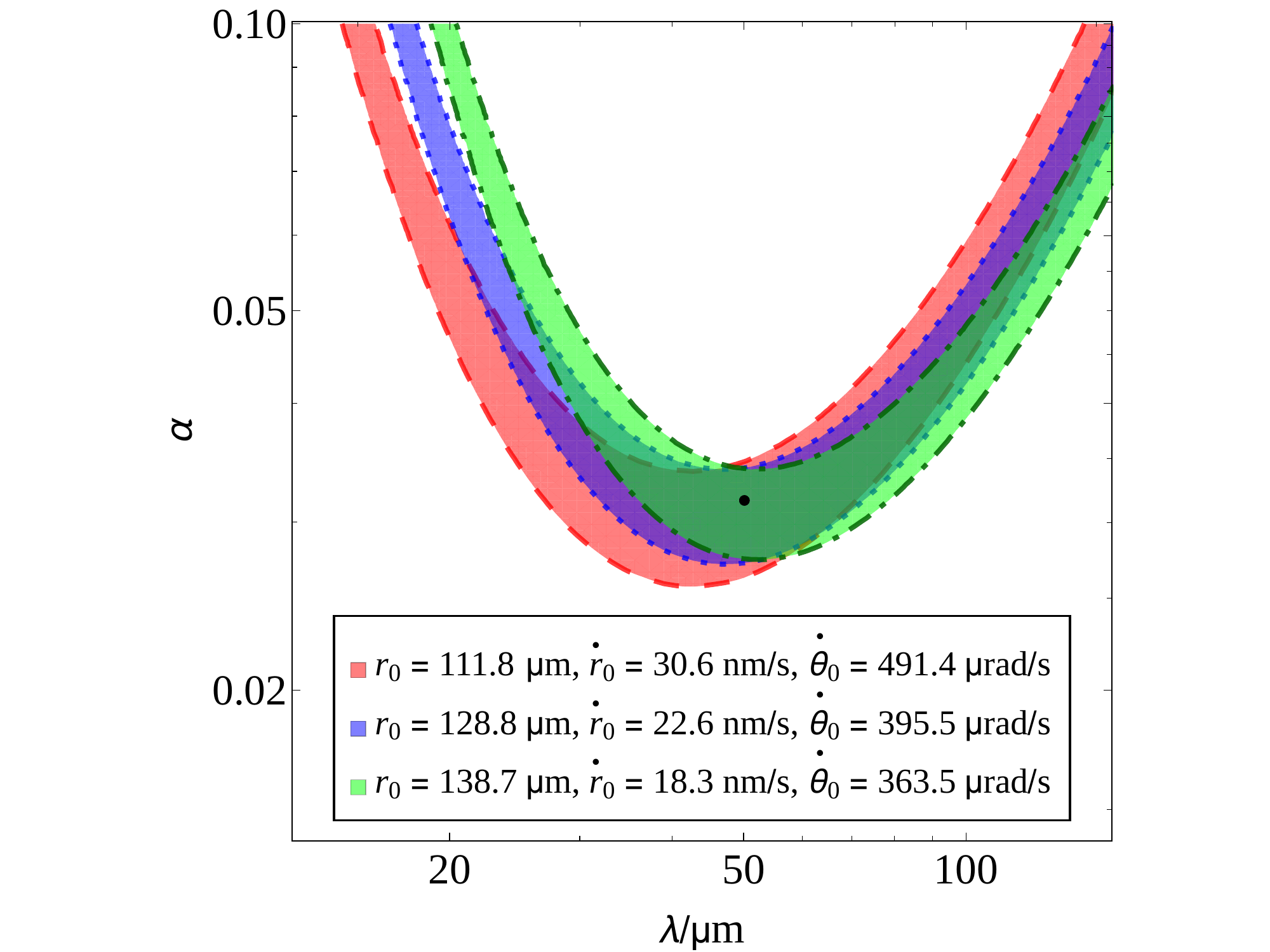} 
    \end{minipage}\hfill
    \begin{minipage}{0.5\textwidth}
        \centering
        \includegraphics[width=\textwidth]{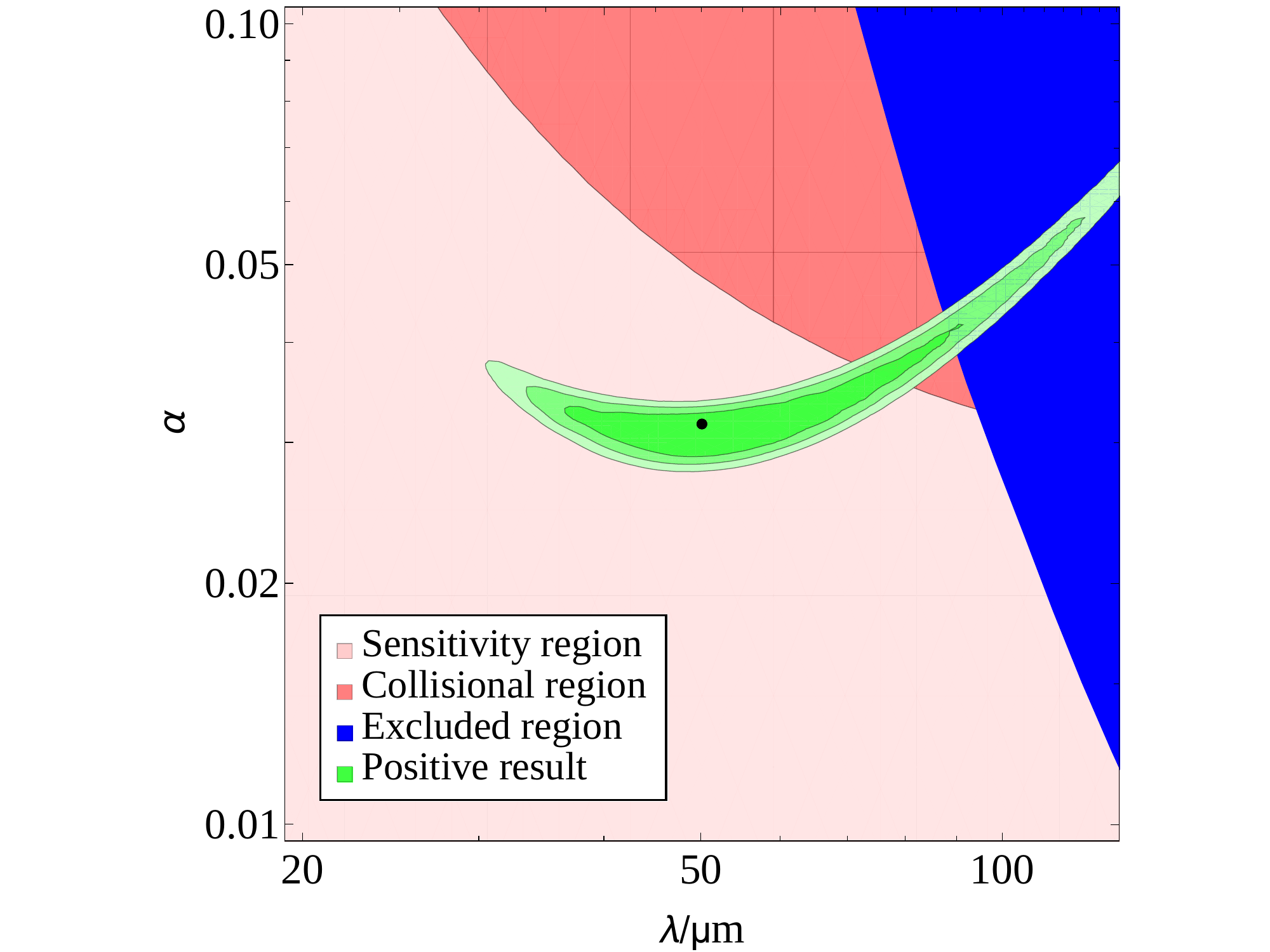} 
    \end{minipage} 
\vspace{0.2 cm}    
    
    \begin{minipage}{0.5\textwidth}
        \centering
        \includegraphics[width=\textwidth]{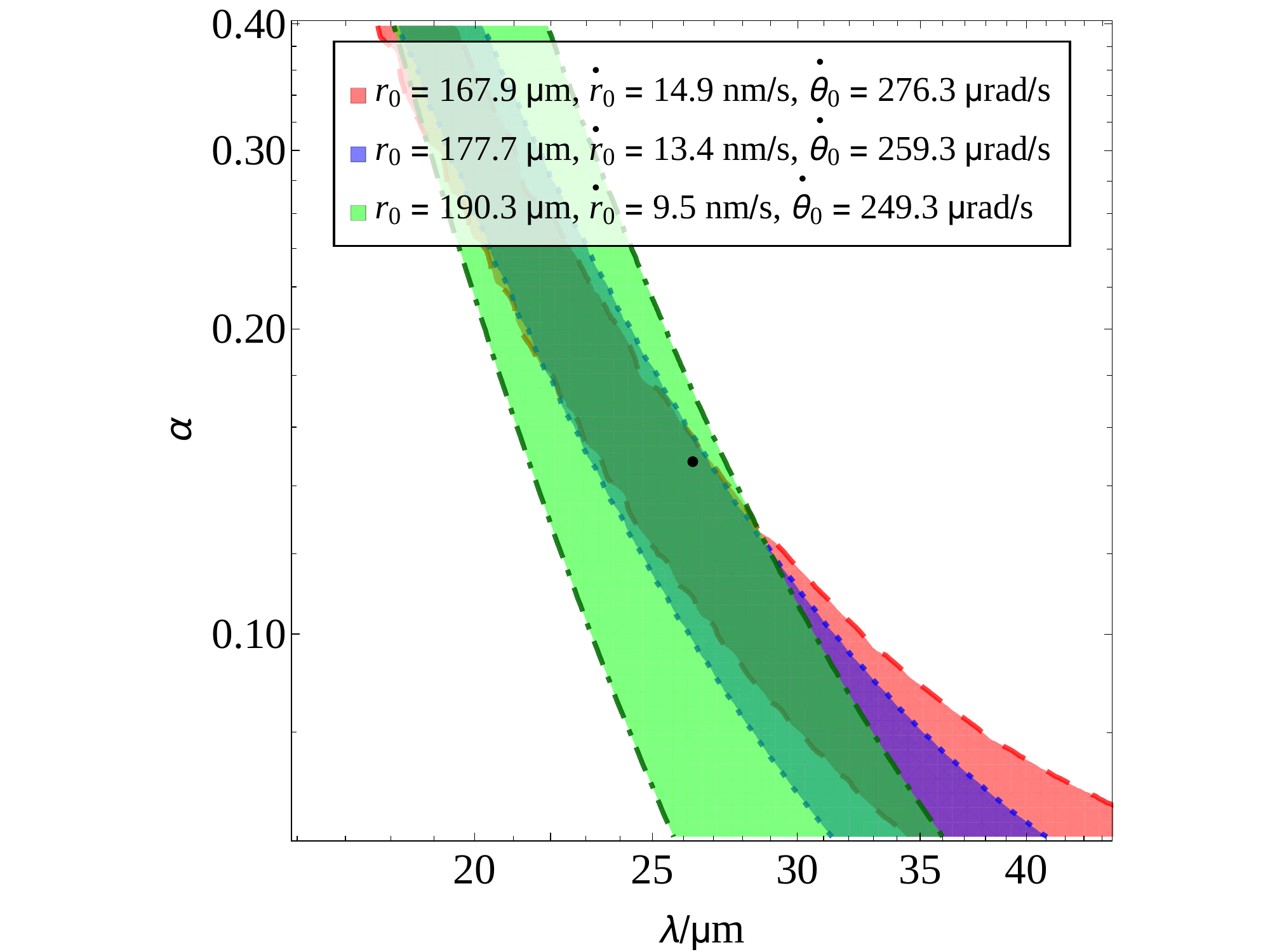} 
    \end{minipage}\hfill
    \begin{minipage}{0.5\textwidth}
        \centering
        \includegraphics[width=\textwidth]{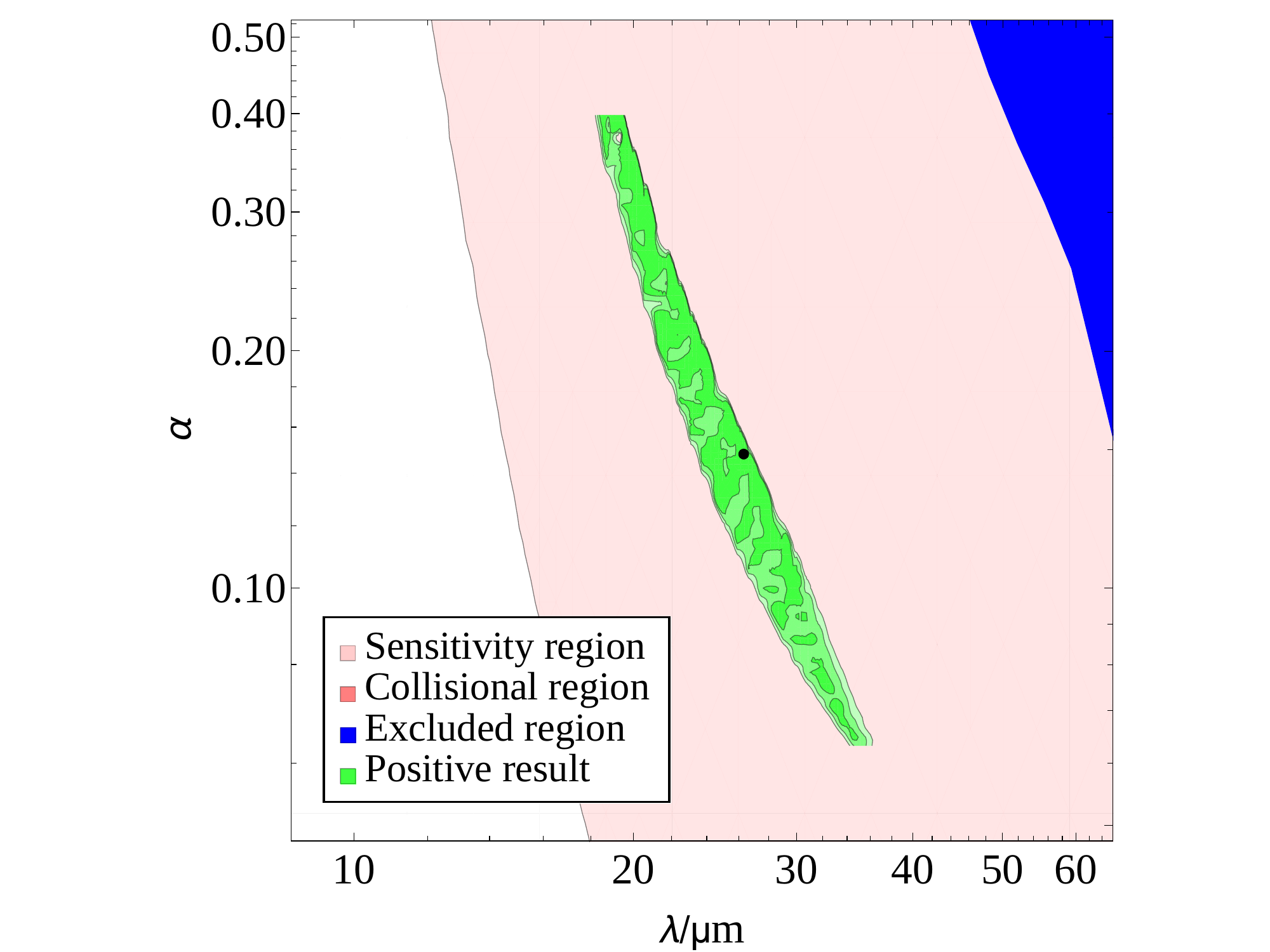} 
    \end{minipage} 
    \caption{
    \it
	Region in the ($\lambda$, $\alpha$) plane compatible with an artificially generated signal that includes a fixed $Q_3$ background (black dot).  
	Top row (\textbf{\textit{Case} 1}): the signal has been generated for $\bar \lambda=5.12$ $\upmu$\textup{m} and $\bar \alpha=0.031$, with $\bar Q_3=0.005$ $\upmu$\textup{m} (and we have considered $Q_2 = Q_4 = 0$ for simplicity). For the fits we have considered $Q_3\leq 0.05$ $\upmu$\textup{m}.
	Bottom row (\textbf{\textit{Case} 2}): the signal has been generated for $\bar \lambda=26.3$ $\upmu$\textup{m} and $\bar \alpha=0.148$, with $\bar Q_3=0.05$ $\upmu$\textup{m}. For the analysis we have used $Q_3\leq 0.5$ $\upmu$\textup{m}.
	Left panels: $2\sigma$ contours for $ \chi^2$ as defined in eq.~(\ref{eq:chi2precession}) for three different choices of the initial conditions (indicated in the legends). 
	Right panels: $1 \sigma, 2 \sigma$ and $3\sigma$ contours for $\chi^2$ as defined in eq.~(\ref{eq:chi2precession}) for the combined fit of all three choices in the left panels. In the same plot, we also depict the presently excluded region (in blue)
	and the collisional region (dark meshed-red).
	  }
        \label{fig:ChiSquare1}
\end{figure*}

  

Our results for the case of a positive signal in the non-collisional region are shown in Fig.~\ref{fig:ChiSquare1}. We generated a set of {\em theoretical expectations} over 30 revolutions  to replace the experimental measurements, 
for two benchmark points of the parameter space, ($\bar \lambda, \bar \alpha$), depicted by a black dot in all panels.  These points are chosen so they can be studied with initial conditions similar to those adopted in {\bf Case 1} and {\bf Case 2} when discussing the experiment sensitivity. Moreover, as $Q_2$ does not affect precession measurements and $Q_4$ is sub-dominant, they have both been set to zero, while some particular values of $Q_3$, denoted as $\bar Q_3$, have been taken into account. In the posterior analysis, we consider $Q_3$ to be positive and set an upper bound for its value, ${\cal Q}_3$, as done in the previous section.  The four panels in Fig.~\ref{fig:ChiSquare1} are defined as follows:
\begin{itemize}
\item Top row ({\bf Case 1}): 
We have $\bar \lambda = 5.12$ $\upmu$m, $\bar \alpha = 0.031$ and $\bar Q_3 = 0.005$ $\upmu$m, and have set $ {\cal Q}_3 = 0.05$ $\upmu$m for the analysis. Clearly, the minimum of the $\chi^2$ distribution defined in eq.~(\ref{eq:chi2precession}) coincides with $\lambda = \bar \lambda$ and $\alpha = \bar \alpha$, and $\chi^2_{\rm min} = 0$, as we are not taking into account statistical fluctuations. In the left panel, we show the $2 \sigma$ contour for $\mathrm{\Delta} \chi^2= \chi^2-\chi^2_\text{min}$ for three different choices of the initial conditions (as defined in the plot legend). 
The change of the initial conditions when we observe a positive deviation from Newton's $1/r^2$ law is needed to significantly reduce the correlation between the determinations of $\lambda$ and $\alpha$. This is depicted in the right panel, 
where we show $1\sigma,2 \sigma$ and $3\sigma$ contours for the combined fit of the three measurements. As a reference, we represent the current excluded region (in blue) and the collisional region in the absence of backgrounds (in meshed red). 
We can see for this particular benchmark point and {\bf Case 1} initial conditions that the correlation is successfully removed, and a very precise measurement of $\bar \alpha$ is possible 
(with a larger error in $\bar \lambda$, though).
\item Bottom row ({\bf Case 2}):
We have: $\bar \lambda = 26.3$ $\upmu$m, $\bar \alpha = 0.148$ and $\bar Q_3 = 0.05$ $\upmu$m, and use a prior ${\cal Q}_3 \leq 0.5$ $\upmu$m. Again, in the left panel we show $2\sigma$ contours for three different choices of the initial conditions (as indicated in the plot legend), 
whereas in the right panel we combine the three measurements and give $1\sigma,2 \sigma$ and $3\sigma$ contours for the combined fit. In this case, we can see that a very good measurement of $\bar \lambda$ is possible, albeit
we can have an error on $\bar \alpha$ larger than 100\%.
\end{itemize}
The results we presented here could be repeated for other benchmark points in the parameter space, where we expect to obtain similar conclusions as long as these points are not too near the sensitivity limit of the experimental setup. 
In order to improve the precision of the measurement of the worst-measured parameter (either $\bar \lambda$ or $\bar \alpha$), the experiment could furthermore be repeated for more choices of initial conditions (something
that, in turn, would require a better control of them). 

We must eventually stress that, in case of a positive result, repeated measurements using different initial conditions could allow for something extremely interesting: 
by measuring simultaneously the Yukawa term, $\alpha \, \exp (-r/\lambda)$, and the $Q_i$ coefficients, we could pin down which particular modification of gravity we are dealing
with. For example, extra-dimensional extensions of gravity can be cast in the form of a series of Yukawa terms $\exp [-f(m_n) \,  r ]$ whose exponent depends on the KK-graviton mass spectrum $m_n$ of each particular model (being it LED, CW/LD or RS). Differences between the specific extra-dimensional model arise in the dependence of the coupling of
the Yukawa terms, $\alpha$, that in general is a function of the parameters that define the geometry of the extra-dimension (the compactification radius $R$ and the curvature $k$). 
On the other hand, other modifications of gravity imply a different dependence of $F_{\rm BN} (r)$ on the spatial distance $r$ (see, for example, 
Refs.~\cite{Perivolaropoulos:2016ucs,Edholm:2016hbt}). 
This specific dependence on $r$ of each model can be usually cast in form of a polynomial of central potentials, whose coefficients are precisely our $Q_i$ terms. Therefore, 
constrining $\alpha$ and $\lambda$ together with several $Q_i$ (instead of treating them as nuisances, as we have done here) could help in distinguishing different models. 
Clearly, a very good control over backgrounds and sources of systematics would be mandatory in order to carry on this task.

\section{Conclusions}
\label{sec:concl}

Extensions of the Standard Model of particle physics aiming at solving one or more of its open problems  
(the existence of Dark Matter in the Universe, the asymmetry between baryons and anti-baryons, or the origin of neutrino masses, to name just a few) 
 have been proposed from the very beginning, most popular among them being Supersymmetry and Technicolor. These extensions were suggested to take care of one of the theoretical longstanding problem of the SM: the hierarchy problem. This is nothing more that the annoyingly large hierarchy existing between the (by now observed) Higgs mass and the fundamental scale at which New Physics should replace the SM. From a field theoretical point of view the scalar mass should get corrections of the order of that fundamental scale. Therefore, the observed ``light'' Higgs mass asks for some cancellation mechanism to keep these additive corrections small, or that the scale of New Physics is indeed around the corner.
Unfortunately, no hints of New Physics have been detected at the LHC, and models in which new particles are added to the SM spectrum with relatively light masses (within $[100, 1000]$ GeV) are currently in trouble.

 During the the nineties, a different class of extensions was proposed to explain the hierarchy problem and, at the same time, tackle another longstanding problem: our inability to understand
 gravity as a quantum field theory.  These models consider the existence of one or more extra spatial dimensions, that could be compact or not. In the market there are currently three frameworks: Large Extra-Dimensions, Randall-Sundrum and Clockwork/Linear Dilaton models. 
 Of course, non-observation of tensions with the SM predictions at the LHC also puts severe constraints to these models, when considered as possible solution to the hierarchy problem, as they all imply TeV-scale New Physics.
 As a consequence, a fundamental scale of gravity at the scale of few TeV is mostly excluded.
 However, these models also lead to deviations from known physics in a completely different realm: as they assume the existence of one or more new spatial dimensions, they also imply that Newton's $1/r^2$ law
 for gravitational force is modified to take into account propagation of gravity in the new dimensions. 
 
 This is the problem that we have studied: how we can improve present bounds
 on deviations from Newton's law, currently certifying that gravity behaves 4-dimensionally down to distances of $\lambda \sim 50$ $\upmu$m (for a magnitude of the corrections of $\alpha\sim 1$). Experimental setups aiming at the study of deviations from Newton's law usually measure the absolute value of the force acting between a source of gravitational field and a test mass. 
 These experiments, though, are significantly affected by electrically-induced backgrounds, what limits their sensitivity to New Physics.
 In a previous paper \cite{Donini:2016kgu}, it was suggested to study the dynamical behaviour of a microscopic gravitational system made of a Satellite orbiting around a bigger body, the Planet. 
 New Physics effects would induce precession in the orbit of the former, what could be used as an unambiguous signal for terms of the effective potential
 that differ from the Newtonian one and from those generated by the main electrically-induced backgrounds. In spite of being independent of such backgrounds, an equivalent signal could nevertheless be generated by sub-dominant noises in a realistic setup.
 
 In this paper we have explored in detail the expected results for a realistic experiment 
 on the line of what proposed in Ref.~\cite{Donini:2016kgu}, whilst adding two major improvements: 
 first, we have reoptimized the setup in order to 
 improve its sensitivity, understanding the dependence of the experimental observables on tunable parameters of the setup such as  the mass of the gravitational source or the choice of initial conditions; 
 second, and most important, we have studied the impact of attractive background sources that could be cast in terms of central potentials. 
 To do so we have used a modified Newtonian potential, adding terms proportional to $1/r$, $1/r^2$ and $1/r^3$, and computed the amount of precession they induce. Eventually, we have been able to derive constraints
 on the maximal amount of backgrounds of the three types that our experimental setup can allow without significant loss in the sensitivity to $\lambda$ and $\alpha$. Our final conclusion is that the proposed setup should be able to improve
 present bounds on $\lambda$ by a factor 5 to 10 for any value of $\alpha \in [5\times10^{-3}, 10^3]$.
 
 The range of the 
 improvement over present bounds depends significantly on the value of background terms in the potential, specially on those proportional to $1/r^2$, like the dominant contribution from the electric Casimir effect between two spheres. However, for this particular source it comes out that, taking a (mostly) conducting
 sphere as the Planet and a (mostly) insulating one as the Satellite (as in the setup we are proposing),  the electric Casimir effect will be repulsive and induce precession in the opposite sense to New Physics 
 for positive $\alpha$. This means that the sensitivity of the experimental setup is actually much less affected by this particular background source in the positive $\alpha$ half-plane than if it was attractive. The case of negative $\alpha$ would be much more influenced, since both electric Casimir effects and New Physics would produce precession in the same sense, meaning the signal would be much harder to isolate. In any case, we have left the analysis of this scenario for a future study.
 
 Last, but not least, we have performed the exercise of studying the attainable precision with the proposed experimental setup in case of a positive signal of deviation from Newton's law. We have shown that  repeated measurements for slightly different choices of the initial conditions allow to break the correlations between $\lambda$ and $\alpha$, leading to a very good measurement of at least one of the two variables in the Beyond-Newtonian  potential (which of them depending on the particular point of the parameter space). 
 
To conclude, we think to have shown beyond any reasonable doubt the capability of a dynamical measurement of deviations from Newton's law to improve over present bounds. We hope this paper may help in passing 
 from a theoretical study of a {\em gedanken} experiment to a new phase in which this proposal may be implemented in a real table-top experiment.
 
\begin{acknowledgement}
We acknowledge helpful discussions with A.~Cros and A. Molina that gave us advise on the levitation setup. We also thank S. G. Marim\'on, who contributed in the early stages of this project. 
This work has been supported by the Spanish grants SEV-2014-0398 and FPA2017-84543-P, the European project H2020-MSCA-ITN-2019//860881-HIDDeN, and 
by the Generalitat Valenciana  through the grant PROMETEO/2019/083. JBB is also supported by the Spanish grant FPU19/04326 of MU. SNG is supported by the Atracció de Talent PhD fellowship of UV.
\end{acknowledgement}

\appendix

\section{Magnets configuration}
\label{app:MC}

In this Appendix we give some basic ideas regarding a magnet configuration that may be used to levitate the pyrolitic graphite Satellite (for more a more elaborate levitating setup, see Ref.~\cite{Berry_1997}). In order to make things as simple as possible, we
consider two commercial cylindrical neodymium (Sintered NdFeB) magnets, with remanent magnetic field of 1.25 T and 15 mm of radius. This yields within the range attainable with commercial magnets (usually between 1 ans 1.4 T). We set them with their axes aligned and parallel to Earth's gravitational field, with opposite orientations for their north/south poles, and separated by a distance of 4 mm. The orbital plane will be located between them with the planet on their common axis. The maximum operating temperature for these magnets is $T_{\rm max} \in [80,220] \,^\circ$C, ensuring they will present the correct behavior under the conditions at which the proposed experiment would take place. 

\begin{figure}[!h]
\centering
\includegraphics[angle=0,width=0.49\textwidth]{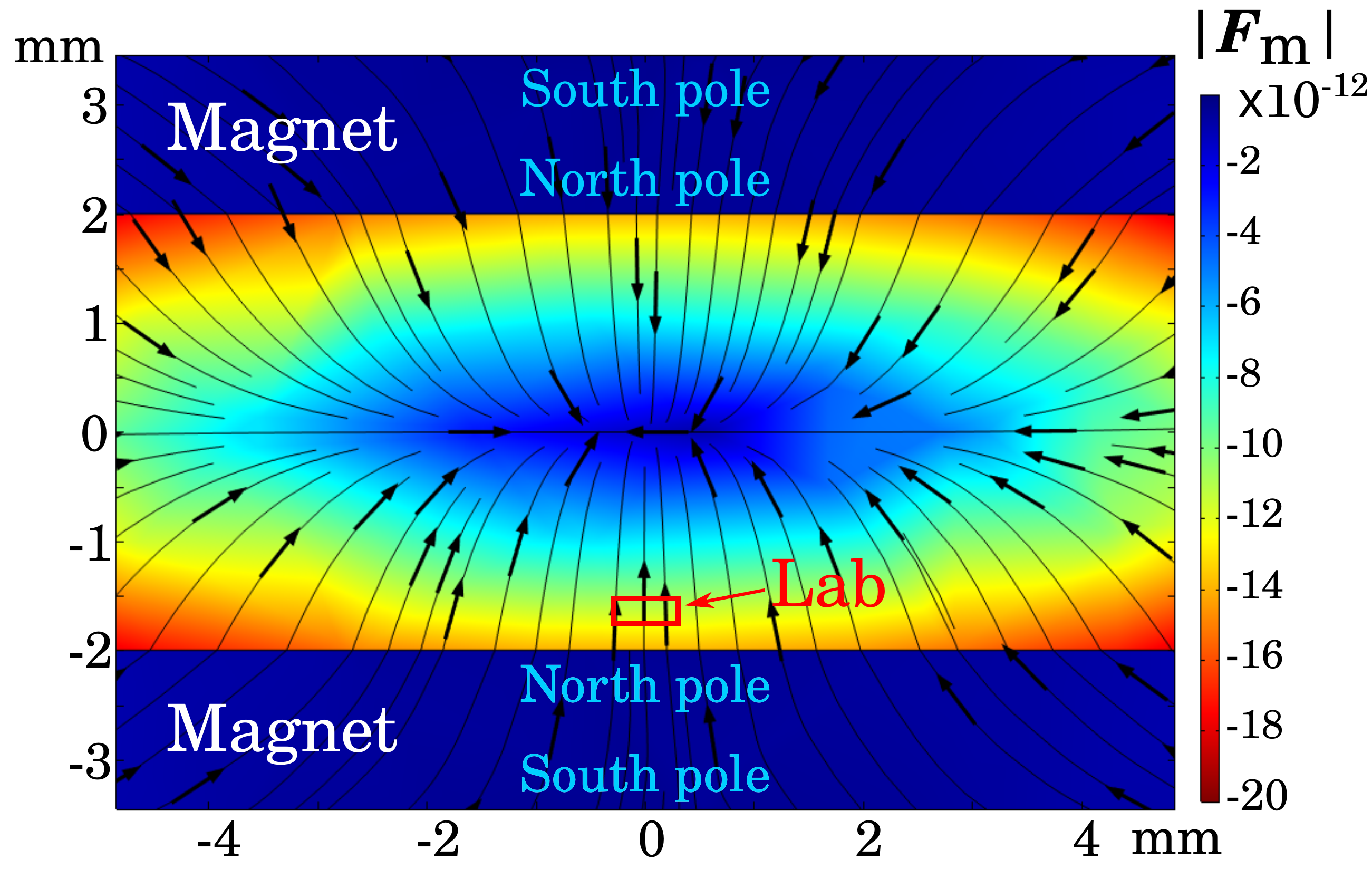}
    \caption{
    \it
The magnetic force field of the considered setup for  two 1.25 \textup{T} neodymium magnets, in units of $10^{-12}$ \textup{N}. The separation between the two magnets (depicted in dark blue) is 4 \textup{mm}. The radius of the magnets on the plane parallel to the orbits is 15 \textup{mm}. The equilibrium position of the Satellite is approximately 300 $\upmu$\textup{m} above the north pole of the lower magnet. The Lab zone is represented by a red rectangle. We used the COMSOL Multiphysics® software to perform the modeling \cite{comsol}.
	  }
        \label{fig:magnets}
\end{figure}

This configuration has been modeled using the COMSOL Multiphysics® software \cite{comsol}. The magnetic force exerted on the Satellite is depicted in Fig. \ref{fig:magnets}, where the region geometrically inside the magnets is represented in dark blue and the region in between the two is depicted via a color palette that gives the strength of the magnetic force in units of $10^{-12}$ N. The arrows on each line give an idea of the force field, once we take into account that the pyrolitic graphite diamagnetic susceptibility ($\chi = -1.6 \times 10^{-4}$) is negative. For a Satellite of mass $m_\text{S} = 1.2 \times 10^{-12}$ kg, the equilibrium position of the system is approximately 300 $\upmu$m above the surface of the lower magnet. 
The \textit{Lab} zone (the region where the Satellite orbits around the Planet) is located within the red rectangle, that we have drawn to give a graphical understanding of the setup. 
Remind that the typical orbits considered in this paper are a few hundreds of $\upmu$m at most and, therefore, safely within the Lab. In this region, effects due to inhomogeneities of the surface of the lower magnet, of typical size smaller than 1 $\upmu$m, depending on the coating of the magnets,  are completely negligible.

The position of the Lab is a stable equilibrium region for the Satellite, as the magnetic force will cancel the Earth gravitational force. Should it be introduced slightly above the orbital plane, it will start to fall down and then oscillate around the orbit plane. These oscillations are orthogonal to the orbit plane, and so will have no impact on the determination of a precession signal. Similarly, border effect will generate magnetic forces tangential to the plane. These are expected to be negligible, since the radius of the magnets is much larger than the Lab size. Also, as the orbit is located near the axis of the magnets, these effects will produce a force linear with the distance to the axis, not inducing any precession, and so the signal will remain unspoiled.
Nevertheless, these subdominant magnetic effects may affect the expected value of the revolution period or lead to unexpected collision between the two bodies. Therefore a careful calibration phase of the experiment must be considered. During this, any magnetically-induced electrical force should also be studied, even though the Satellite material has been particularly chosen in order to minimize such electrical effects.

\section{Uncertainties in the fixing of initial conditions}
\label{app:IC}

One of the key aspects of the experimental setup proposed in Sect.~\ref{sec:setup} is the fact that, by choosing an appropriate set of initial conditions for our microscopic planetary system we can optimize
the signal-to-background ratio.  
 However, our capability to fix a given set of initial 
conditions is also affected by uncertainties. First of all, the starting position of the Satellite is only known with a given error. On top of this, as we suggest in Sect.~\ref{sec:setup} and in Ref.~\cite{Donini:2016kgu}, 
putting into motion our Satellite at a given velocity may be achieved by the means of  photo-irradiation \cite{Kobayashi:2012}, a procedure which also implies an uncertainty. 

 A central point which is worth reminding is that a variation in the initial conditions does not affect whether the Satellite precedes or not around the Planet. Therefore, if we were able to detect 
 precession, any error in the  fixing of the initial conditions would be irrelevant. However, these uncertainties do play a role when  trying to optimize the setup
 or to determine its sensitivity ({\em i.e.,} its capability to exclude New Physics at a given confidence level), due to different reasons. 
 First, errors in the choice of initial conditions eventually imply a wrong prediction of the Newtonian period. In Tab ~\ref{tab:icvariations} we can observe the size of the variation for percent level uncertainties.
\begin{table}[h]
 \centering
 \caption{ \it 
Newtonian period $T_{\rm N}$ (in seconds) for the proposed setup (with $m_\text{P}=0.75\times 10^{-4}$) for variations of the initial conditions at the percent level, with the benchmark corresponding to
$r_0 = 111.8$ $\upmu$\textup{m}, $\dot{r}_0 = 30.6$\textup{ nm s}$^{-1}$ and $\dot{\theta}_0 = 491.1$ $\upmu$\textup{rad s}$^{-1}$, ({\it i.e.,} {\rm {\bf \textit{Case} 1}} as explained in the main text).
}
\label{tab:icvariations}
  \begin{tabular}{cc|ccc}
   &  & \multicolumn{3}{c}{$\mathrm{\Delta} r_0/r_0$}  \\
    &  & -1\% & 0\% & 1\% \\ \hline
 & -1\% & 8475.8 \; & 8673.4 \;& 8881.0 \\
 $\mathrm{\Delta} v_0/v_0$ & 0\% & 8703.4 \; & 8910.0 \; & 9127.2 \\ 
  & 1\% & 8936.5 \; & 9152.6 \; & 9379.8 \\   
\end{tabular}

\end{table}
 
\begin{figure}[!h]
\centering
\includegraphics[angle=0,width=0.52\textwidth]{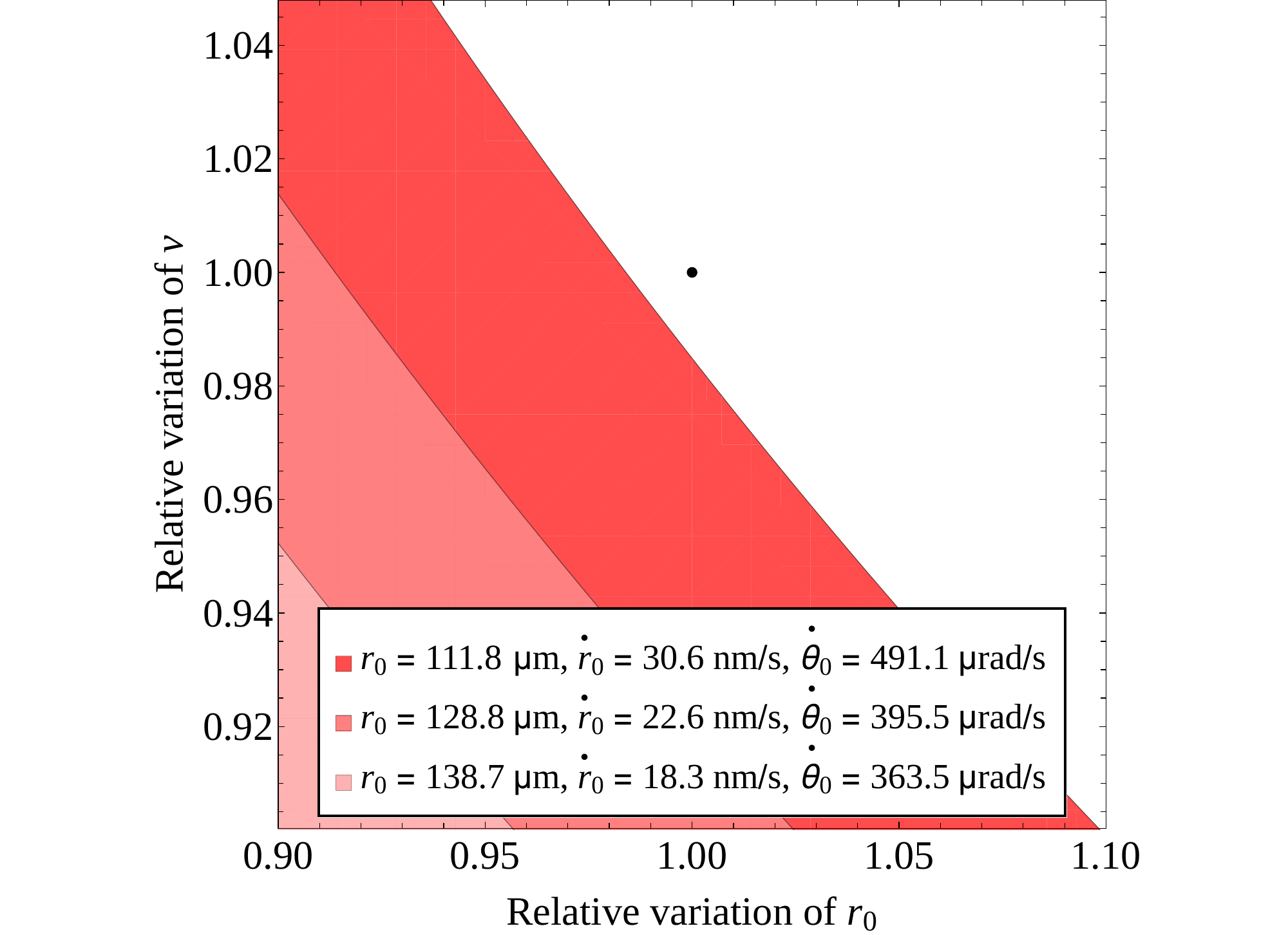}
\caption{\it Impact of relative error on the choice of the initial conditions on the qualitative behaviour of the trajectory. The black dot represents three possible choices of initial conditions (as reported in the legend)
for which \textup{S} is expected to perform a closed non-collisional orbit around \textup{P} in the Newtonian case. The three coloured regions represent the region of the parameter space of initial conditions for which we expect collision 
between \textup{S} and \textup{P} in the Newtonian case in the three cases (where the color code correspond to the different choices of initial conditions, as reported in the legend). In each case, we can see that in order to qualitatively preserve the expected closed orbit, 
we should keep errors on the initial conditions fixing below some value that ranges between 1\% for $r_0=111.8$ $\upmu$\textup{m} to  $\sim$10\% for $r_0=138.7$ $\upmu$\textup{m}. 
}
\label{fig:Coll}
\end{figure}

Second, uncertainties also affect whether the Satellite collides with the Planet or not. This latter aspect is analyzed in the Newtonian scenario (without New Physics or backgrounds) in Fig.~\ref{fig:Coll}: for a given set of initial conditions that give a closed non-collisional orbit
 of S around P (where the black dot corresponds to three of these, as explained in the Figure legend), we show the range of relative variations of the initial radius, 
 $\mathrm{\Delta} r_0/r_0$, and of the absolute value of the initial total velocity\footnote{Notice that $v_0 = (\dot{r}_0^2 + r_0^2 \, \dot{\theta}_0^2)^{1/2}$, so it does not only depend on $\dot{r}_0$ and $\dot{\theta}_0$, but also on $r_0$. 
 },  
 $\mathrm{\Delta} v_0/v_0$, for which we would  observe collision. For a given initial condition  that would produce a closed non-collisional orbit, the corresponding colored region represents the  relative changes  $\mathrm{\Delta} r_0/r_0$ and $\mathrm{\Delta} v_0/v_0$  for which we instead expect a collisional trajectory. Each color refers to one of the different choices of 
 $( r_0,  v_0)$, as explained in the Figure legend. In order to preserve the closed orbit of S around P, 
 we find that the initial conditions should differ from the theoretical choice less than a value that ranges  between 1\% for $r_0=111.8$ $\upmu$m to $\sim$ 10\% for $r_0=138.7$ $\upmu$m. 
 If we are able to keep the uncertainties on the initial conditions below this value, observing collision when the expected Newtonian trajectory should be non-collisional would
 unambiguously point to Beyond-Newtonian physics (or significant backgrounds).

\begin{figure*}[!h]
\centering
\begin{tabular}{cc}
\includegraphics[angle=0,width=0.5\textwidth]{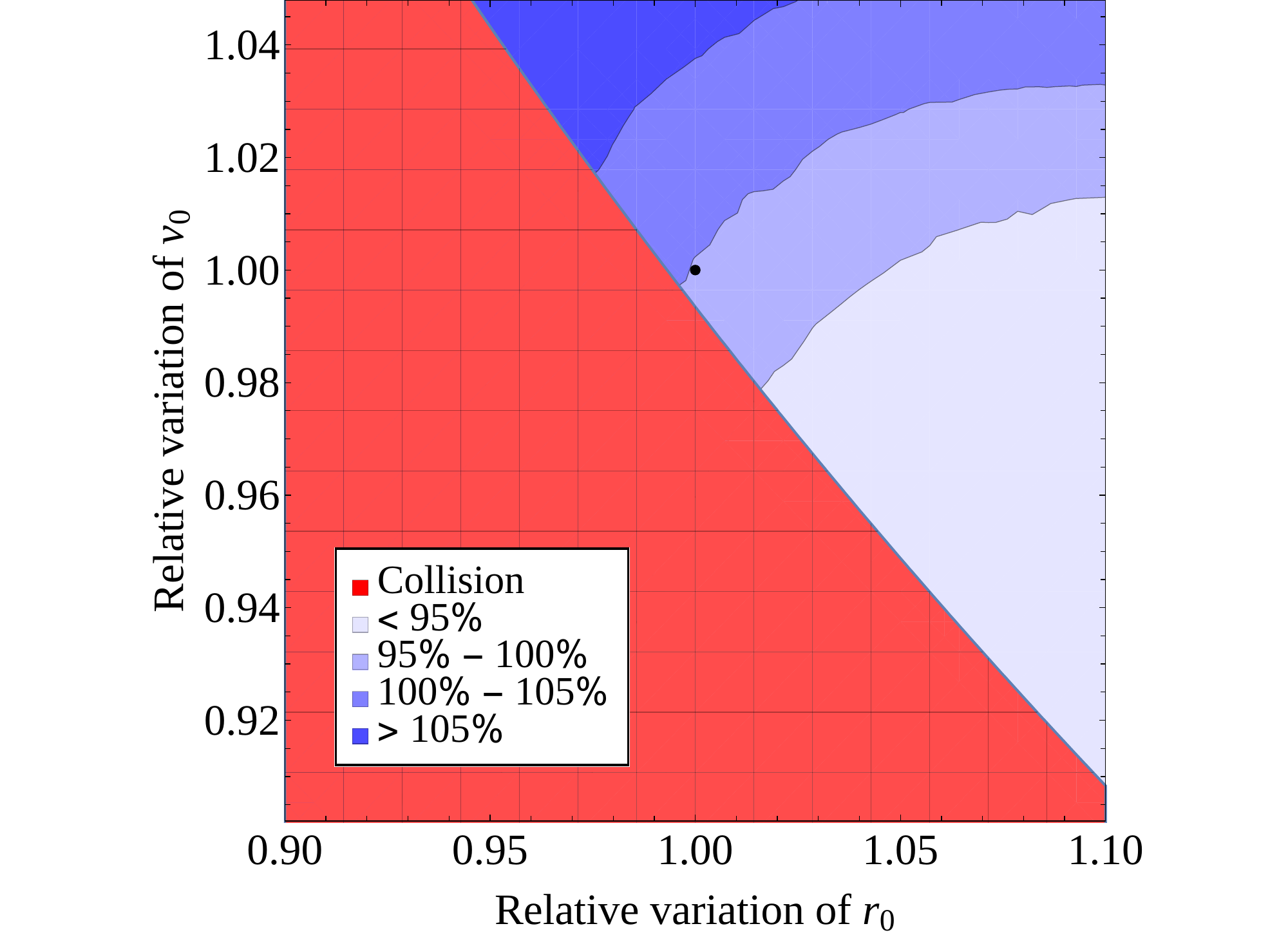} &
\includegraphics[angle=0,width=0.5\textwidth]{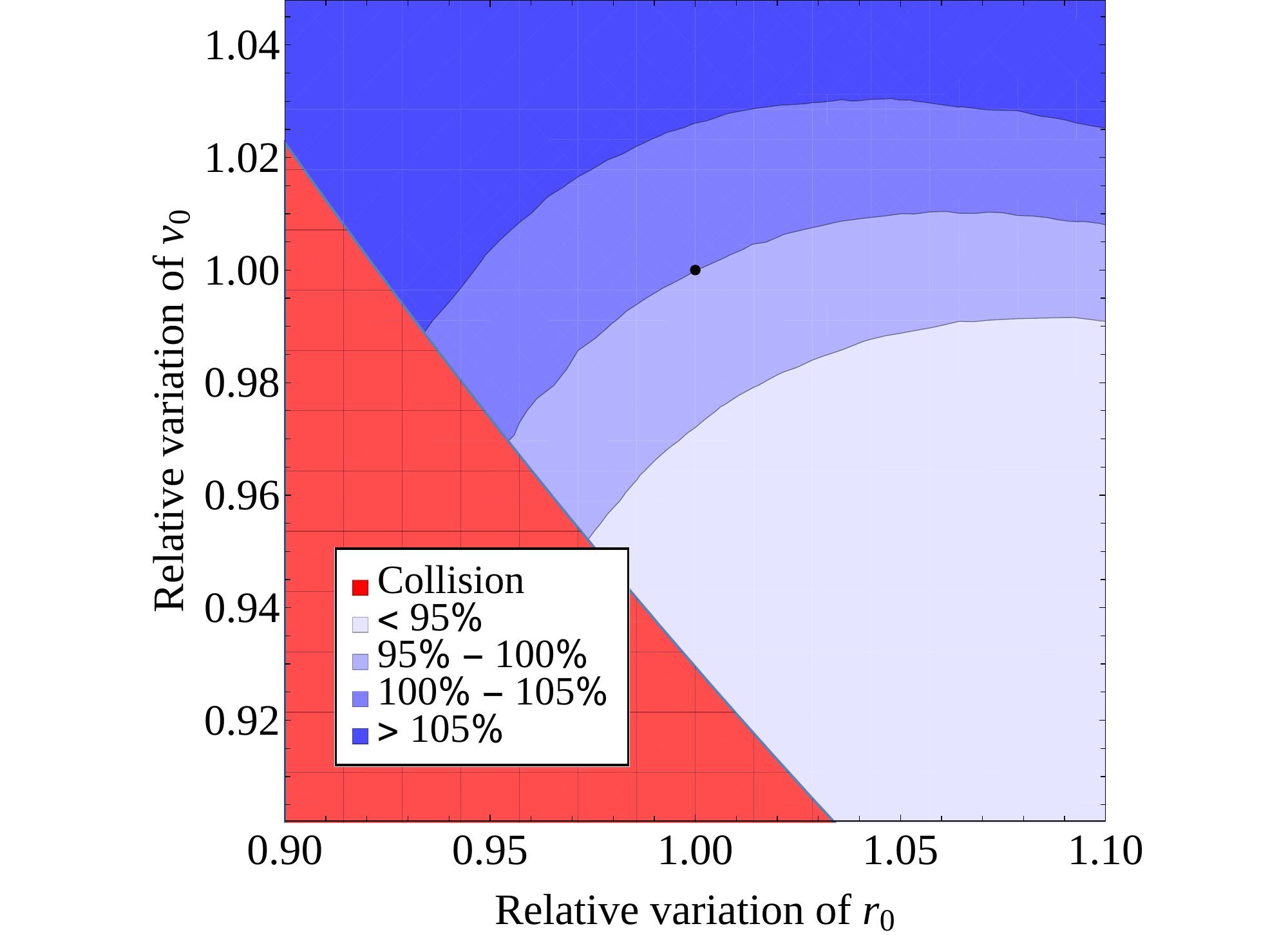}
\end{tabular}
\caption{\it 
Impact of variations in the initial conditions on the observable defined in eq.~\ref{eq:TotalPeriodVariation} measured over 30 revolutions, for $\lambda=50$ $\upmu$\textup{m} and $\alpha=0.03$. 
Left panel: $r_0=111.8$ $\upmu$\textup{m}, $\dot{r}_0=30.6$ \textup{nm s}$^{-1}$ and $\dot{\theta}_0=491.1$ $\upmu$\textup{rad s}$^{-1}$; Right panel:
$r_0=128.8$ $\upmu$\textup{m}, $\dot{r}_0=22.6$ \textup{nm s}$^{-1}$ and $\dot{\theta}_0=395.5$ $\upmu$\textup{rad s}$^{-1}$. Variations which will  generate collision are represented in meshed red. We observe that the second set of initial conditions is more ``robust'' against uncertainties in their fixing.
}
\label{fig:VarIC}
\end{figure*}

Finally, even if an error in the initial conditions fixing does not modify the qualitative behaviour of the motion of S (as it was in the previous case, changing from a closed non-collisional orbit to a collisional trajectory), 
it has a quantitative impact on the observed signal. In Fig.~\ref{fig:VarIC} we show how the observable described in eq.~(\ref{eq:TotalPeriodVariation}) measured over $N_{\rm rev}=30$,  for $\lambda=50$ $\upmu$m and  $\alpha=0.03$, changes due to an error in the initial conditions. In the left panel we consider $r_0=111.8$ $\upmu$m, $\dot{r}_0=30.6$ nm s$^{-1}$ and  $\dot{\theta}_0=491.1$ $\upmu$rad s$^{-1}$ (represented by a black dot). The meshed red-shaded region is the one for which change in the initial conditions (either in $r_0$ or $v_0$) would induce collision. 
On the other hand, the blue-shaded regions represent relative changes of $( r_0,  v_0)$ which are small enough to preserve the qualitative behaviour of the trajectory of S. 
Given the chosen initial conditions,  for $\lambda=50$ $\upmu$m and  $\alpha=0.03$ we expect to measure $\mathrm{\Delta} T_{\rm BN}^{\rm max} = 72.9$ s, 
with an average revolution period (over 30 revolutions) of $\langle T_{\rm BN}\rangle_{\text{30 rev}} \sim 8750$ s. 
In the darkest blue-shaded region (corresponding orientatively to larger initial velocity and smaller initial distance) we have an increase in the signal of more than 5\%, whereas in the lightest
blue-shaded region (smaller velocity and larger initial distance) we expect a decrease of more than 5\%. In the right panel, we slightly modify the initial conditions by increasing the initial distance 
$r_0$ and decreasing the initial velocity $v_0$: the black dot stands for $r_0=128.8$ $\upmu$m, $\dot{r}_0=22.6$ nm s$^{-1}$ and $\dot{\theta}_0=395.5$ $\upmu$rad s$^{-1}$, so that the expected signal is  
$\mathrm{\Delta} T_{\rm BN}^{\rm max}=90.5$ s, with an average revolution period of around $\langle T_{\rm BN}\rangle_{\text{30 rev}} \sim 9700$ s. Also for this particular choice of initial conditions, a much larger relative error on the initial conditions is required to induce collision (roughly around 5\%). These modifications thus make the initial conditions much more "robust" against uncertainties in their fixing.


\section{Electric Casimir force and other noises}
\label{app:Casimir}


As it is commented in the main text, in order to get a (maximally) noiseless situation, we propose using  a platinum Planet of mass $m_\text{P} = 0.75 \times 10^{-5}$ g and radius $R_\text{P}=43.7$ $\upmu$m, and a pyrolytic graphite Satellite of mass $m_\text{S}=1.2 \times 10^{-9}$ g and radius $R_\text{S}=5$ $\upmu$m. 
Also, two different sets of initial conditions are studied. First, {\bf Case 1}: $r_0=111.8$ $\upmu$m, $\dot{r}_0=30.6$ nm s$^{-1}$ and $\dot{\theta}_0=491.1$ rad s$^{-1}$, corresponding to maximal sensitivity in the noiseless case. Second, the more conservative {\bf Case 2}, with $r_0=177.7$ $\upmu$m, $\dot{r}_0=13.4$ nm s$^{-1}$ and $\dot{\theta}_0=259.2$ rad s$^{-1}$, which is less affected by backgrounds.
As it is explained in Sect.~\ref{sec:backgrounds}, in order to study the effects of attractive backgrounds, we consider a modified Newtonian potential in the form of eq.~(\ref{eq:mNpotential1}), 
with $Q_2$, $Q_3$ and $Q_4$ parameterizing the main noise sources (within the assumption that they can be cast in the form of central potentials). 
In Sect.~\ref{sec:backgrounds}, it is shown  that, in order to achieve maximal sensitivity for the setup, the noise sources should be kept within the following bounds: 
$Q_2 < 10^{-2}$, $Q_3 < 5 \times 10^{-2}$ $\upmu$m and $Q_4 < 1$ $\upmu$m$^2$ for {\bf Case 1}, and $Q_2 < 5 \times 10^{-1}$, $Q_3 \leq 0.1$ $\upmu$m and $Q_4 \leq 5$ $\upmu$m$^2$ for {\bf Case 2}.

In this Appendix we analyze some physical noises focusing on the most significant  $Q_3$-like one, the electric Casimir effect between two spheres (note that for two conductive planes, the electric Casimir effect would instead
induce a $Q_4$-like term). In Sect.~\ref{sec:Casimir} the expected Casimir force for our setup will be computed; and in Sect.~\ref{sec:Noises} some other possible noise sources are briefly discussed.

\subsection{Electric Casimir force between spheres}\label{sec:Casimir}

The electric Casimir force is a macroscopic manifestation of quantum field theory. It is related to the vacuum energy of the electromagnetic field in the space between two or more macroscopic bodies as, for example, conducting planes
(the case for which it was first computed in Ref.~\cite{Casimir:1948dh}). However, in a general case, some characteristics of the bodies such as their geometry or their temperature, $T$, play a significant role, producing leading corrections to the simplest case.

In the case of two spheres of radii $R_1$ and $R_2$ with a surface-to-surface distance $d$, several results have been obtained using different methods and limits \cite{Teo:2012,Bulgac:2006}. In the case in which the separation $d$
is much smaller then the two radii, $d\ll R_1, R_2$, and the temperature is finite but low with respect to $1/d$, $T \times d \ll 1$, the Casimir force in (3+1)-dimensions between two ideally conducting (CC) or insulating (II) spheres is:
\begin{equation}
\label{eq:CClowTlowd}
F^{\text{CC, II}}_\text{Cas} =  -\frac{\pi^3}{360} \, \frac{R_\text{red}}{d^3}-  \zeta(3) \, R_\text{red}\,  T^3 + \frac{2 \pi^3 }{45} R_\text{red} \, d \, T^4 + \dots \, ,
\end{equation}
where $R_\text{red} = R_1 R_2/(R_1 + R_2)$ is the reduced radius. On the other hand, for one ideal conductor and one ideal insulator (CI) we have, in the same limits: 
\begin{equation}
\label{eq:CIlowTlowd}
F^{\text{CI}}_\text{Cas} =  \frac{7 \pi^3}{2880} \, \frac{R_\text{red}}{ d^3} + \frac{2\pi^3}{45} R_\text{red} \, d \, T^4 + \dots \, ,
\end{equation}
with the main difference being the change of sign of the leading $1/d^3$ term and the absence of the $T^3$ one. For zero temperature a expression with a wider range of applicability can be extracted from Ref. \cite{Bulgac:2006},
\begin{multline}\label{eq:Caszerotemp}
F^{\text{XY}}_\text{Cas}=\kappa^\text{XY} R_\text{red}(R_1+R_2)\left[\frac{1}{d^3(d+R_1+R_2)}\right. \\ \left.+\frac{1}{2d^2(R_1+R_2+d)^2}\right]+ \dots,
\end{multline}
where $\kappa^\text{CC}=\kappa^\text{II}=-\uppi^3/360$ and $\kappa^\text{CI}=7\uppi^3/2880$\footnote{This coefficient has been obtained by matching the result of Ref.~\cite{Bulgac:2006} to eq. \ref{eq:CIlowTlowd} in the $d\ll R_1, R_2$ limit.}.

On the other hand, for still small separation, but large temperature ($T\times d \gg 1$), we have instead:
\begin{equation}
\label{eq:highTlowd}
F^{\text{XY}}_\text{Cas} = \tilde{\kappa}^\text{XY} \frac{\zeta(3)}{4} \, \frac{R_\text{red}}{d^2} \, T + \dots \, , 
\end{equation}
where $\tilde{\kappa}^\text{CC}=\tilde{\kappa}^\text{II}=-1$ and $\tilde{\kappa}^\text{CI}=3/4$. In this case we can again see that the force  between two conducting or insulating spheres is attractive, whereas it becomes repulsive for two spheres of same nature (notice that, in a realistic case neither a platinum sphere is a perfect conductor, nor pyrolytic graphite is a perfect insulator, and so an intermediate behaviour is expected). A significant difference between low and high $T$ limits is that in the former case $T$-dependence starts at ${\cal O}(T^3)$ and the leading distance
dependence is ${\cal O}(d^{-3})$, whereas in the latter case the leading term is $T$-dependent and goes as ${\cal O} (T/d^2)$.

We can now use these results to compute the magnitude of the electric Casimir effect in the two cases considered in the main text. In {\bf Case 1}  the value of the surface-to-surface separation between S and P in our setup varies between $d_{\rm a} \sim 100 \, \upmu$m at the apoapsis to just $d_{\rm p} = 1.7 \, \upmu$m at the
periapsis in the Newtonian case. Then, as S approaches P we have $d \ll R_{\rm P}, R_{\rm S}$, and we can apply eq.~(\ref{eq:CIlowTlowd}) or eq.~(\ref{eq:highTlowd}), depending on the temperature at which the experiment is developed.
If the temperature of the system\footnote{Remind that even at room temperature we would have $T \times d \lesssim 0.2$ and therefore, low separation formul\ae \, should still apply.}
 is $T=100$ K, we have $T \times d = 0.074$ and are safely within the conditions needed to apply eq.~(\ref{eq:CIlowTlowd}). We find that at the periapsis of the Newtonian orbit, the Newtonian gravitational force is
 $|F_{\text{N,p}}| = 2.36 \times 10^{-22}$ N, whereas at the same point the Casimir force would be as big as $|F^{\rm CI}_{\text{Cas, p}}| = 2.17 \times 10^{-15}$ N to leading order, with higher order corrections modifying it by $\sim20\%$ \cite{Teo:2012}.  This means that  it completely dominates the system at very short distances. Using eq.~(\ref{eq:mNpotential1}), we can compute that the correction to the Newton force at the periapsis 
 corresponds to a $Q_3$-like term with $Q_3 \sim 9000$ $\upmu$m, far beyond the limits established in Sect.~\ref{sec:backgrounds}. 
 
In {\bf Case 2} the system is less affected by this effect. The separation between S and P varies from $d_{\rm a}\sim 200$ $\upmu$m to $d_{\rm p}=52.1$ $\upmu$m in the Newtonian case. However, more strict conditions on the temperature would be required to be in the low $T$ limit. Assuming this has been achieved we can compute $|F_{\text{N,p}}| = 5.90 \times 10^{-23}$ N and $|F^{\rm CI}_{\text{Cas, p}}| = 7.66 \times 10^{-20}$. Again, the Casimir force would be dominant at this point, with the same $Q_3$ as before, since the characteristics of P and S have not been changed.

Nevertheless, some comments are in order: First, we have compared the Newtonian and Casimir forces at the periapsis of the Newtonian orbit. However, when combining both, the repulsive Casimir force would prevent S from approaching that much to P, meaning the movement would be restricted to distances at which both forces are, at most, of comparable magnitude. 

Second, the Casimir force changes from being attractive in the conductor-conductor or insulator-insulator setup to being repulsive in the  conductor-insulator case that corresponds to our proposal. For real
materials that are neither perfect conductors nor ideal insulators, we expect the Casimir force to be at some intermediate point, and so to be somewhat smaller (in absolute value) than what we would get from eqs.~(\ref{eq:CClowTlowd}) 
and (\ref{eq:CIlowTlowd})  in the case of ideal materials. 

Third, several methods to reduce the Casimir force by orders of magnitude have been studied in the literature. For example, we could paint the planet with a 
thin aerogel \cite{Esquivel_Sirvent:2007}  or with an oxide layer \cite{Esquivel_Sirvent:2019}, or cover its surface with a properly designed nanostructure \cite{Intravaia:2013}. The adoption of one of these methods is something that should
be studied in detail, but it is clearly out of the scope of this paper. 

\begin{figure*}[ht]
\centering
    \begin{minipage}{0.5\textwidth}
        \centering
        \includegraphics[width=0.75\textwidth]{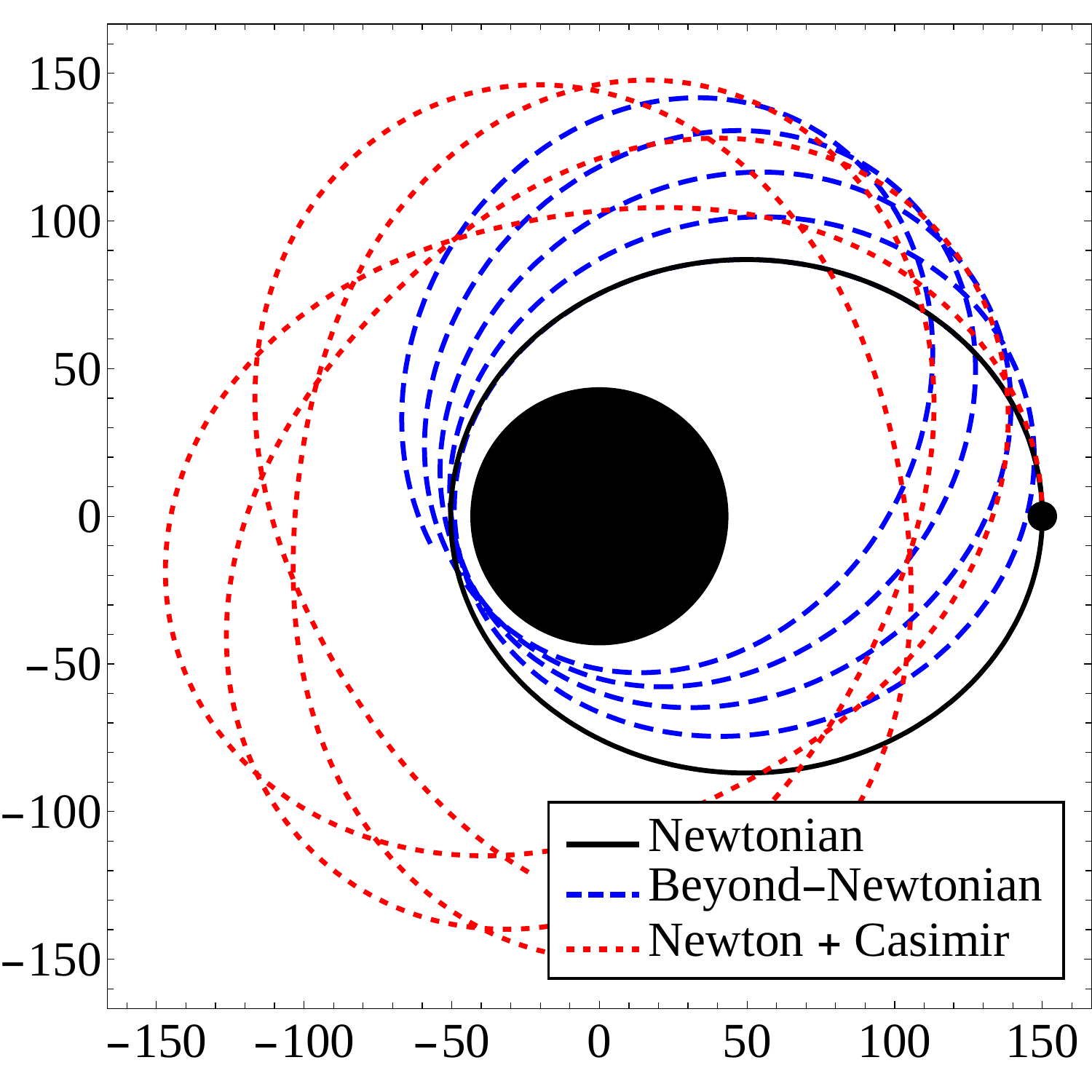} 
    \end{minipage}\hfill
    \begin{minipage}{0.5\textwidth}
        \centering
        \includegraphics[width=0.75\textwidth]{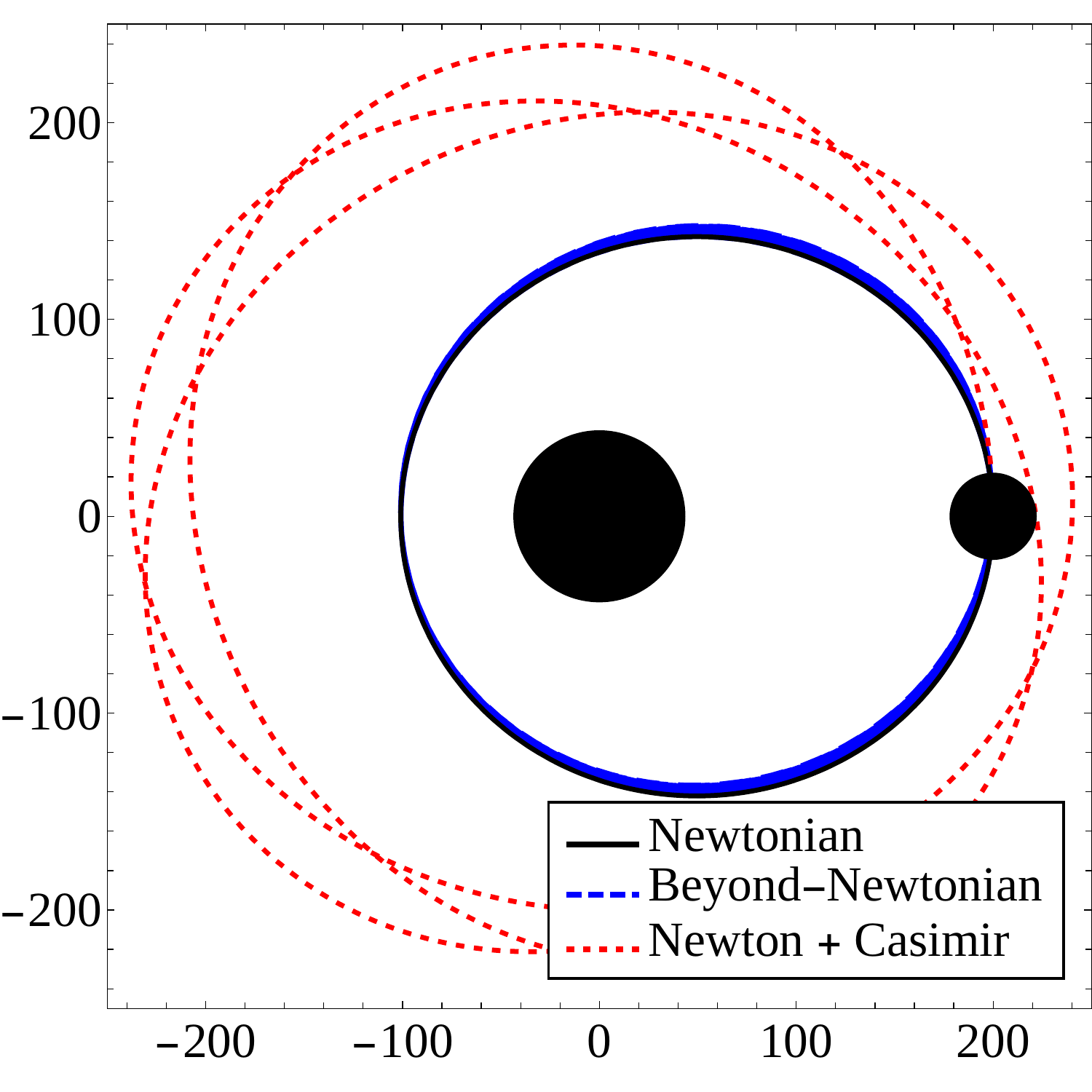} 
    \end{minipage}
\caption{\it Trajectory of \textup{S} around \textup{P} under the effect of Newton's gravitational force (black solid line), Newton's  force plus the electric Casimir force for conductor-insulator spheres (red dotted line) ,
and a Beyond-Newtonian force with $\lambda = 10$ $\upmu$\textup{m} and $\alpha = 2$ (blue dashed line). The Planet and the Satellite are depicted by black dots with $R_\textup{P}=43.7$ $\upmu$\textup{m} and  
$R_\textup{S}=5$ $\upmu$\textup{m} ($R_\textup{S}=22$ $\upmu$\textup{m}) for the left (right) panel, respectively. The initial conditions used are the ones corresponding to {\bf \textit{Case} 1} ({\bf \textit{Case} 2}) as explained in the main text, and the Casimir force has been artificially  reduced by a factor of $10^{-3}$ ($10^{-1}$). All distances are represented in micrometers.
}
\label{fig:Casimir}
\end{figure*}

Last but not least, in {\bf Case 2} a simple modification can be done which would decrease the relative importance of the electric Casimir effect. It is based on incresing the mass of S, for example, from $m_\text{S}=1.2\times 10^{-9}$ g to $m_\text{S}=10^{-7}$ g, what also increases its radius to $R_\text{S}=22.1$ $\upmu$m. Then $d_\text{p}=35.0$ $\upmu$m at the periapsis of a Newtonian orbit, and the Newtonian and Casimir forces would be $|F_{\text{N,p}}| = 4.92 \times 10^{-21}$ N and $|F^{CI}_{\text{Cas, p}}| = 8.17 \times 10^{-19}$ N, respectively. The background factor would  thus be reduced to $Q_3\sim 100$ $\upmu$m. Note that, although being much smaller than before, this value is still above the limits determined in Sect. \ref{sec:backgrounds}.

In Fig.~\ref{fig:Casimir} we draw the expected orbits in the case of a Newtonian force (solid black), of a Newtonian force in presence of the Casimir effect computed from eq. \ref{eq:Caszerotemp} (dotted red), and of the Beyond-Newtonian force of eq.~\ref{eq:YukawaPotential} with  $\lambda = 10$ $\upmu$m and $\alpha = 2$ (dashed blue). In order to obtained closed orbits for the $\text{Newton}+\text{Casimir}$ trajectories, we have multiplied the Casimir force by a reducing factor which we expect may represent the second and third points discussed above. We have considered the two cases introduced in the main text (including the increased $M_\text{S}$ for {\rm Case 2} as discussed above) and computed the relevant forces in the orbit which includes both Newtonian and Casimir effects. Note that these numerical results have been computed without using the mentioned artificial factors: they are only applied for the orbit representations.
\begin{itemize}
\item {\bf Case 1} (Left panel):
Radius at the apoapsis: $r_{\text{a}}=150$ $\upmu$m;  Satellite mass: $m_{\rm S} = 1.2 \times 10^{-9}$  g (corresponding to $R_{\rm S} = 5$ $\upmu$m); 
Casimir force at the apoapsis: $|F_{\rm Cas,a}  |= 4.54 \times 10^{-21}$ N; Casimir force at the periapsis: $|F_{\rm Cas,p} | = 4.04 \times 10^{-20}$ N; 
Newton force at the apoapsis: $|F_{\rm N,a} |= 2.67 \times 10^{-23}$ N; Newton force at the periapsis: $|F_{\rm N,p}| = 5.65 \times 10^{-23}$ N. 
In this case, the distance at the periapsis is $r_{\rm p} = 103$ $\upmu$m, due to the repulsive Casimir force, to be compared with the expected distance $r_{\rm p} \sim 50$ $\upmu$m for a pure Newtonian force. In the figure (but not in the previous computations) a reducing factor of $10^{-3}$ has been used.

\item {\bf Case 2} (Right panel): Radius at the apoapsis: $r_{\text{a}}=200$ $\upmu$m;  Satellite mass: $m_{\rm S} = 10^{-7}$  g (corresponding to $R_{\rm S} = 22$ $\upmu$m); 
Casimir force at the apoapsis: $|F_{\rm Cas,a}  |= 2.43 \times 10^{-21}$ N; Casimir force at the periapsis: $|F_{\rm Cas,p} | = 6.38 \times 10^{-21}$ N; 
Newton force at the apoapsis: $|F_{\rm N,a} |= 8.61 \times 10^{-22}$ N; Newton force at the periapsis: $|F_{\rm N,p} | = 1.25 \times 10^{-21}$ N. 
In this case, the initial position corresponds to the periapsis of the orbit, and the apoapsis lies at $r_{\rm a} = 241$ $\upmu$m. A reducing factor of $10^{-1}$ has been used in the Casimir force to represent the Newton + Casimir orbit.
\end{itemize}
 The platinum Planet and the pyrolytic graphite Satellite are depicted by black dots with their real size, $R_\text{P}=43.7$ $\upmu$m and  $R_\text{S}=5$ $\upmu$m ($R_\text{S}=22$ $\upmu$m) for the left (right) panel, respectively.

It can be seen in Fig.~\ref{fig:Casimir} (left)  that, since Casimir force is repulsive, it induces precession of the trajectory,  
but in the opposite direction to that of the Beyond-Newtonian case: whilst the New Physics signal, for the chosen initial conditions, corresponds to a counter-clockwise precession (for positive $\alpha$), 
the Casimir force induces a clockwise precession that is easily distinguishable from the former. 
Therefore, thanks to the choice of materials for our gravitational system, the impact of the electric Casimir effect on our ability to measure in the ($\lambda$, $\alpha$) plane for positive $\alpha$ is small. The reasons are that it does not produce collision between the P and S, and generally induces a signal which is different to the one expected from New Physics.
 However, extracting a New Physics signal in such situation would still be hard. It would first require a calibration phase that provides a thorough understanding of the Casimir force in the system, and then a new analysis of sensitivity which explicitly includes the this effect in the potential.
The case of negative $\alpha$ would be much more complicated, and would probably require  a new optimization of the initial conditions and of the setup itself, as well as the reduction of the magnitude of the Casimir force by some carefully studied coating or microstructuring.

\subsection{Other sub-dominant noise sources}
\label{sec:Noises}

We discuss here other sub-dominant noise sources. To do so, we make use of the background noise estimates from different sources from Ref.~\cite{westphal:2020}, where a millimetre-sized Cavendish torsion pendulum 
was used to measure the gravitational force between two conducting spheres. We expect that most sub-dominant noise sources will coincide with those studied there. For example, the gravitational attraction on the Satellite produced by elements of the setup different than the Planet, the imperfect sphericity of Satellite and Planet, or errors in the mass measurements. All of these issues are expected to have a negligible effect. 
Also the gravitational effect of other celestial bodies (like the Moon) and corrections from General Relativity (as it was studied in Ref.~\cite{Donini:2016kgu}) should not be significant for our setup. 

Other noise sources are those related to measuring the period of the Satellite. In the cited reference, a camera is used to get a continuous image of the position of the two spheres. In our case, this may be more difficult to perform, due to the smaller size of the setup. For this reason, in this work we have only proposed to measure the time for which the Satellite completes a $2 \pi$-revolution around the Planet. However, having a continuous image of the setup would help to increase the sensitivity, as it would then be possible to fit the orbit to the one predicted by the Beyond-Newtonian model taking multiple benchmark points for comparison.

In addition, in Ref.~\cite{westphal:2020} it was stressed the relevance of performing the measurements in an environment as isolated as possible:  the setup can be easily perturbed by effects as tiny as traffic or seismic episodes happened thousands of kilometres away. In the paper, they performed most of their runs at night during the Christmas period and used a seismometer to be able to detect seismic perturbations. This limited the stability of their experimental 
setup and so a run could last up to around 13 h. We are  not  sure of what could be the impact of these effects on our own proposal, but we expect it to be significantly smaller than that in Ref.~\cite{westphal:2020}, since the size of the relevant part of our setup can be smaller than 1 mm$^2$. In any case, an optimal solution to this issue, which would probably allow for the experiment to stay in stable conditions for much longer periods, would be to perform it in space.

Further possible noise sources are other electromagnetic forces different from the main electrical ones already discussed in Sect.~\ref{sec:backgrounds}. In our setup, we proposed to use a paramagnetic platinum Planet and a 
diamagnetic pyrolytic graphite Satellite. This should make negligible the magnetic dipole attraction between both bodies generated by Earth's magnetic field or by anthropogenic magnetic noise.

\bibliographystyle{h-elsevier}
\bibliography{references}

\end{document}